\documentclass[twocolumn, trackchanges]{aastex63}
\usepackage{amsmath}
\usepackage{graphicx}
\usepackage{grffile}
\usepackage[utf8x]{inputenc}
\usepackage[encapsulated]{CJK}
\usepackage{ucs}
\newcommand{\cntext}[1]{\begin{CJK}{UTF8}{gbsn}#1\end{CJK}}

\newcommand{\mAA}{\mbox{\normalfont\AA}}
\newcommand{\sii}{\rm Si{\textsc{ii}}}
\newcommand{\siistar}{\rm Si{\textsc{ii}}^*}
\newcommand{\oii}{\rm O{\textsc{ii}}}

\newcommand{\feii}{\rm Fe{\textsc{ii}}}
\newcommand{\feiistar}{\rm Fe{\textsc{ii}}^*}

\newcommand{\ha}{{\rm{H}}\alpha}
\newcommand{\hb}{{\rm{H}}\beta}
\newcommand{\lya}{{\rm{Ly}}\alpha}

\newcommand{\sfrir}{{\rm SFR(IR)}}
\newcommand{\sfrha}{{\rm SFR(\ha)}}

\shorttitle{Fluorescence emission-line study of galactic outflows}
\shortauthors{Wang et al.}

\begin{document}

\title{A systematic study of galactic outflows via fluorescence emission:\\implications for their size and structure}

\correspondingauthor{Bingjie Wang}
\email{bwang@jhu.edu}

\author[0000-0001-9269-5046]{Bingjie Wang (\cntext{王冰洁}\!)}
\affiliation{Department of Physics \& Astronomy, Johns Hopkins University, Baltimore, MD 21218, USA}
\author{Timothy M. Heckman}
\affiliation{Department of Physics \& Astronomy, Johns Hopkins University, Baltimore, MD 21218, USA}
\author{Guangtun Zhu}
\affiliation{Department of Physics \& Astronomy, Johns Hopkins University, Baltimore, MD 21218, USA}
\author{Colin A. Norman}
\affiliation{Department of Physics \& Astronomy, Johns Hopkins University, Baltimore, MD 21218, USA}

\received{2020 February 17}
\accepted{2020 April 11}
\submitjournal{\apj}

\begin{abstract}
Galactic outflows play a major role in the evolution of galaxies, but the underlying physical processes are poorly understood. This is mainly because we have little information about the outflow structure, especially on large scales. In this paper, we probe the structure of galactic outflows in low-$z$ starburst by using a combination of ultra-violet spectroscopy and imaging of the fluorescence emission lines (associated with transitions to excited fine-structure levels) and spectroscopy of the corresponding strongly blue-shifted resonance absorption lines. We find that in the majority of cases the observed fluorescence emission lines are much weaker and narrower than the absorption lines, originating in the star-forming interstellar medium and/or the slowest-moving part of the inner outflow. In a minority of cases, the outflowing absorbing material does make a significant contribution to the fluorescence emission. These latter systems are characterized by both strong $\lya$ emission lines and weak low-ionization absorption lines (both known to be empirical signs of Lyman-continuum leakage). We argue that the observed weakness of emission from the outflow seen in the majority of cases is due to the missing emission arising on scales larger than those encompassed by the aperture of the {\it{Hubble Space Telescope}}. This implies shallow radial density profiles in these outflows, and suggests that most of the observed absorbing material must be created/injected at radii much larger than that of the starburst. This has important implications for our understanding of both the physics of galactic outflows and for our estimation of their principal properties.
\end{abstract}


\section{Introduction\label{sec:intro}}

Galactic outflows are invoked as the principal feedback mechanism in models of galaxy formation and evolution. Despite of a general agreement on their importance in regulating the galactic baryonic cycle, outflows are often parametrized in models or simulations by descriptions based on simple theoretical arguments and/or empirical relations. This is because the interplay among the relevant physical processes operates at scales below the resolution of any current simulation (\citealt{Somerville2015} and references therein).

Detailed observations of outflows spanning multiple scales would provide valuable inputs for ``sub-grid" physics in simulations as well as help toward a complete understanding of feedbacks, but they too have been proven to be difficult (see \citealt{Veilleux2005,Heckman2017} for recent reviews). Most of our knowledge has come from data on resonance lines seen in absorption in ``down-the-barrel" spectra, from which characteristic outflow speeds and column densities can be inferred (e.g., \citealt{Heckman2000,Heckman2015,Chisholm2017}). Unfortunately, those data alone provide little information on the radial structure of outflows, since they result from the integrated absorption along the line of sight. This makes it difficult to assess how and where outflows impact the rest of the galaxy's gas supply.

One problem in particular is that the estimation of the outflow rates derived from the absorption lines depend directly on the effective size that is assumed for the absorbing material. That is, simple dimensional analysis implies that the mass outflow rate will be proportional to the column density times the outflow velocity times the outflow size. The first two quantities can be estimated from the absorption-line data, but the outflow size cannot. It is often assumed to be of-order a few times the radius of the starburst (e.g., \citealt{Heckman2015}). Thus, one of the most important missing pieces of data would be an estimate of the size of the region of outflowing absorbing material.

This requires mapping the outflow in emission. While this is commonly done using nebular emission lines like H$\alpha$ (e.g., \citealt{Armus1990}), these nebular lines may not fairly trace the same material seen in absorption (they preferentially trace the highest density gas). Ideally, the outflow could be mapped using the resonance lines in themselves in emission. Many of these resonance transitions also have associated transitions to excited fine-structure levels that can produce fluorescence emission lines (e.g., \citealt{Rubin2011,Jones2012,Erb2012,Tang2014,Finley2017a}).
These fluorescence lines offer a major advantage over the resonance lines for probing the outflow structure. This is because it is not straightforward to disentangle the separate contributions of emission and absorption to the observed properties of the resonance lines. In contrast, the fluorescence lines are always optically-thin, and (in most cases) sufficiently offset in wavelength from the associated resonance lines that the profiles of the fluorescence and resonance lines are unaffected by the other. For this reason, the fluorescence emission lines will be the focus of this paper.

According to standard outflow models \citep{Prochaska2011,Scarlata2015,Zhu2015}, the observed resonance absorption takes place in material located directly along the line-of-sight toward a background light source. For each absorbed resonance photon, a photon will be re-emitted isotropically. For ions with fine-structure splitting, this can either be a resonantly-scattered photon or a fluorescent photon. In the simple case of a spherically-symmetric outflow with no dust, the number of photons is conserved, so that the net absorbed flux should be equal to the sum of the resonantly-scattered and fluorescence emission. In contrast, observations typically show that the emission lines are significantly weaker than the absorption lines (e.g., \citealt{Erb2012,Kornei2013,Tang2014,Alexandroff2015,Finley2017a,Finley2017b,Steidel2018}).
There are a number of ways in which weaker emission can occur. The most straightforward explanation is that the angular size of the absorbing/emitting region significantly exceeds the size of the spectroscopic aperture (``slit-loss"). This has been directly established for the $\lya$ emission line in Lyman Break Galaxies (LBGs) \citep{Steidel2011}. We are particularly interested in testing this possibility more generally, since it would have direct implications for the size and radial structure of outflows. In this paper, we therefore undertake a systematic comparison of the properties of the fluorescence emission lines and resonance absorption lines for a sample of star-forming galaxies driving outflows. We will utilize a combination of information coming from both ultra-violet (UV) spectroscopy and imaging.

The structure of this paper is as follows. In Section~\ref{sec:data} we summarize the observational data sets, including both the new data taken by the {\it{Hubble Space Telescope}} (HST) and archival data, as well as their analysis.
In Section~\ref{sec:res} we present results on the relative strengths and widths of the fluorescence emission lines from the Cosmic Origins Spectrograph (COS; Green et al. (2012)), and discuss imaging of the fluorescence emission from the Wide Field Camera 3 (WFC3). In Section~\ref{sec:discuss} we highlight correlations between line strengths and galaxy/starburst properties, and discuss implications for wind structure. We then conclude our findings in Section~\ref{sec:conclusion}.

All the transitions considered in this paper are listed in Table~\ref{tab:atom}. Also, when applicable, we adopt the best-fit cosmological parameters from the Planck 2018 analysis: $H_{0}=67.66$ ${\rm km \,s^{-1} \,Mpc^{-1}}$, $\Omega_{M}=0.311$, and $\Omega_{\Lambda}=0.690$ \citep{Planck2018}.

\begin{deluxetable}{lccccccc}
\tablecolumns{8}
\tablewidth{0pc}
\tablecaption{Atomic data for the transitions considered\label{tab:atom}}
\tablehead{
    \colhead{} &
    \colhead{$\lambda$} &
    \colhead{$A_{ul}$} &
    \colhead{$f_{lu}$} &
    \colhead{$E_{l}$} &
    \colhead{$E_{u}$}\\
    \colhead{} &
    \colhead{($\mAA$)} &
    \colhead{(s$^{-1}$)}&
    \colhead{}&
    \colhead{(cm$^{-1}$)}&
    \colhead{(cm$^{-1}$)}}
\startdata
  $\sii$ & 1190.42 & 6.53$\times10^8$ & 0.277 & 0 & 84004.26\\
  & 1193.29 & 2.69$\times10^9$ & 0.574 & 0 & 83801.95\\
  $\siistar$ & 1194.50 & 3.45$\times10^9$ & 0.738 & 287.24 & 84004.26\\
   & 1197.39 & 1.40$\times10^9$ & 0.150 & 287.24 & 83801.95\\
  $\sii$ & 1260.42 & 2.57$\times 10^9$ & 1.224 & 0 & 79338.50\\
  $\siistar$ & 1265.00 & 4.73$\times 10^8$ & 0.113 & 287.24 & 79338.50 \\
  $\sii$ & 1304.37 & 3.64$\times 10^8$ & 0.093 & 0 & 76665.35 \\
  $\siistar$ & 1309.28 & 6.23$\times 10^8$ & 0.080 & 287.24 & 76665.35 \\
  $\sii$ & 1526.72 & 3.81$\times 10^8$ & 0.133 & 0 & 65500.47\\
  $\siistar$ & 1533.45 & 7.52$\times 10^8$ & 0.134 & 287.24 & 65500.47 \\
  \hline
  $\feii$ & 2586.65 & 8.94$\times 10^7$ & 0.072 & 0 & 38660.05\\
  & 2600.17 & 2.35$\times 10^8$ & 0.239 & 0 & 38458.99\\
  $\feiistar$ & 2612.65 & 1.20$\times 10^8$ & 0.122 & 384.7872 & 38660.05 \\
  & 2626.45 & 3.52$\times 10^7$ & 0.046 & 384.7872 & 38458.99
\enddata
\tablecomments{Data are taken from the NIST Atomic Spectra Database.}
\end{deluxetable}

\section{Data\label{sec:data}}

\subsection{Sample selection}

In the HST program GO-15340 (PI T. Heckman) we observed a sample of five galaxies with COS and WFC3.
They were selected from SDSS based on the following criteria.
\begin{enumerate}
	\item 0.064 \textless\ $z$ \textless\ 0.066. This range is determined by the transmission curve of F280N, which covers $\feiistar$ 2626 at 0.057 \textless\ $z$ \textless\ 0.072. A narrower redshift range is chosen to ensure the coverage of the bluer $\feiistar$ 2613 line, while not to include the $\feii$ 2600 absorption.
	\item Specific star-formation rate (sSFR) greater than $10^{-9} {\rm yr^{-1}}$ and equivalent width (EW) of $[\oii]$ greater than 50 \AA. Together they indicate substantial starbursts and therefore large outflow rate and extended fluorescence emission.
	\item NUV magnitude less than 19 mag. This ensures the feasibility for COS G130M observations.
	\item Diverse morphology. About 50 galaxies that pass the above criteria are categorized based on the optical morphology, including inclination. The galaxy with the highest NUV flux and/or $[\oii]$ EW is selected in each group.
	\item An estimated NUV flux of the star-bursting region inside a 3\arcsec\ aperture less than 19.5 mag. This is inferred by assuming the difference between the total magnitude and the aperture magnitude in the NUV to be about the same as that in the SDSS $u$-band.
\end{enumerate}
The resulting sample is listed in Table~\ref{tab:obs}.

\begin{deluxetable*}{clccc}
\tablecaption{Observation logs\label{tab:obs}}
\tablecolumns{5}
\tablewidth{0pt}
\tablehead{
\colhead{Name} &
\colhead{Galaxy} &
\colhead{$z$} &
\colhead{WFC3 filter/COS grating} &
\colhead{Exposure time}
\\
\colhead{} &
\colhead{} &
\colhead{} &
\colhead{(Central wavelength)} &
\colhead{(s)}
}
\startdata
J0831(S) & GALEX-J083101.8+040317 & 0.065 & F280N (2832.2) & 1560+1560+2400\\
         &                        &       & F343N (3435.3) & 720\\
         &                        &       & F395N (3955.2) & 720\\
         &                        &       & G130M (1291)   & 4928.384\\
         &                        &       & MIRRORA        & 270\\
\hline
J0831(N) & GALEX-J083101.8+040318 & 0.064 & G130M (1291)  & 4928.352\\
         &                        &       & MIRRORA        & 270\\
\hline
J1157    & GALEX-J115747.0+583503 & 0.064 & F280N (2832.2) & 1840+1760+2640\\
         &                        &       & F343N (3435.3) & 720\\
         &                        &       & F395N (3955.2) & 720\\
         &                        &       & G130M (1291)   & 5425.344\\
         &                        &       & MIRRORA        & 270\\
\hline
J1210    & GALEX-J121014.3+443958 & 0.065 & F280N (2832.2) & 1600+1600+2520\\
         &                        &       & F343N (3435.3) & 720\\
         &                        &       & F395N (3955.2) & 720\\
         &                        &       & G130M (1291)   & 2060.384\\
         &                        &       & MIRRORA        & 170\\
\hline
J1618    & GALEX-J161832.6+274352 & 0.065 & F280N (2832.2) & 1600+1560+2400\\
         &                        &       & F343N (3435.3) & 720\\
         &                        &       & F395N (3955.2) & 720\\
         &                        &       & G130M (1291)   & 2060.320\\
         &                        &       & MIRRORA        & 150\\
\enddata
\end{deluxetable*}

\subsection{Archival data}

In addition to the five new observations listed above, we also include the following four samples from the literature.

First, we analyze the individual spectra of the Lyman break analogs (LBAs) which were previously investigated by \cite{Heckman2011a,Alexandroff2015} (HST-GO-11727 and HST-GO-13017; PI T. Heckman).
Second, we consider the $\lya$-emitting galaxies (LAEs) as compiled in \cite{Scarlata2015}.
Lastly, we compare to two samples of high-$z$ ($z \gtrsim 3$) galaxies: the LBGs \citep{Jones2012} and star-forming galaxies in the Keck Lyman Continuum Spectroscopic Survey (KLCS; \citealt{Steidel2018}).

\subsection{Data processing and analysis}

\subsubsection{Spectra\label{sec:data_spec}}

\begin{figure*}
\gridline{
\fig{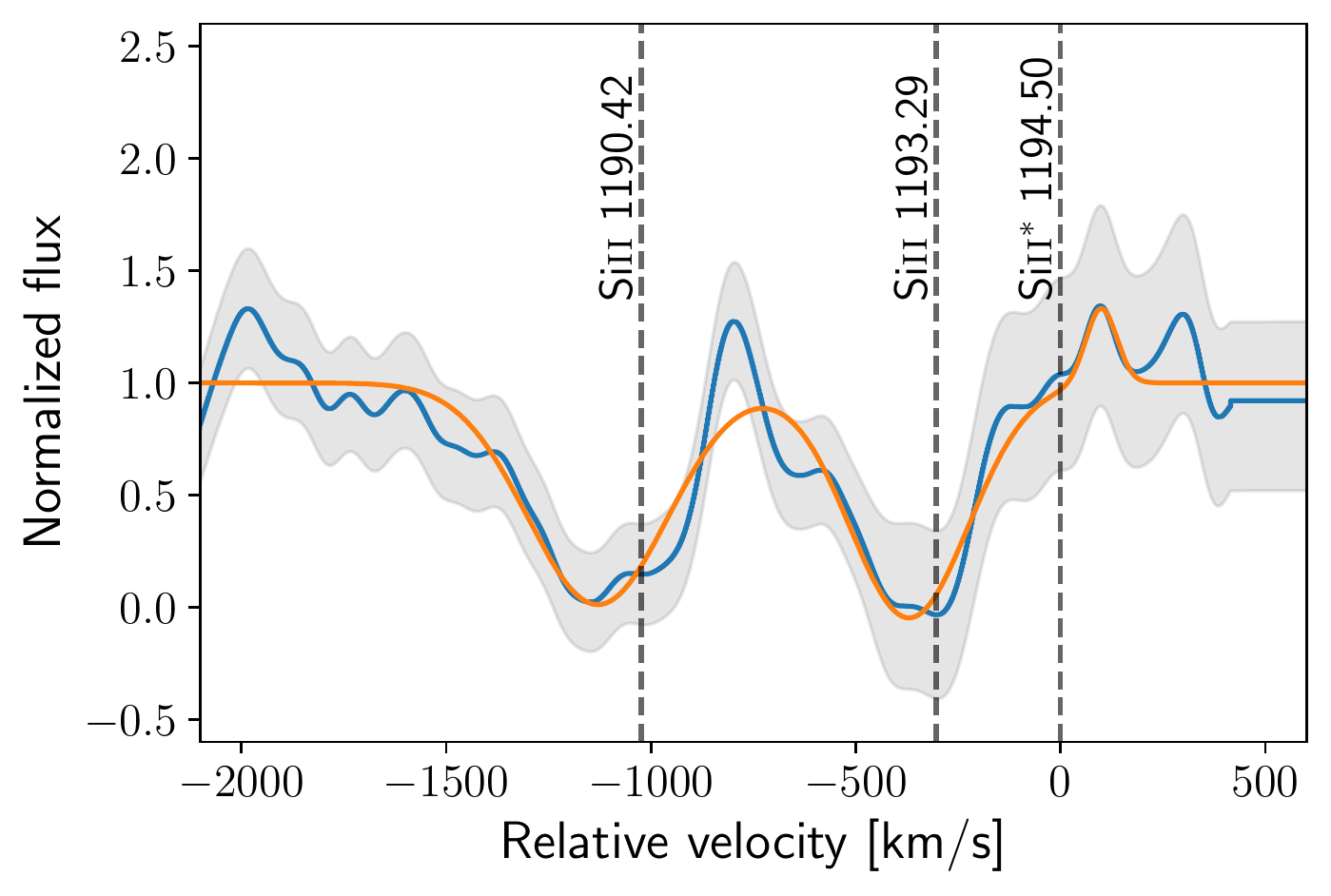}{0.32\textwidth}{}
\fig{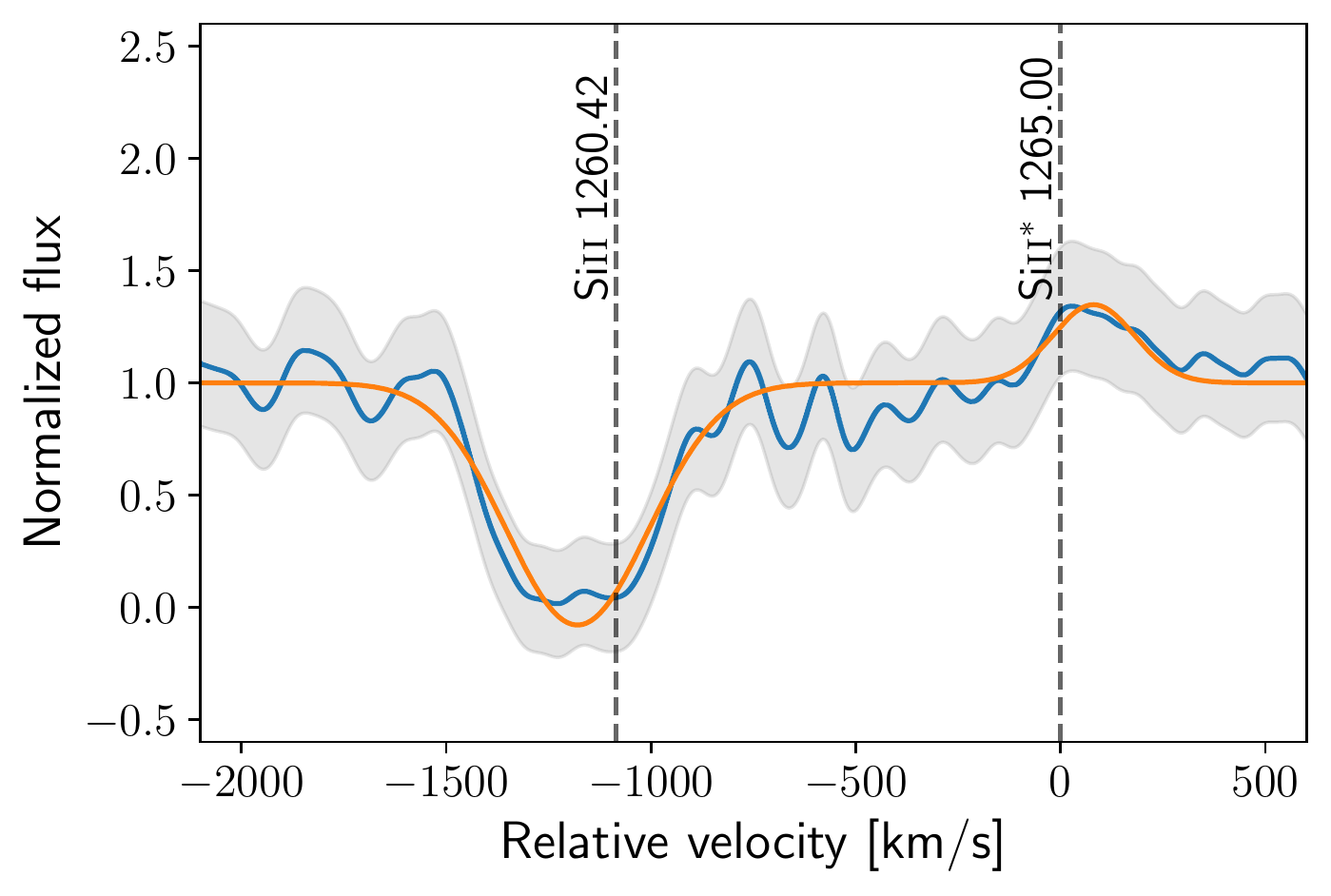}{0.32\textwidth}{}
\fig{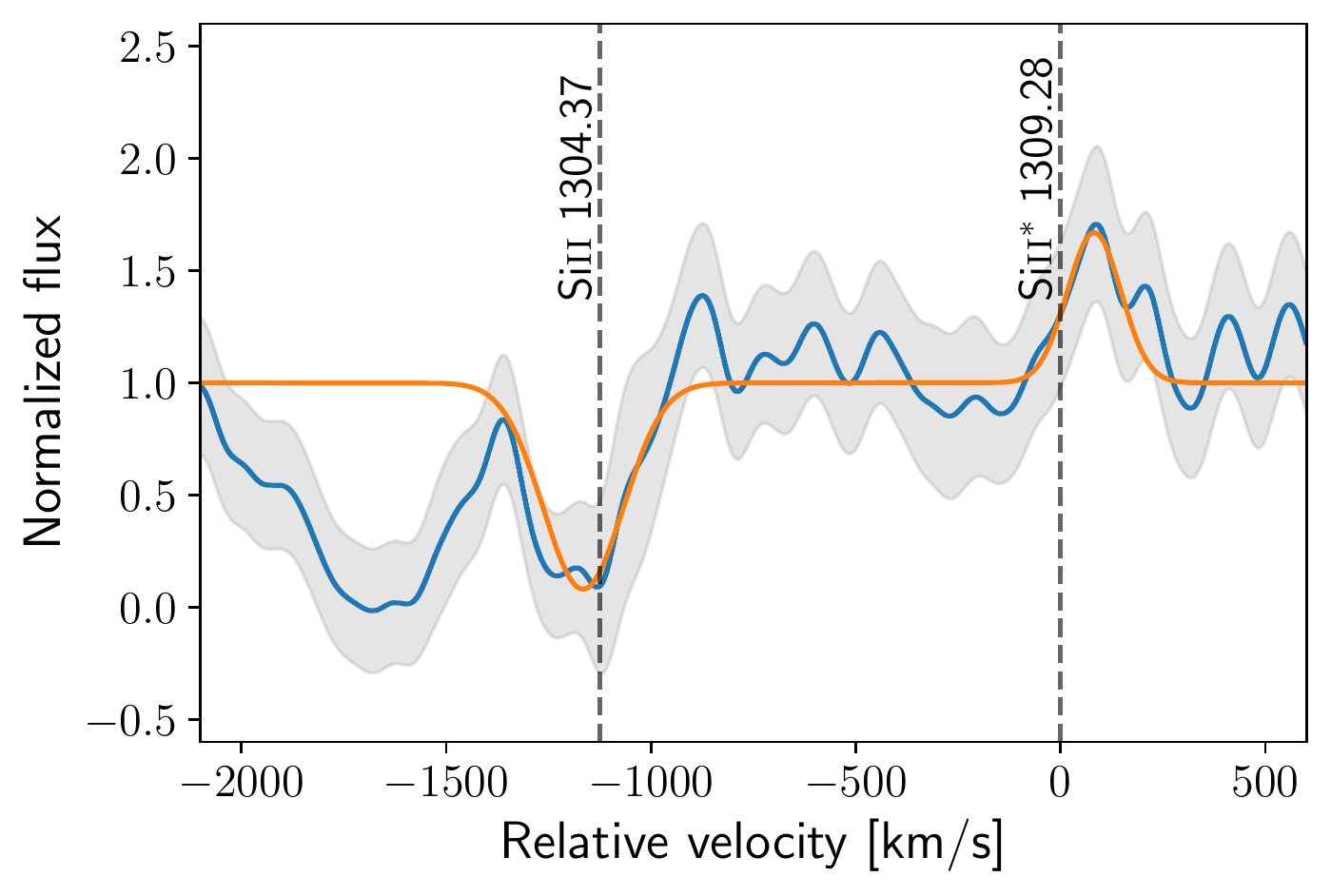}{0.32\textwidth}{}
}
\caption{Continuum-normalized spectra with the stellar features removed, zooming in on lines of interest. The gray shades indicate 1$\sigma$ uncertainty. Over-plotted in orange are the gaussian fits. Each velocity scale is relative to the vacuum wavelength of the fluorescence emission line. The three segments of J0831(S) are shown as examples of the figure set. The complete figure set (15 images) is available in the online journal.\label{fig:spec_lines}}
\end{figure*}

A COS FUV spectrum usually consists of two segments. In our case, one segment covers the observed wavelength range approximately from 1286 \AA\ to 1429 \AA\, while another one covers from 1131 \AA\ to 1274 \AA. We first convert both segments to the rest frame of the galaxy, and then merge and smooth them with gaussian kernels of $\sigma\sim 0.14$ \AA. An average is taken for any overlapping parts, weighted by the respective inverse uncertainties.
We then remove the spectral features produced by stars using synthetic spectra generated from Starburst99 (hereafter SB99; \citealt{Leitherer1999}) in order to focus on the properties of the interstellar gas. The models are produced based on a star formation history of a continuous and constant rate of star formation. The stellar population, parameterized by a Kroupa initial mass function (IMF; \citealt{Kroupa2001}), evolves from the zero-age main sequence using the evolutionary models of the Geneva Group. Both spectra are normalized to some local continuum near the spectral lines of interest.
Those spectra are shown in three segments in Figure~\ref{fig:spec_lines}, each zooming in on: (1) $\sii$ 1190, $\sii$ 1193, $\siistar$ 1195; (2) $\sii$ 1260, $\siistar$ 1265; and (3) $\sii$ 1304, $\siistar$ 1309.

We fit the spectral lines with gaussians, or multi-component gaussians in case of blended lines. We then measure EW, full width at half maximum (FWHM), and velocity centroid ($V_{\rm ctr}$). Since stellar features including the stellar continuum are removed, the average continuum level in the processed spectra lies at zero. For the purpose of defining the EW, we shift each spectrum along the y-axis to have the continuum lie at unit flux.
Unless otherwise stated all measurements can be assumed to have errors on the order of 10-15\% dominated by systematics in the polynomial fit to the continuum and subtraction of the SB99 models.
We provide all the measurements in tabular form in the Appendix.

Unfortunately, at $z \sim 0.065$ the $\siistar$ 1309 line is likely to be contaminated by the Milky Way (MW) absorption line $\rm{Si\textsc{iv}}$ 1394. We check this by looking for another MW absorption line $\rm{Si\textsc{iv}}$ 1402.8.
The $\siistar$ 1309 lines in the spectra of J0831(N), J1157, and J1618 display various degrees of contamination.

\figsetstart
\figsetnum{1}
\figsettitle{Continuum-normalized spectra with the stellar features removed}

\figsetgrpstart
\figsetgrpnum{1.1}
\figsetgrptitle{J0831(S)}
\figsetplot{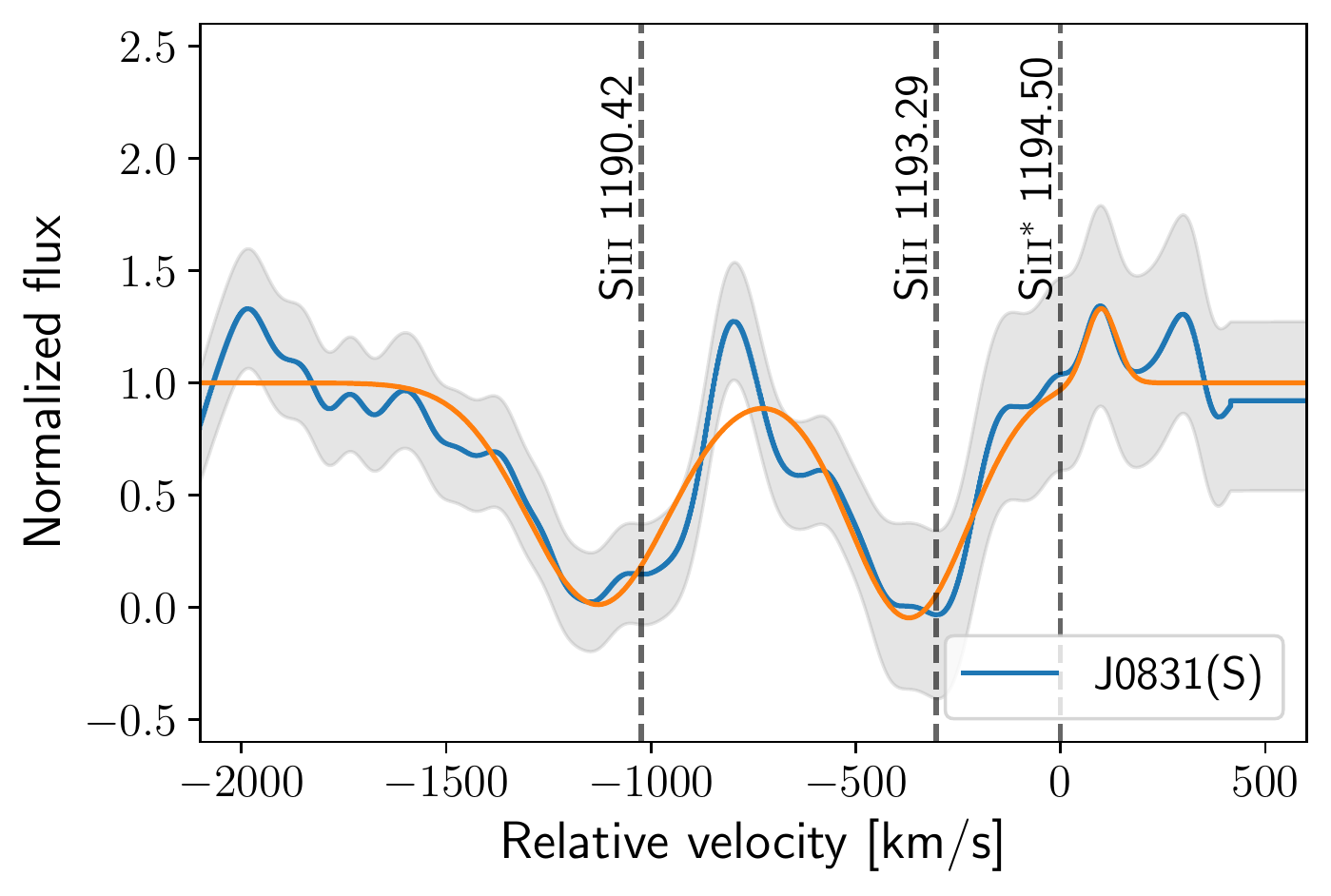}
\figsetgrpnote{Continuum-normalized spectra with the stellar features removed, zooming in on lines of interest. Over-plotted in orange are the gaussian fits.}
\figsetgrpend

\figsetgrpstart
\figsetgrpnum{1.2}
\figsetgrptitle{J0831(S)}
\figsetplot{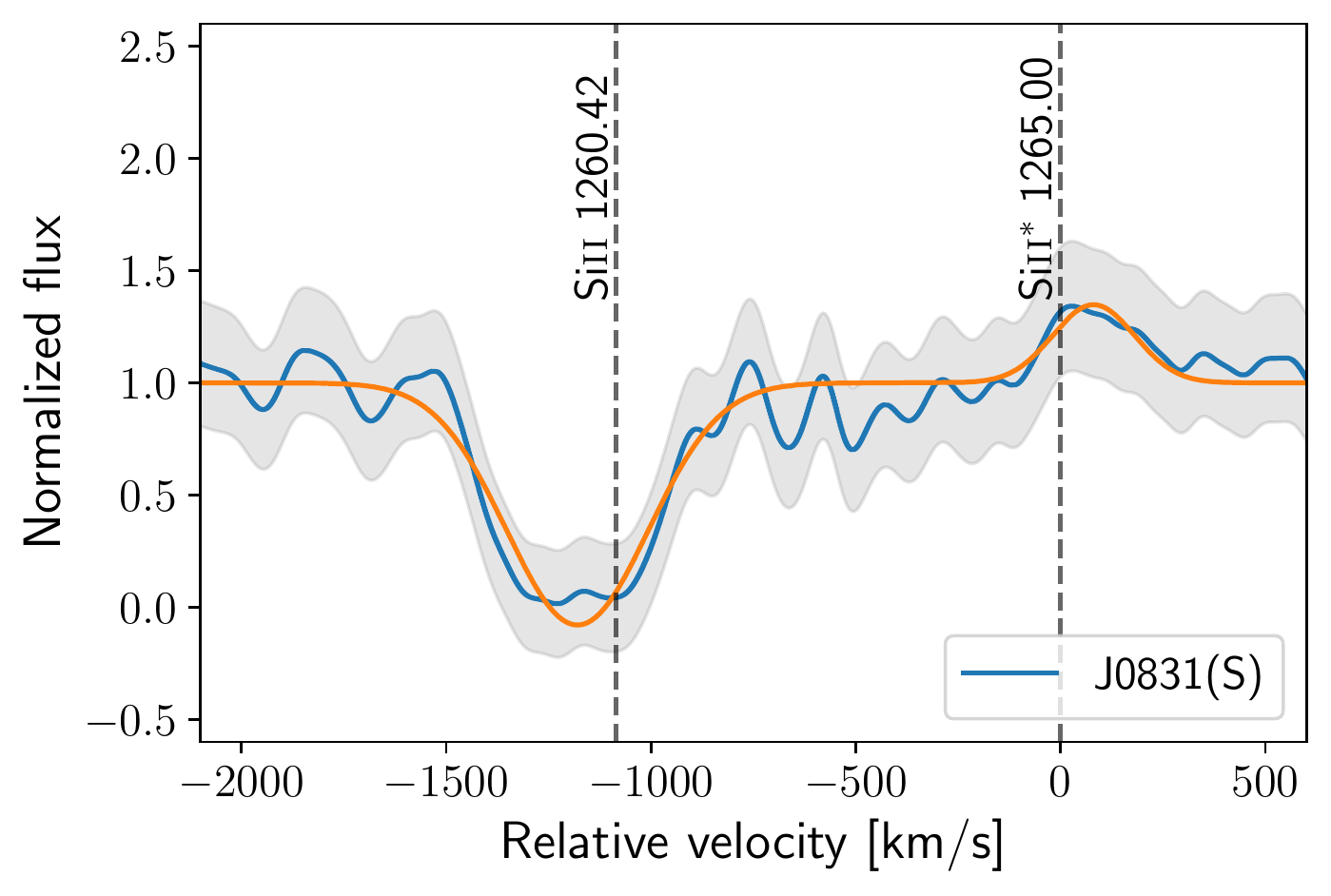}
\figsetgrpnote{Continuum-normalized spectra with the stellar features removed, zooming in on lines of interest. Over-plotted in orange are the gaussian fits.}
\figsetgrpend

\figsetgrpstart
\figsetgrpnum{1.3}
\figsetgrptitle{J0831(S)}
\figsetplot{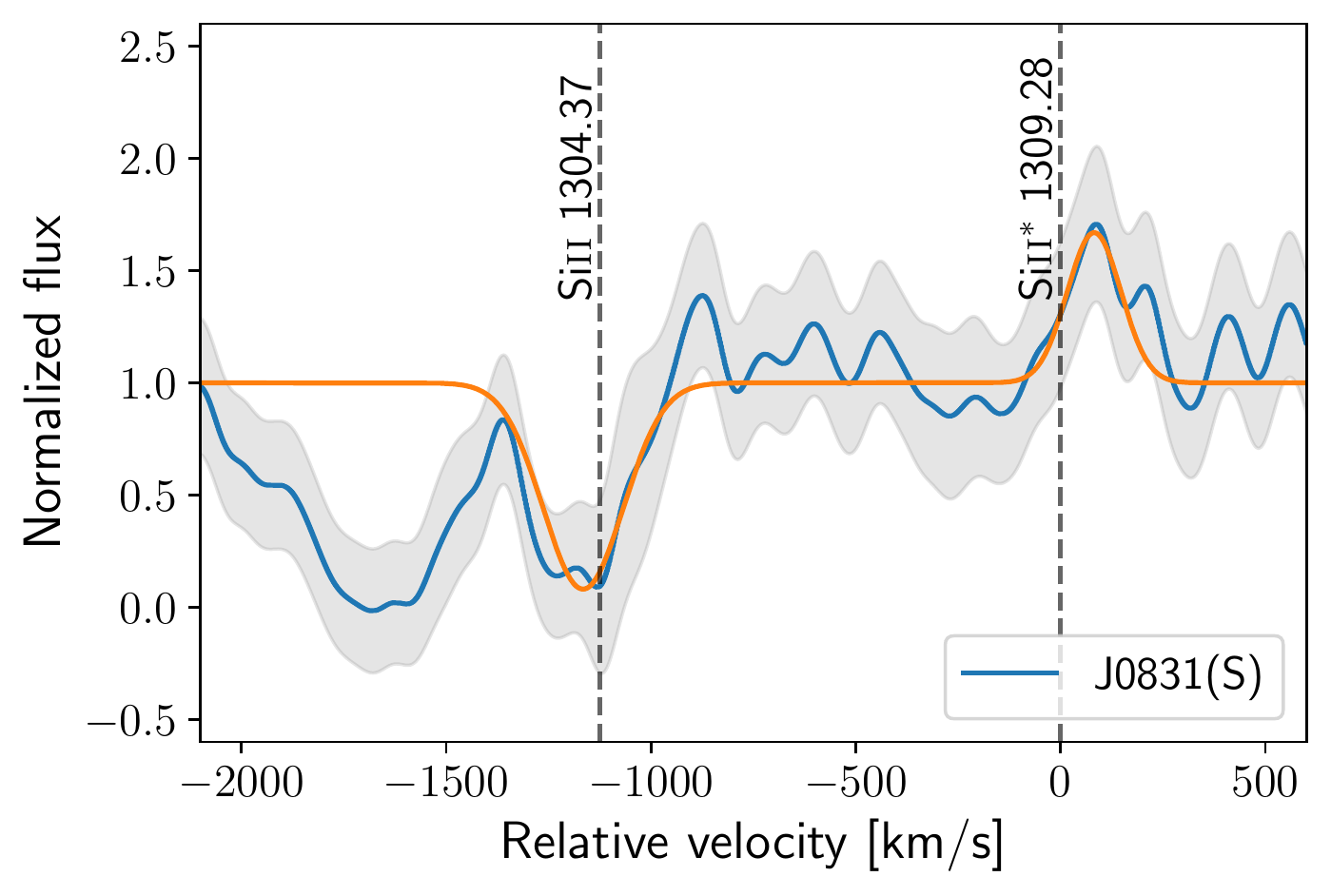}
\figsetgrpnote{Continuum-normalized spectra with the stellar features removed, zooming in on lines of interest. Over-plotted in orange are the gaussian fits.}
\figsetgrpend

\figsetgrpstart
\figsetgrpnum{1.4}
\figsetgrptitle{J0831(N)}
\figsetplot{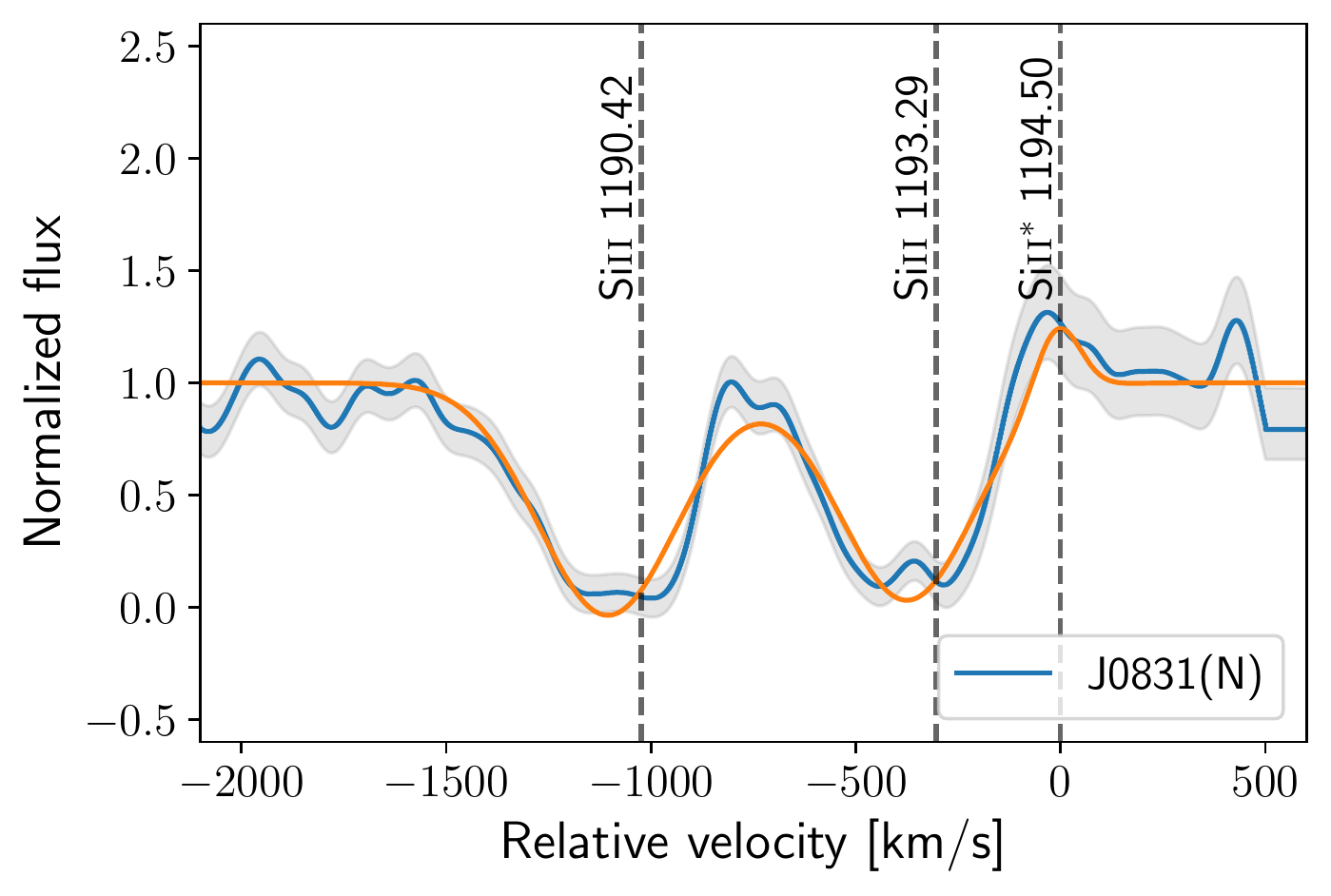}
\figsetgrpnote{Continuum-normalized spectra with the stellar features removed, zooming in on lines of interest. Over-plotted in orange are the gaussian fits.}
\figsetgrpend

\figsetgrpstart
\figsetgrpnum{1.5}
\figsetgrptitle{J0831(N)}
\figsetplot{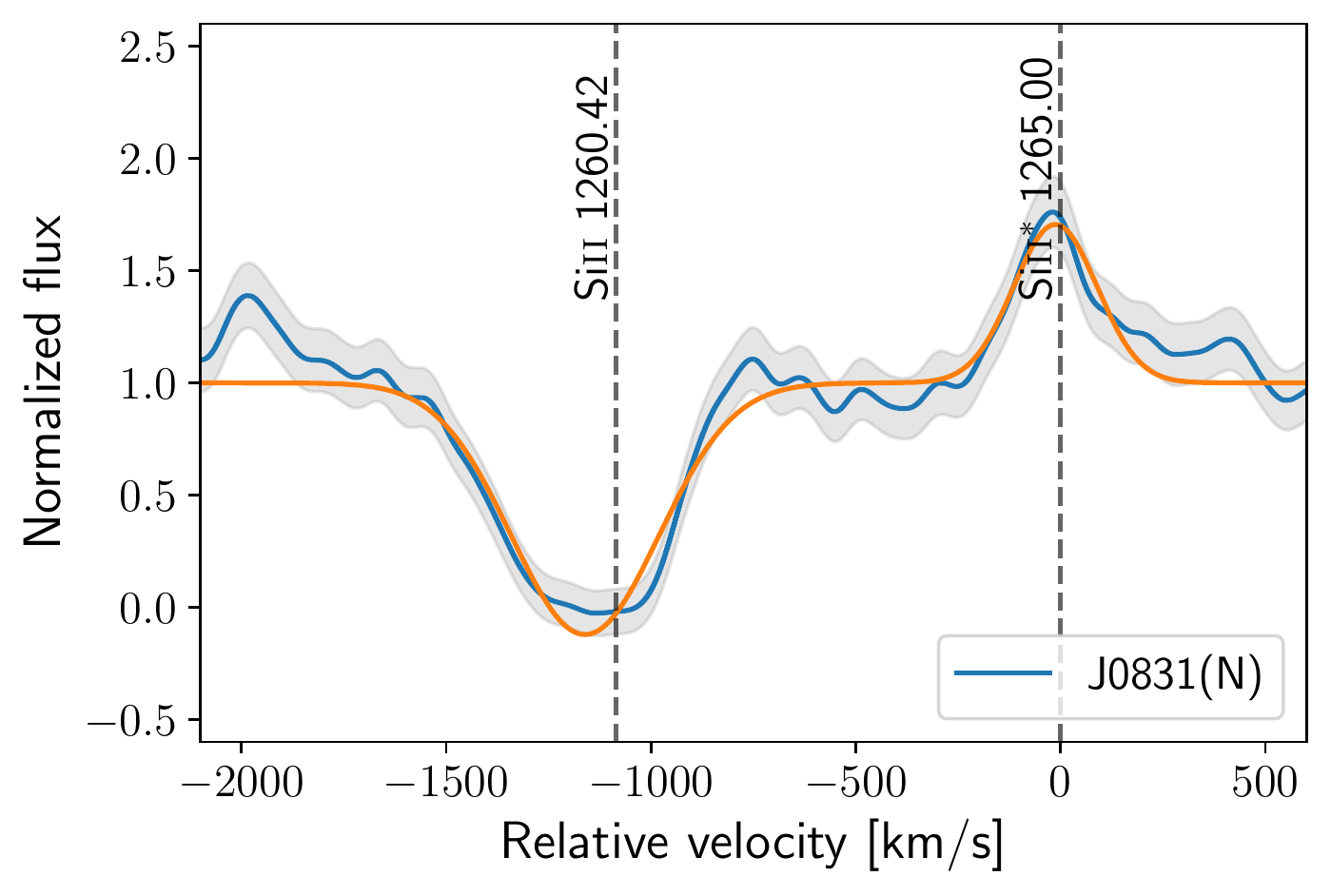}
\figsetgrpnote{Continuum-normalized spectra with the stellar features removed, zooming in on lines of interest. Over-plotted in orange are the gaussian fits.}
\figsetgrpend

\figsetgrpstart
\figsetgrpnum{1.6}
\figsetgrptitle{J0831(N)}
\figsetplot{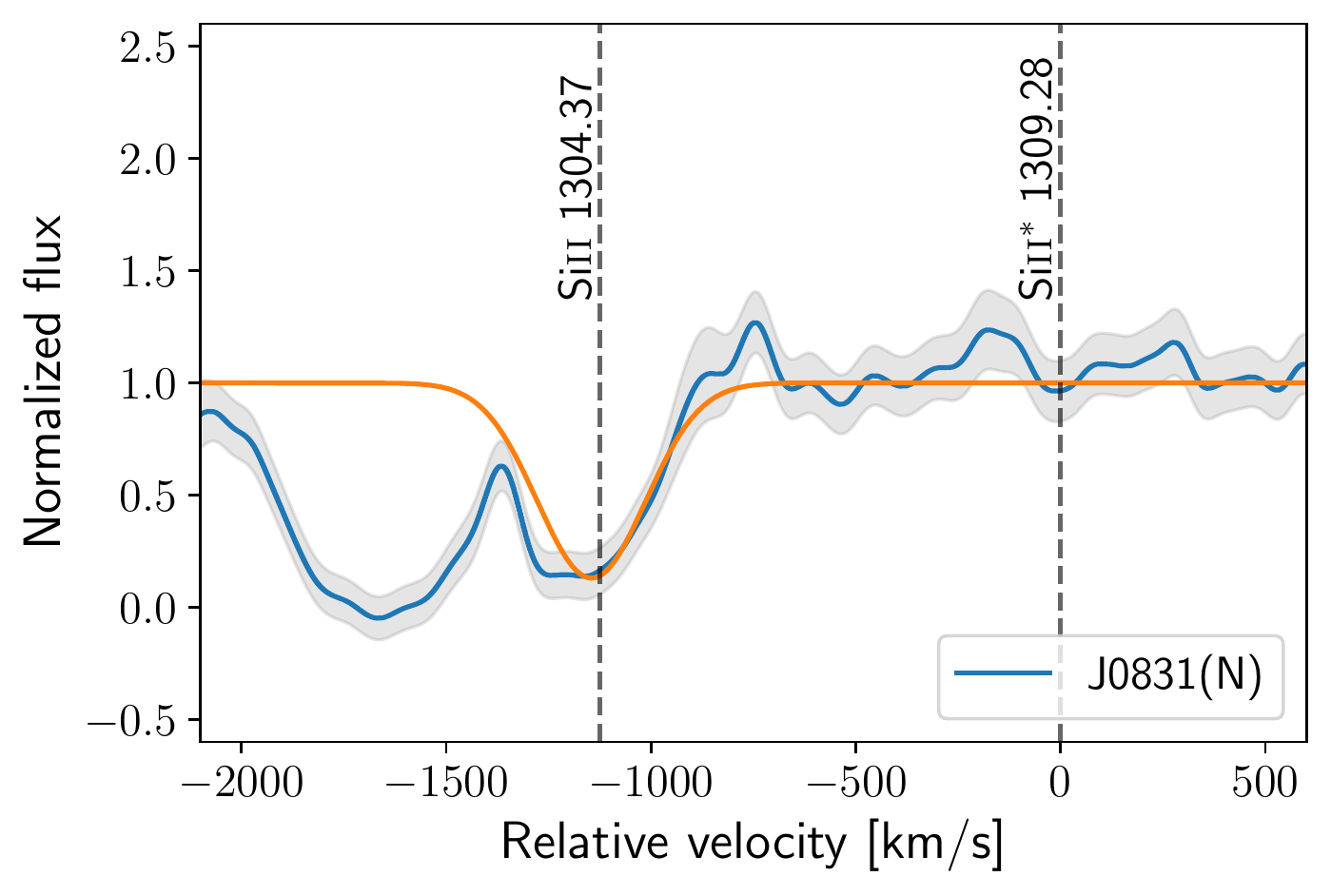}
\figsetgrpnote{Continuum-normalized spectra with the stellar features removed, zooming in on lines of interest. Over-plotted in orange are the gaussian fits.}
\figsetgrpend

\figsetgrpstart
\figsetgrpnum{1.7}
\figsetgrptitle{J1157}
\figsetplot{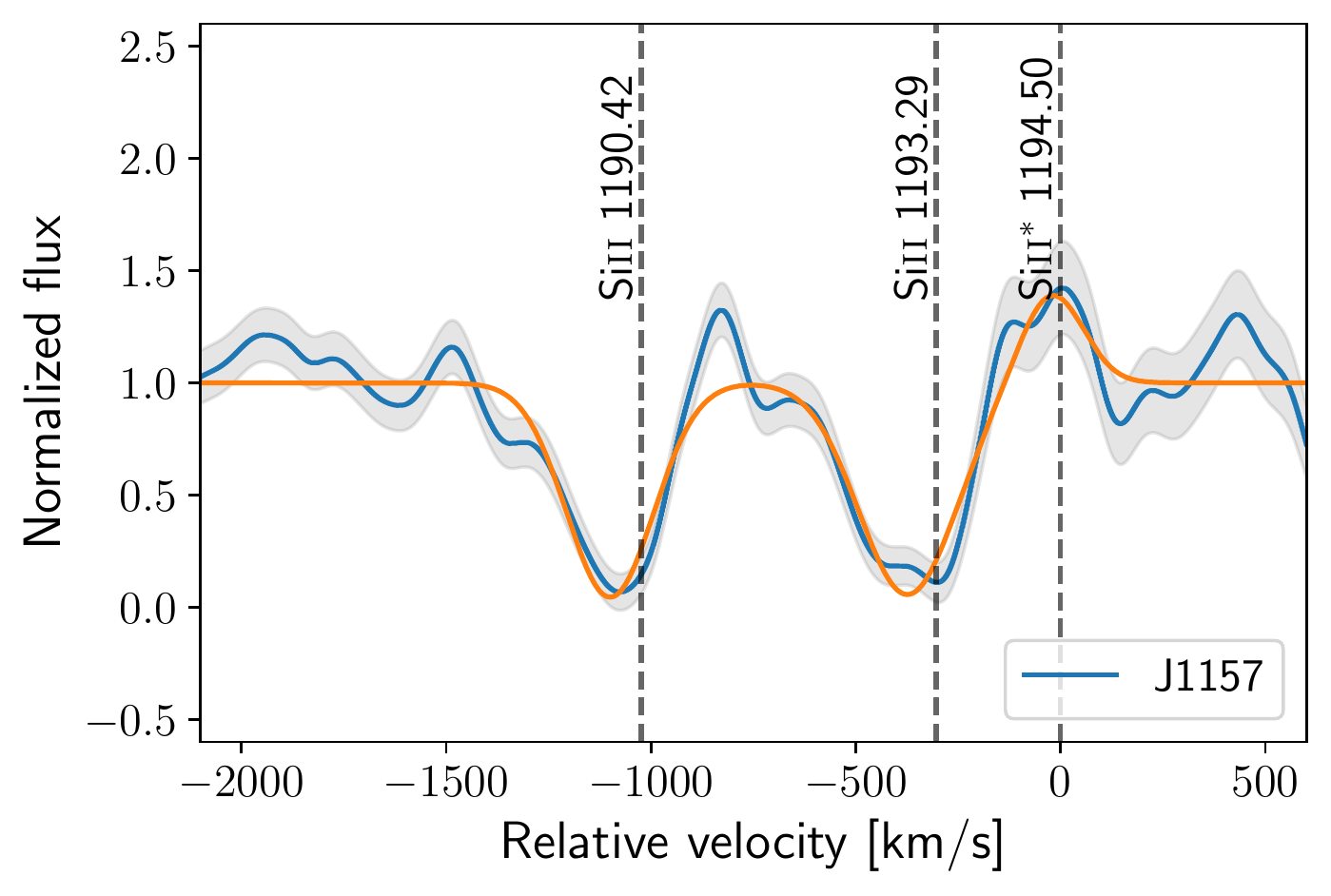}
\figsetgrpnote{Continuum-normalized spectra with the stellar features removed, zooming in on lines of interest. Over-plotted in orange are the gaussian fits.}
\figsetgrpend

\figsetgrpstart
\figsetgrpnum{1.8}
\figsetgrptitle{J1157}
\figsetplot{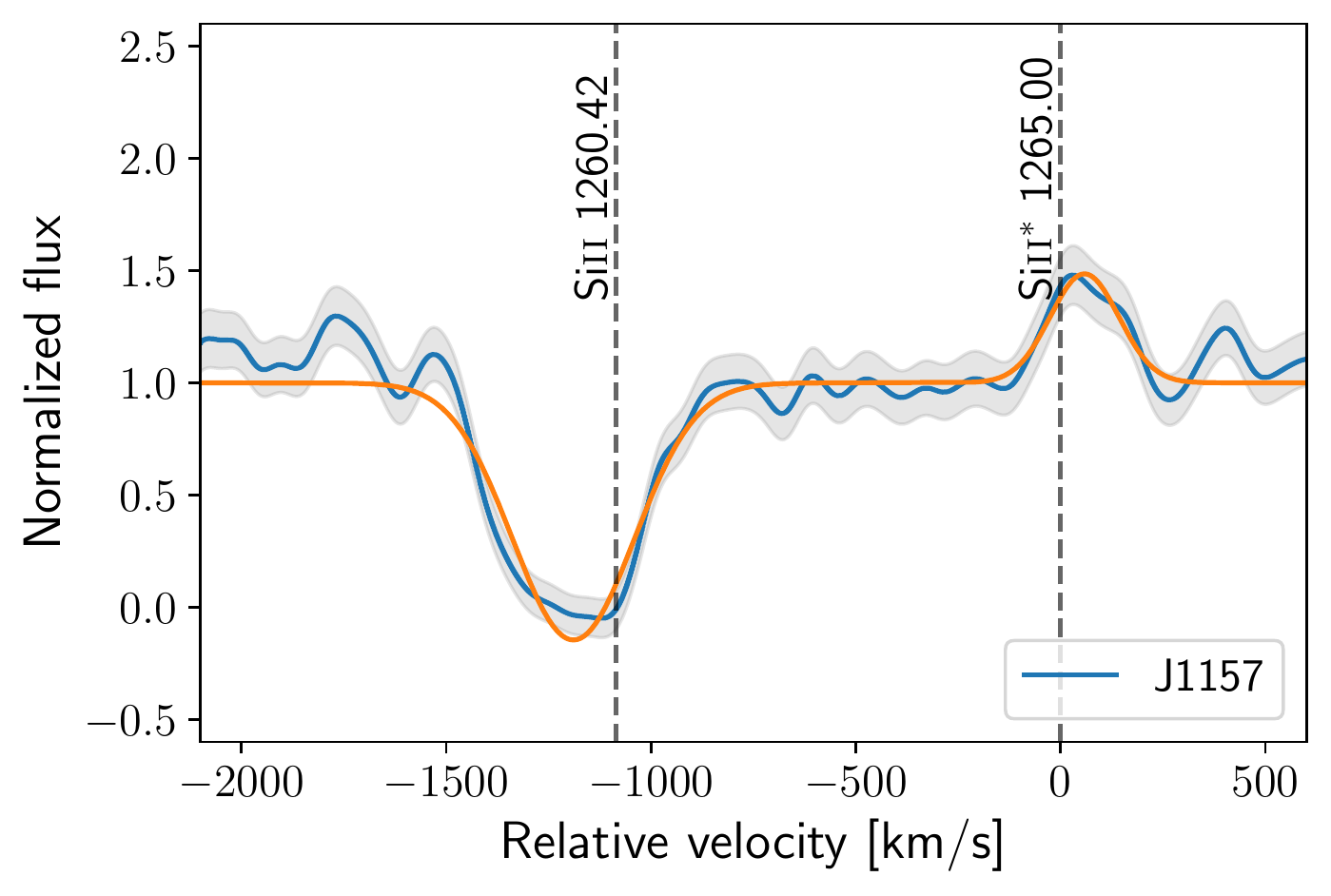}
\figsetgrpnote{Continuum-normalized spectra with the stellar features removed, zooming in on lines of interest. Over-plotted in orange are the gaussian fits.}
\figsetgrpend

\figsetgrpstart
\figsetgrpnum{1.9}
\figsetgrptitle{J1157}
\figsetplot{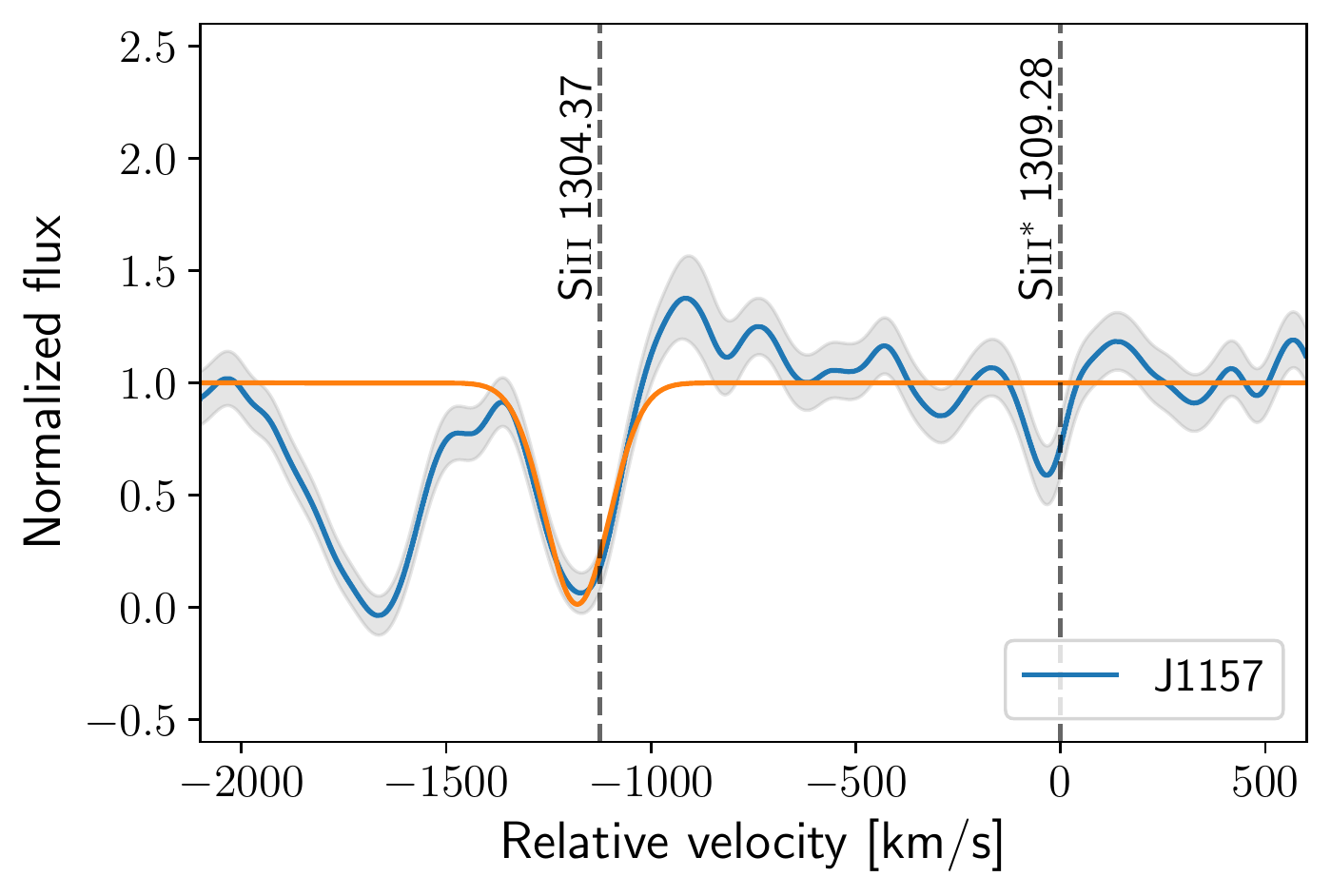}
\figsetgrpnote{Continuum-normalized spectra with the stellar features removed, zooming in on lines of interest. Over-plotted in orange are the gaussian fits.}
\figsetgrpend

\figsetgrpstart
\figsetgrpnum{1.10}
\figsetgrptitle{J1210}
\figsetplot{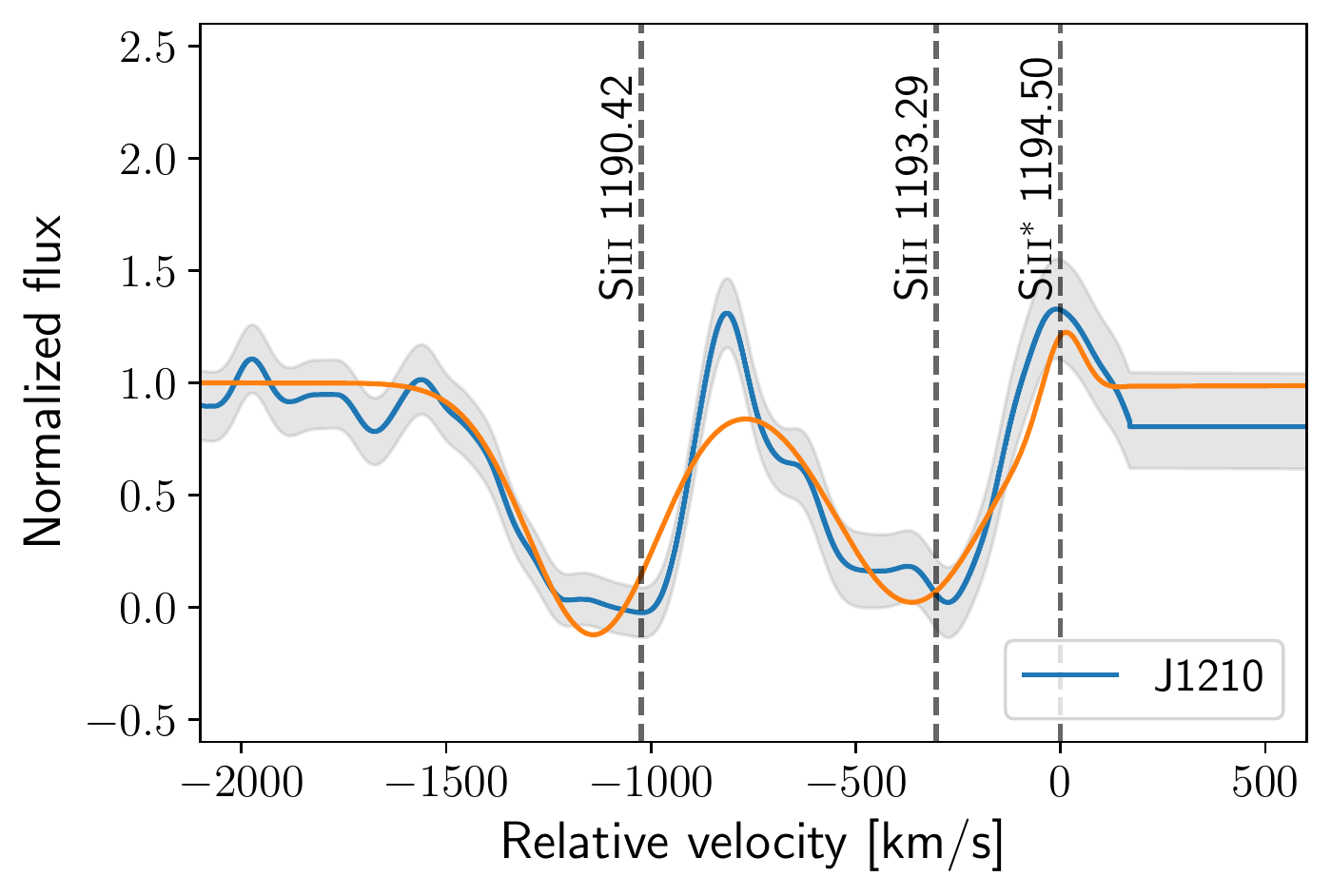}
\figsetgrpnote{Continuum-normalized spectra with the stellar features removed, zooming in on lines of interest. Over-plotted in orange are the gaussian fits.}
\figsetgrpend

\figsetgrpstart
\figsetgrpnum{1.11}
\figsetgrptitle{J1210}
\figsetplot{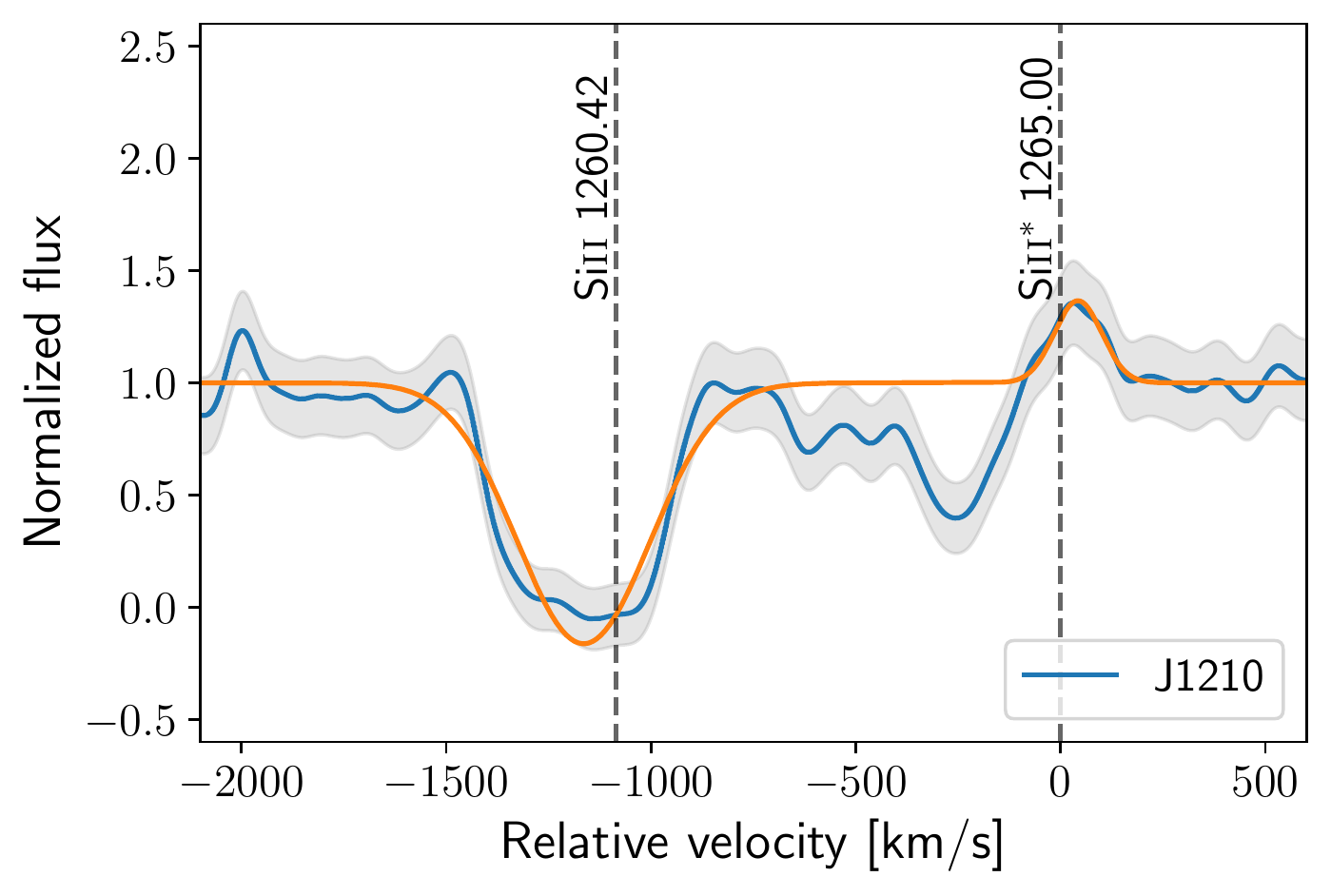}
\figsetgrpnote{Continuum-normalized spectra with the stellar features removed, zooming in on lines of interest. Over-plotted in orange are the gaussian fits.}
\figsetgrpend

\figsetgrpstart
\figsetgrpnum{1.12}
\figsetgrptitle{J1210}
\figsetplot{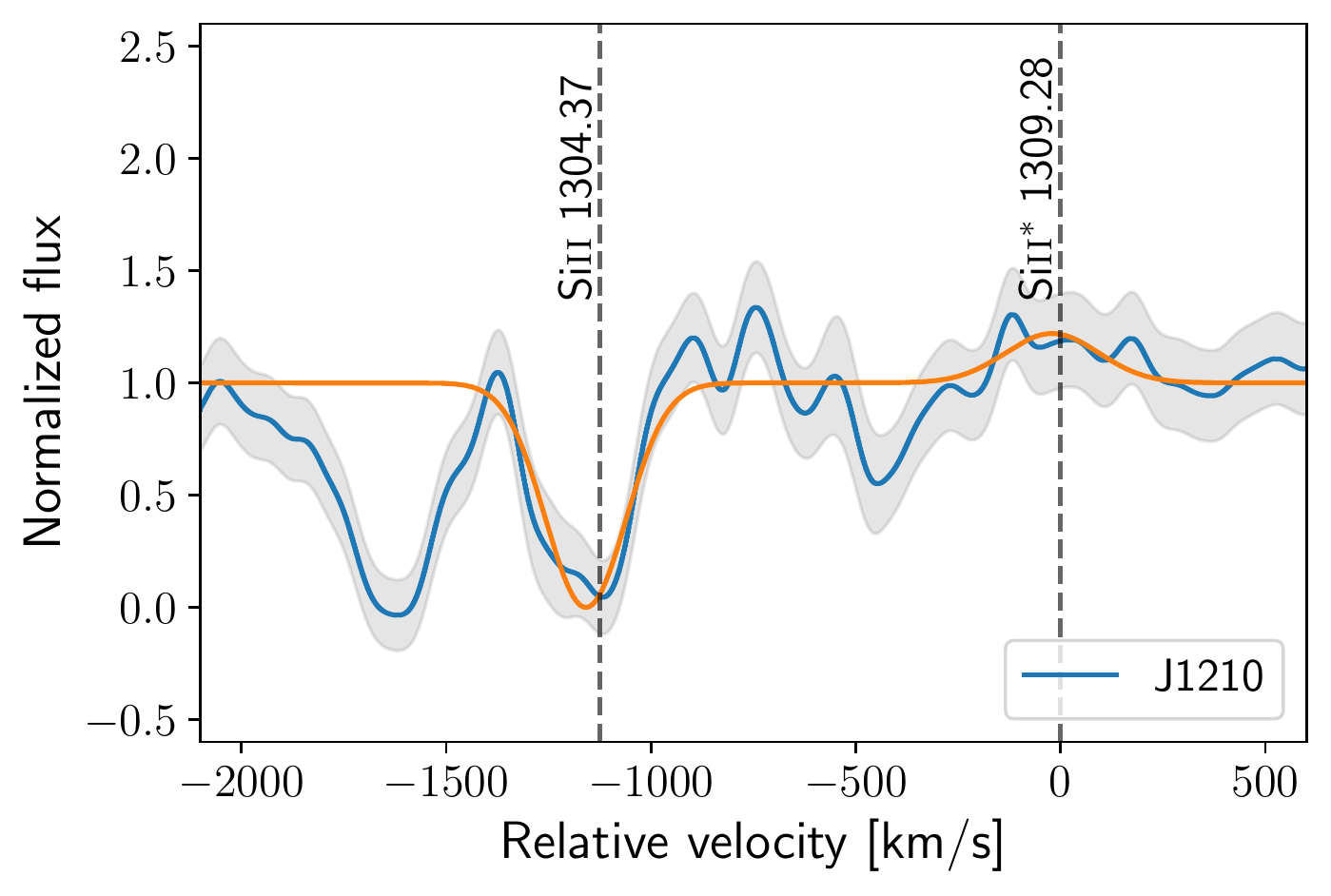}
\figsetgrpnote{Continuum-normalized spectra with the stellar features removed, zooming in on lines of interest. Over-plotted in orange are the gaussian fits.}
\figsetgrpend

\figsetgrpstart
\figsetgrpnum{1.13}
\figsetgrptitle{J1618}
\figsetplot{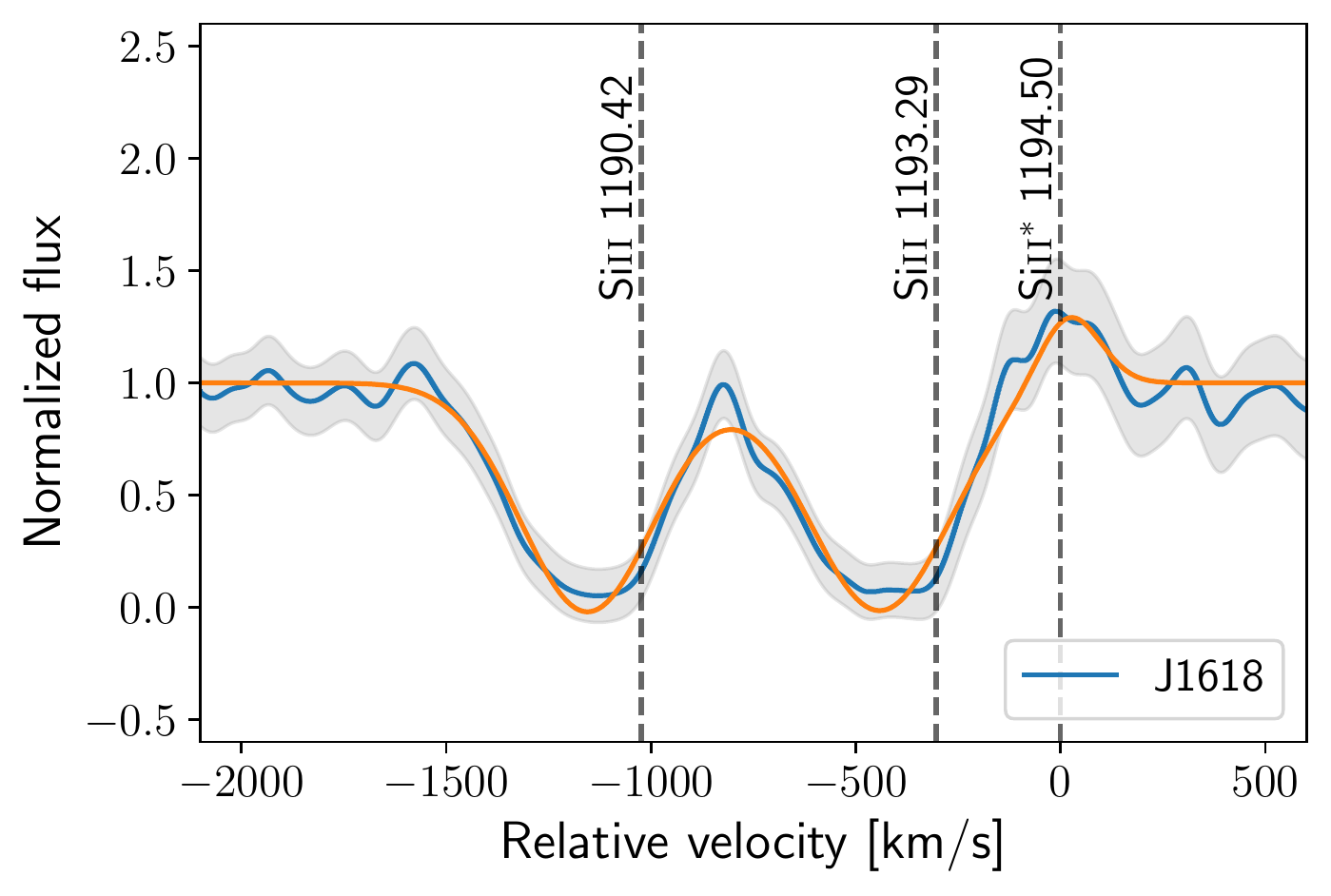}
\figsetgrpnote{Continuum-normalized spectra with the stellar features removed, zooming in on lines of interest. Over-plotted in orange are the gaussian fits.}
\figsetgrpend

\figsetgrpstart
\figsetgrpnum{1.14}
\figsetgrptitle{J1618}
\figsetplot{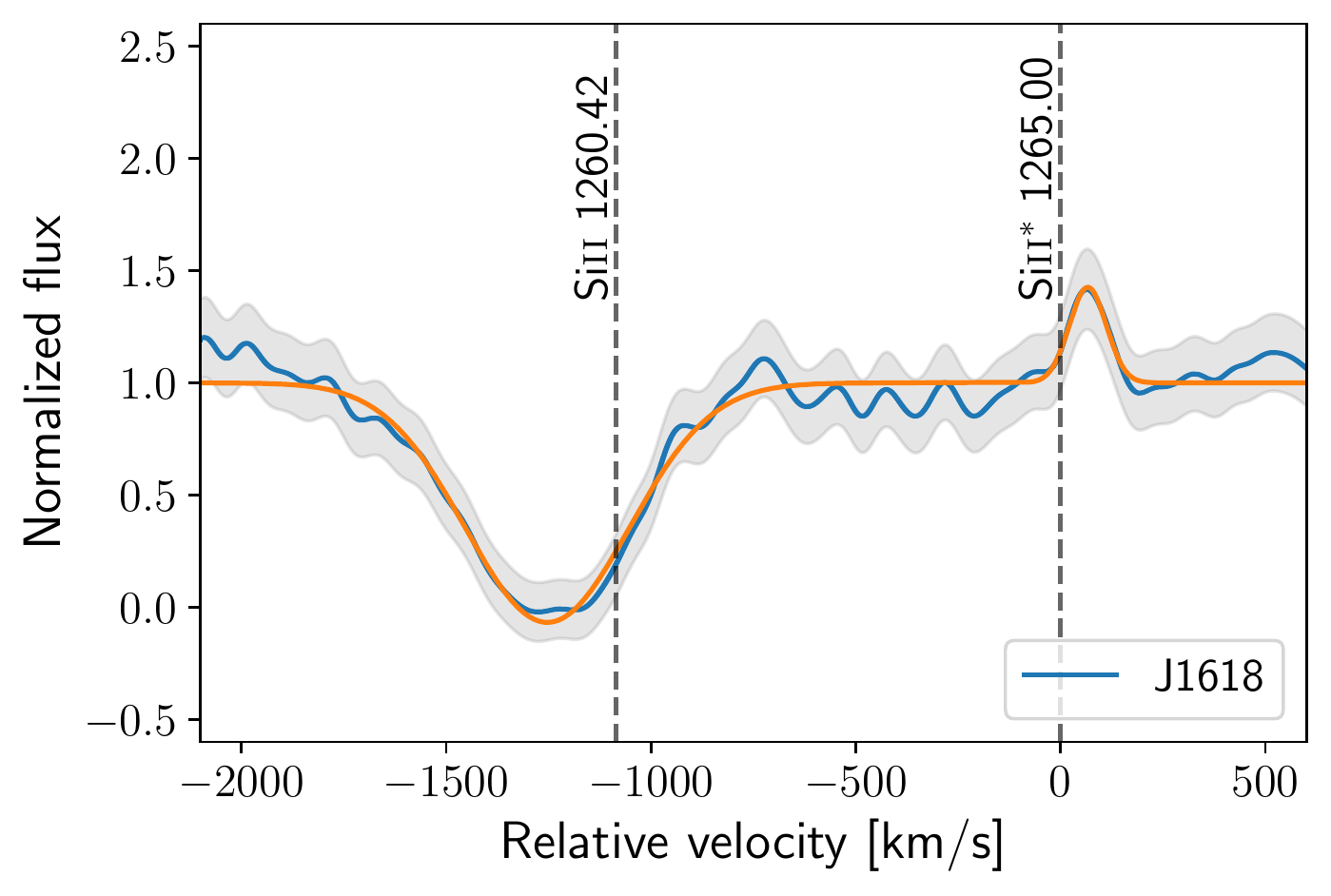}
\figsetgrpnote{Continuum-normalized spectra with the stellar features removed, zooming in on lines of interest. Over-plotted in orange are the gaussian fits.}
\figsetgrpend

\figsetgrpstart
\figsetgrpnum{1.15}
\figsetgrptitle{J1618}
\figsetplot{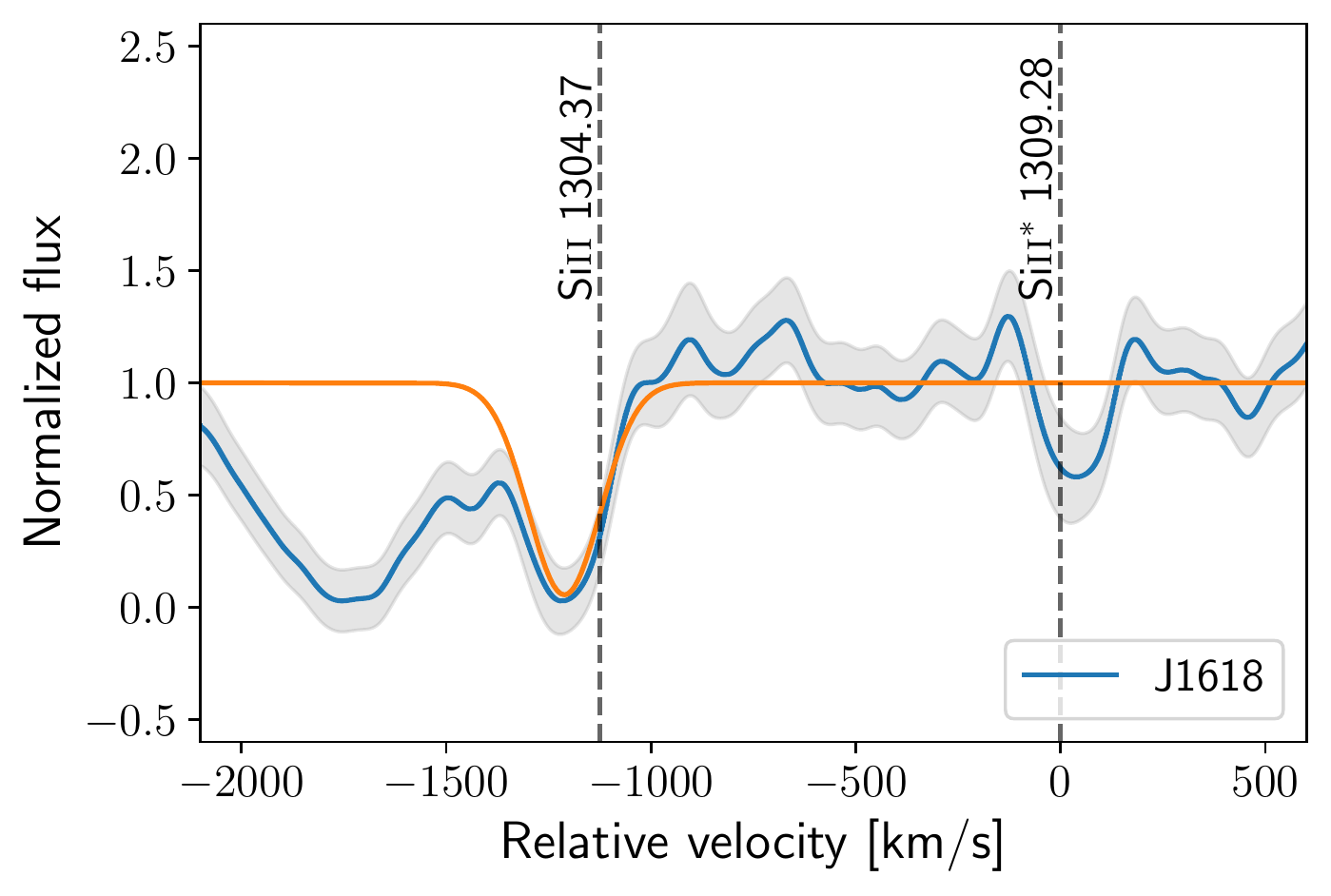}
\figsetgrpnote{Continuum-normalized spectra with the stellar features removed, zooming in on lines of interest. Over-plotted in orange are the gaussian fits.}
\figsetgrpend

\figsetend

\subsubsection{Imaging}

Images were taken with UVIS/F280N, F343N and F395N filters on the WFC3, all of which were reprocessed with the standard pipeline \verb|astrodrizzle|. We align them through spline interpolation to achieve sub-pixel precision. The three exposures of F280N are stacked via exposure-time-weighted average, while the cosmic rays in the single exposures of F343N and F395N are identified using the L.A.Cosmic algorithm \citep{2001PASP..113.1420V,curtis_mccully_2018_1482019} and subsequently masked.
All the images are then converted from units of electron/sec per pixel to that of flux density in $\rm erg \, cm^{-2} \, s^{-1} \, \mAA^{-1}$ per pixel.
As the F343N filter maps the UV stellar continuum, an $\feiistar$ image is obtained by subtracting an F343N image from the stacked F280N one, and an [$\oii$] 3727 image is obtained by subtracting it from the F395N one.

We see small but systematic variations in the residual background in these difference images. We therefore undertake an additional step in subtracting this spatially-varying residual background. We estimate it on a mesh whose cells have scales larger than the source, but small enough to encapsulate the background variations. After subtracting the inferred spatially-varying background, we examine the histogram showing the distribution of individual pixel values after five sigma-clipping, and ensure that it follows a gaussian centered around zero.

\begin{figure*}
\gridline{
\fig{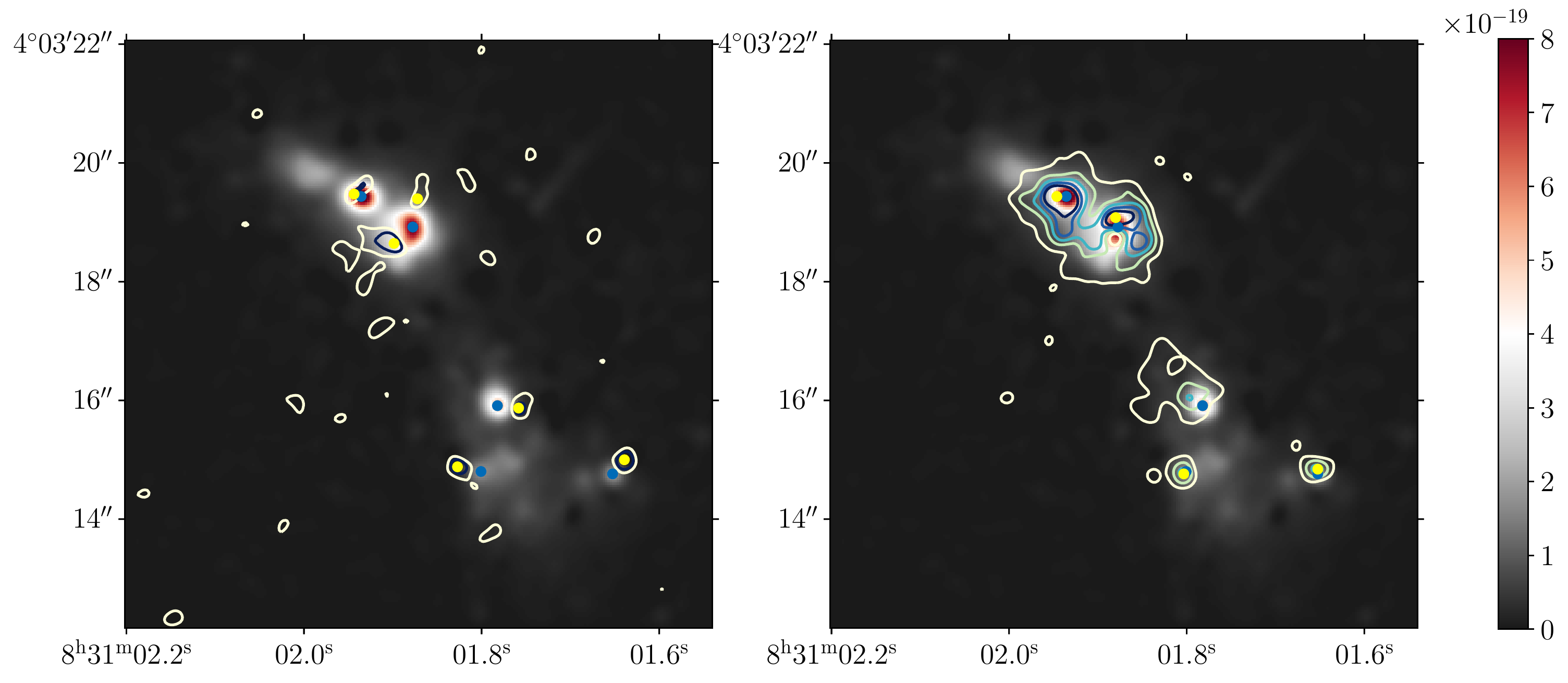}{0.49\textwidth}{(a) J0831}
\fig{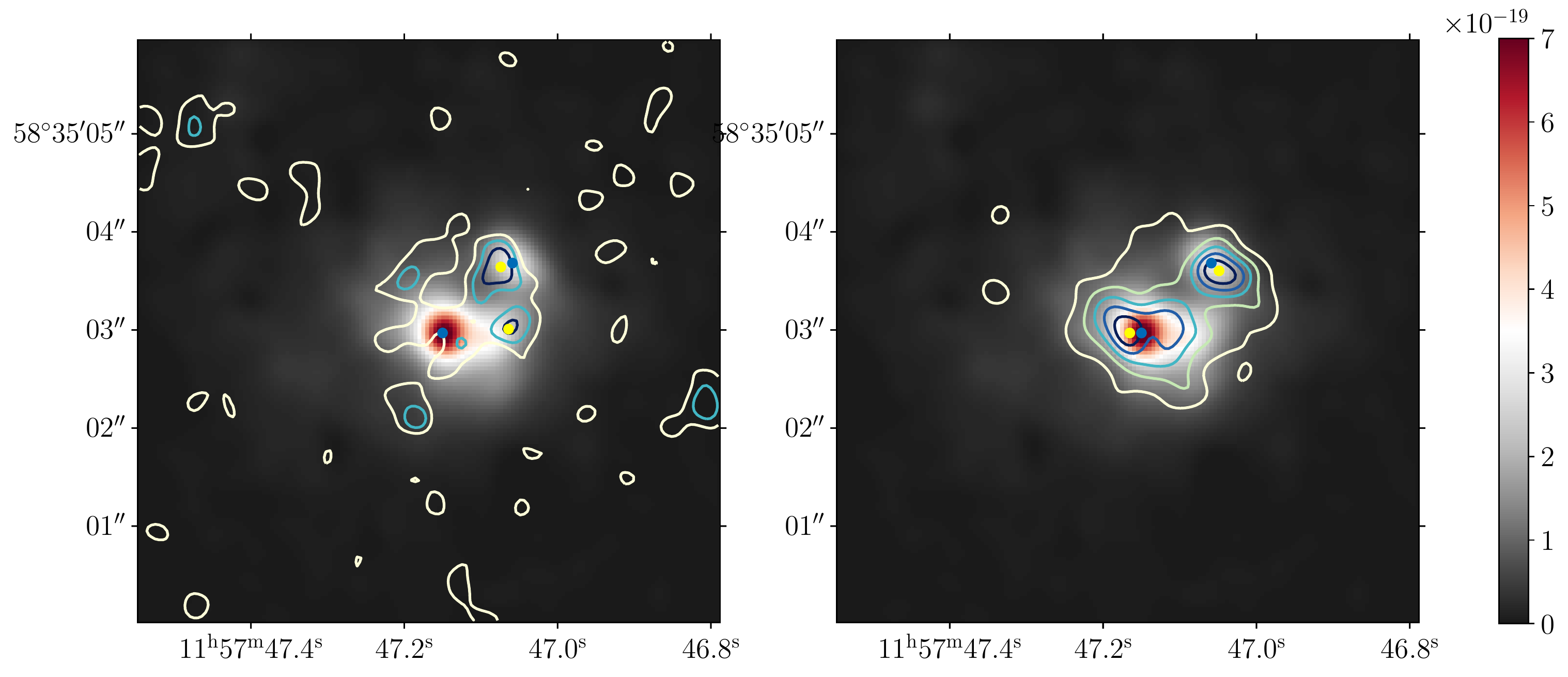}{0.49\textwidth}{(b) J1157}
}
\gridline{
\fig{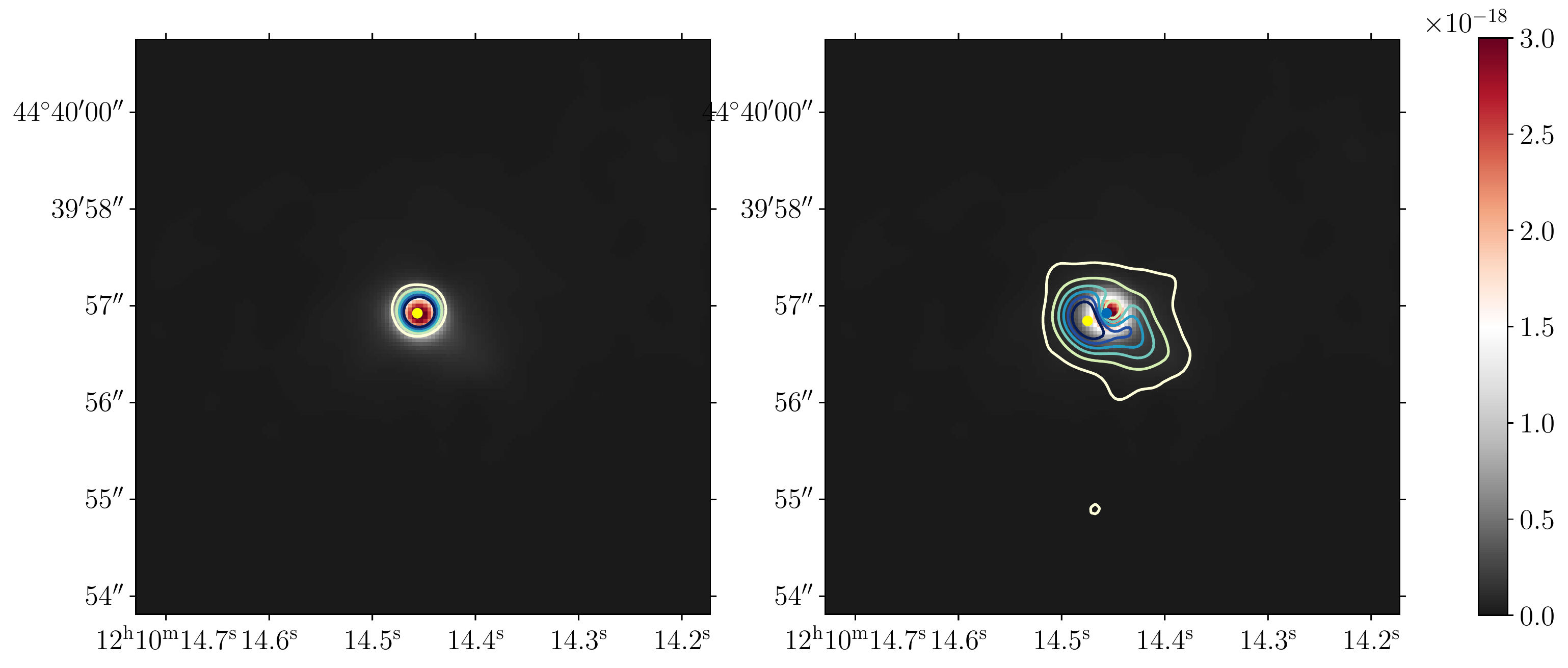}{0.49\textwidth}{(c) J1210}
\fig{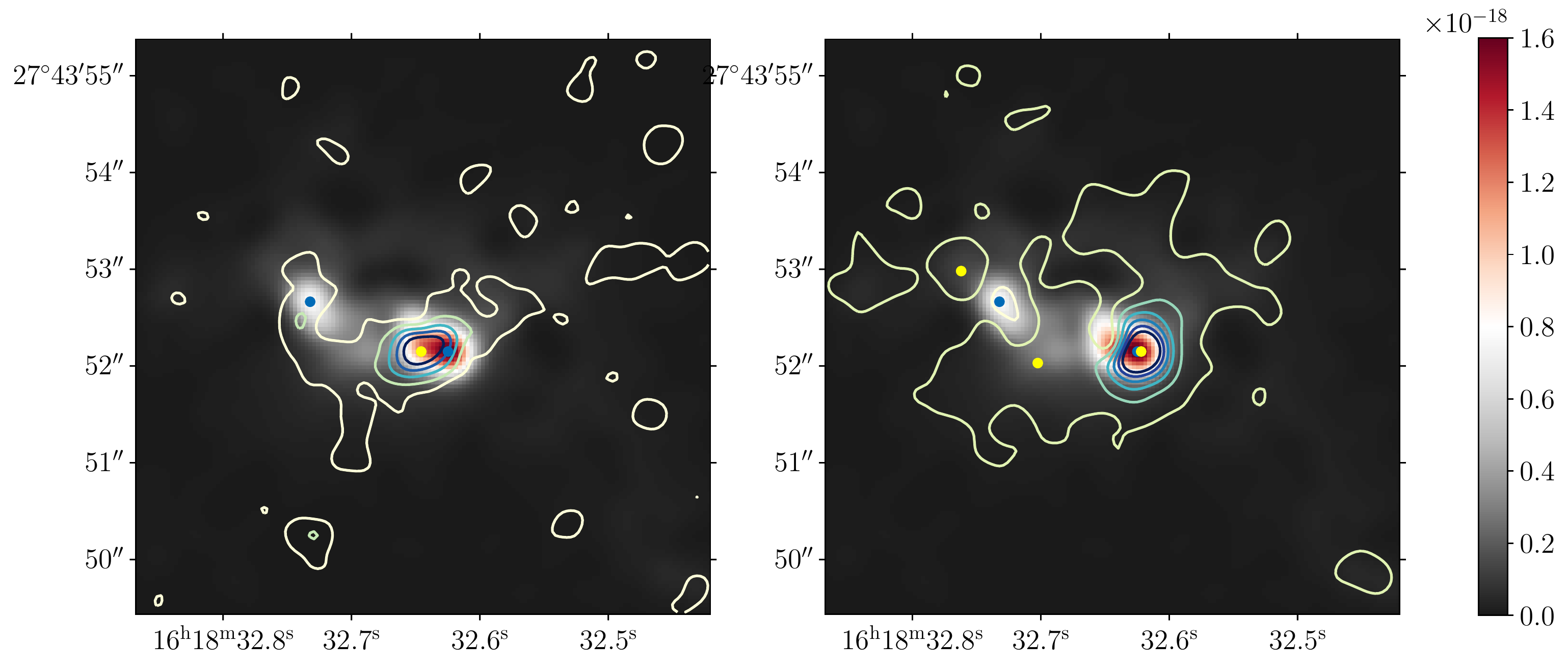}{0.49\textwidth}{(d) J1618}
}
\caption{Contours of the continuum-subtracted $\feiistar$ (left) and [$\oii$] (right) images over-plotted on the continuum images for each galaxy. The peaks of the emission-line images are shown in yellow, while those of the continuum images are shown in blue. Note that all images are smoothed with a gaussian of FWHM=0.01\arcsec. The angular scale is indicated on the y-axis.\label{fig:contour}}
\end{figure*}

We show contours of the continuum-subtracted $\feiistar$ and [$\oii$] over-plotted on the continuum images in Figure~\ref{fig:contour}.

\begin{figure*}
\gridline{
    \fig{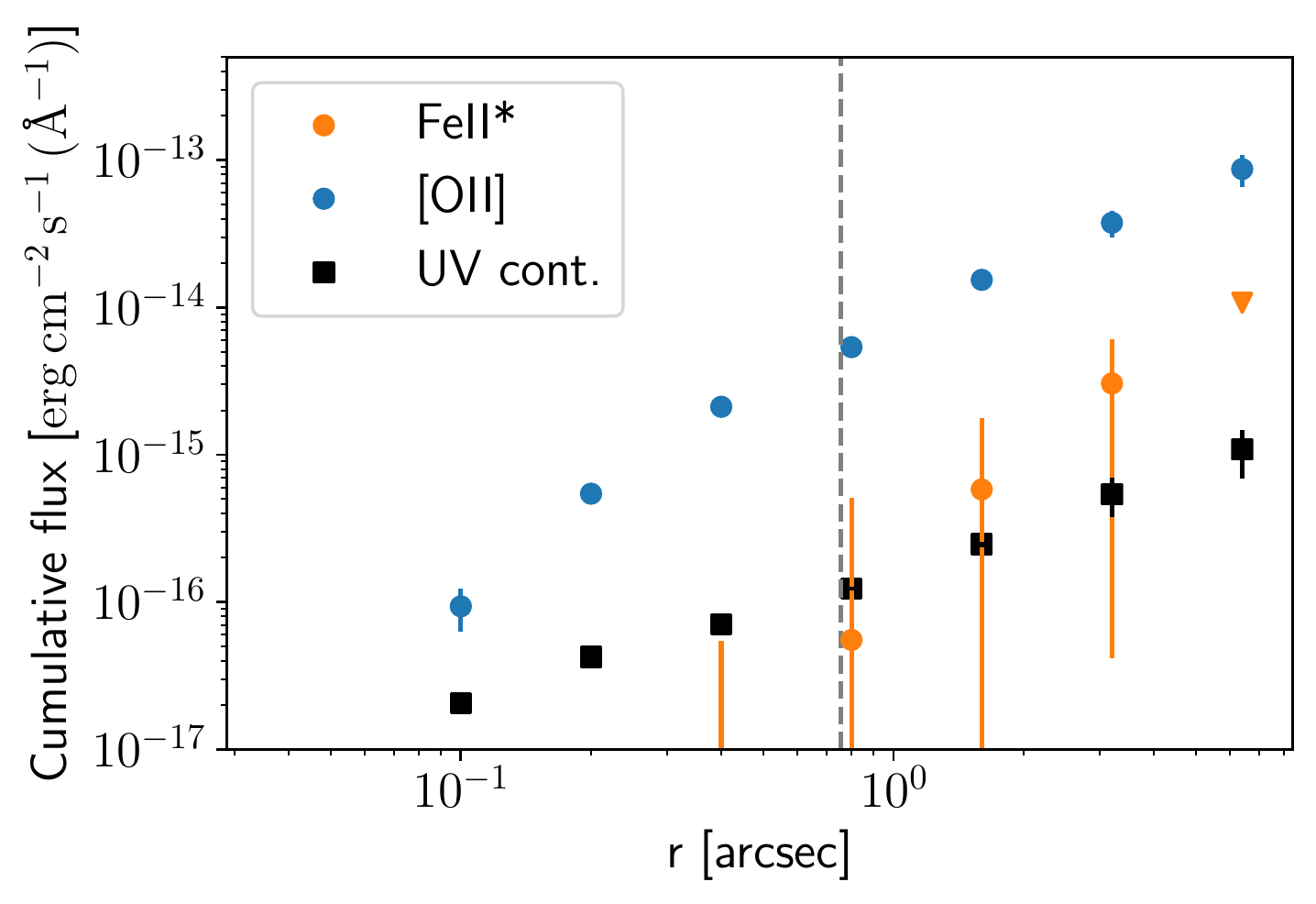}{0.33\textwidth}{(a) J0831(S)}
    \fig{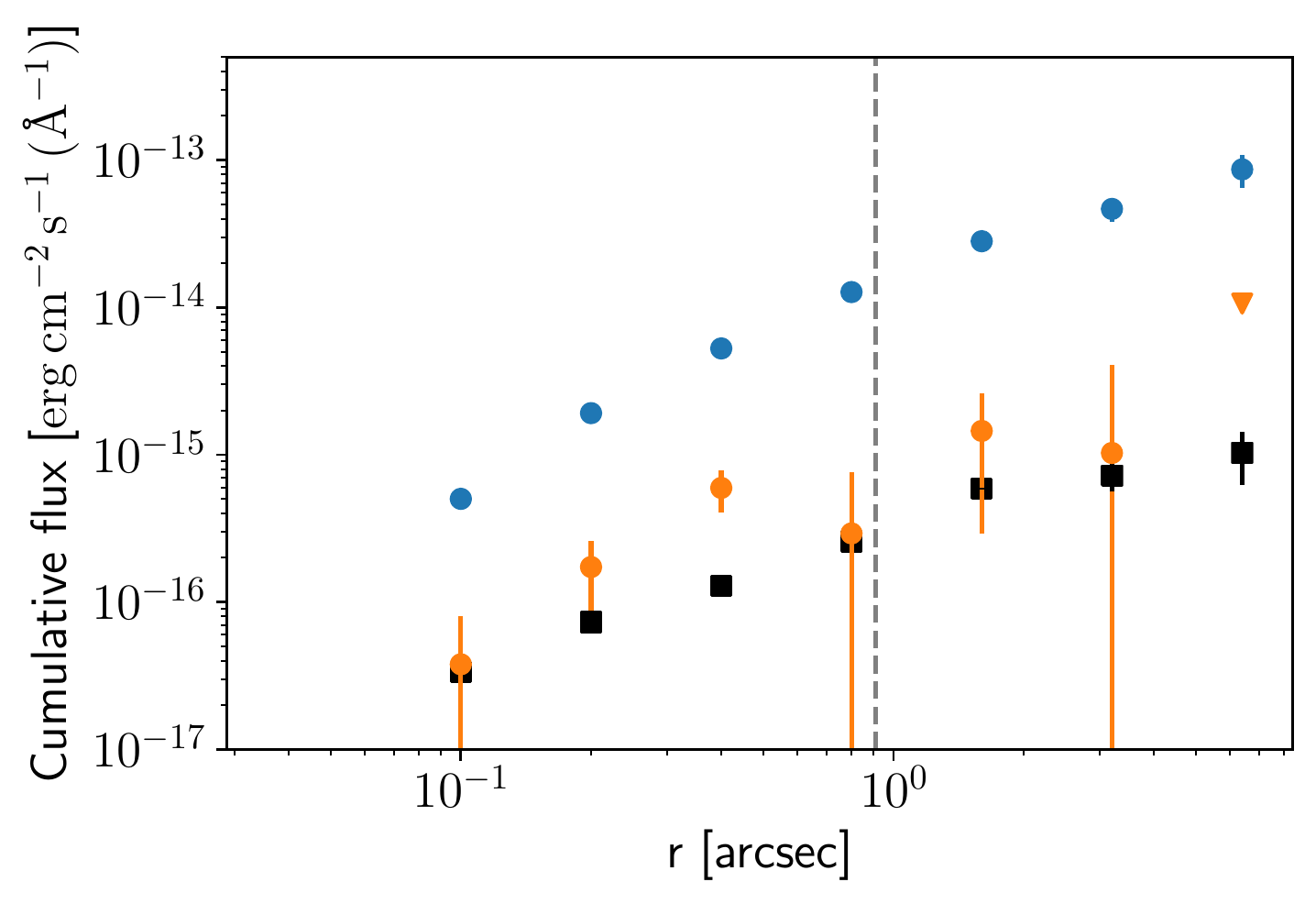}{0.33\textwidth}{(b) J0831(N)}
    \fig{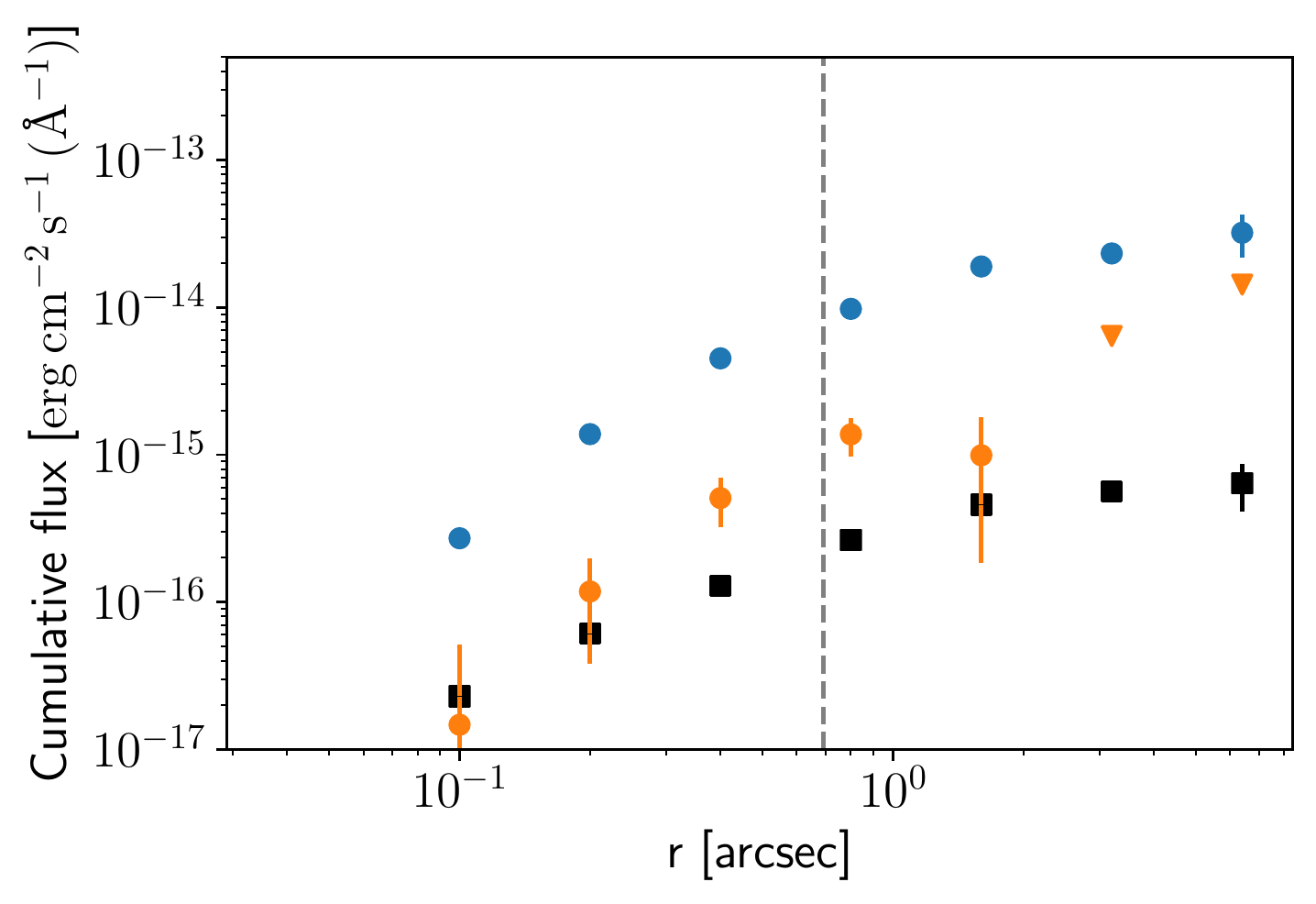}{0.33\textwidth}{(c) J1157}
    }
\gridline{
    \fig{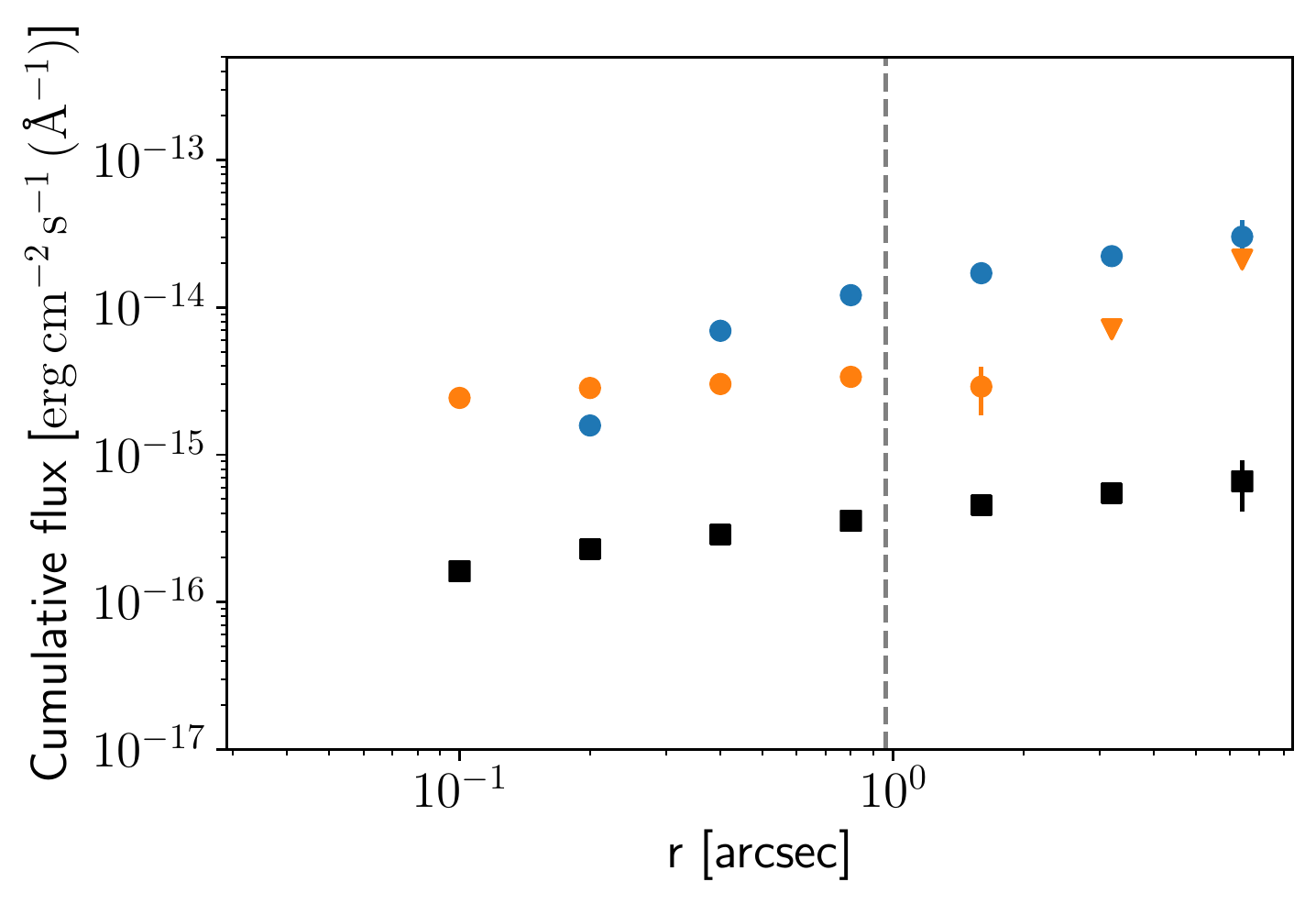}{0.33\textwidth}{(d) J1210}
    \fig{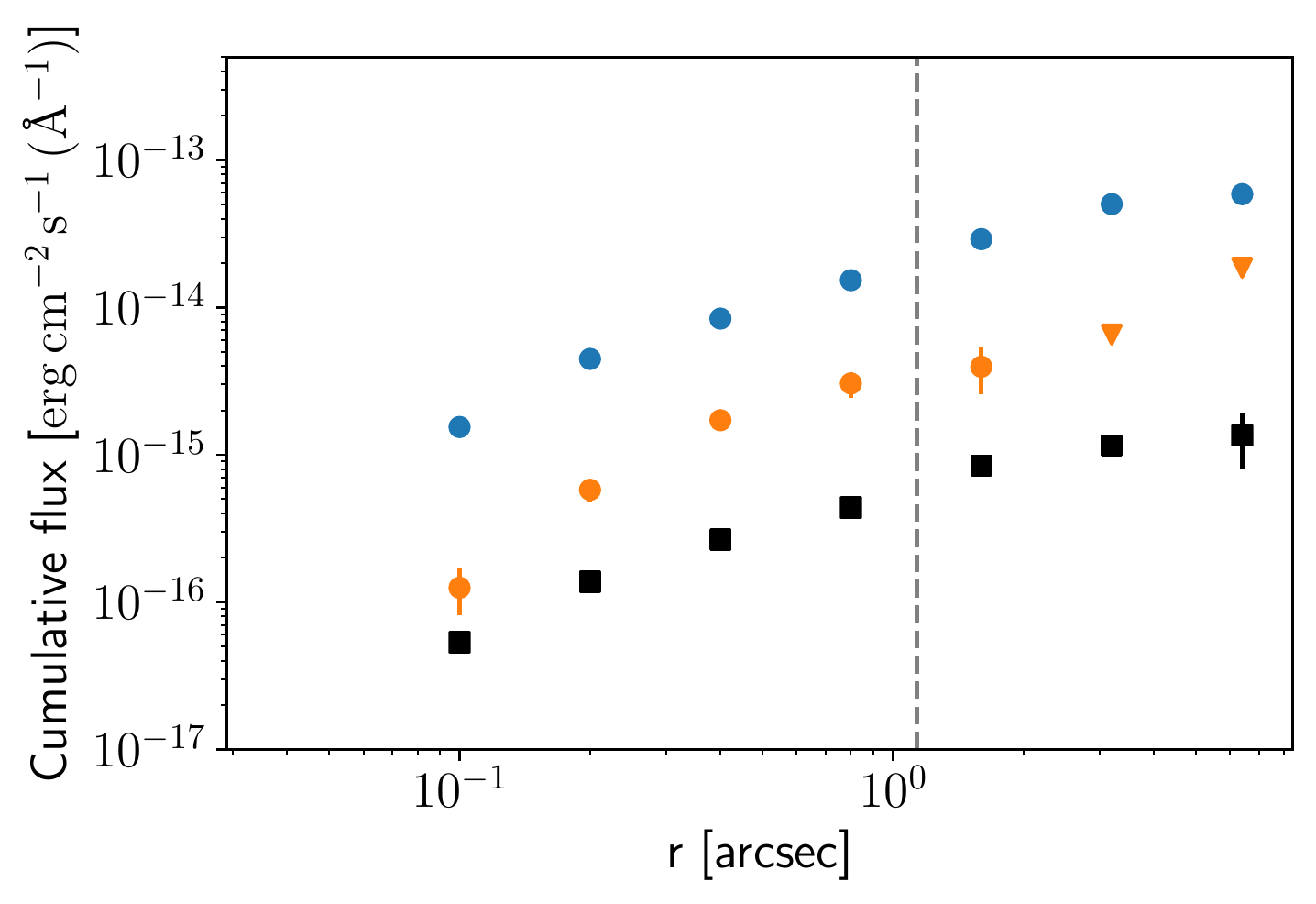}{0.33\textwidth}{(e) J1618}
    }
\caption{Enclosed flux plotted as a function of radius. The data points for $\feiistar$ and [$\oii$] shown in dots are in units of flux, whereas those for the UV continuum shown in black squares are in units of flux density. The triangles are upper limits estimated from background fluctuations. The gray dashed lines indicate $r_{50}$'s of the WFC3 UV continuum images. We infer that nearly all of the detectable $\feiistar$ emission is from inside the starburst region, and is associated with regions of high surface-brightness in [$\oii$] and the UV continuum.\label{fig:cum_flux}}
\end{figure*}

The radial surface brightness as shown in Figure~\ref{fig:cum_flux} is measured in a set of apertures/annuli, which centers on the strongest peak found in the continuum and extends to 6.4\arcsec. We estimate the uncertainty by measuring the flux in the same set of apertures/annuli centered on many locations of the blank sky, and then quote the $1 \sigma$ values of the gaussian fits to the distribution of background fluctuations.

\subsubsection{Archival data}

The individual spectra of the LBAs are re-analyzed according to the description given in Section~\ref{sec:data_spec}. For the sake of consistency, we include only galaxies with COS data and with the set of galaxy parameters that can be measured using the spectra in the SDSS. Their $\sii$ and $\siistar$ line profiles are attached in the Appendix.

We also re-measure the EWs, FWHMs, and velocity centroids of the $\sii$ and $\siistar$ lines in the composite spectrum of 25 LAEs at $z \sim 0.3$ \citep{Scarlata2015}.

As for the LBGs, we take the measurements of the two stacked spectra, which were obtained from two subsamples distinguished by redshifts, from the original paper \citep{Jones2012}. The mean redshifts are 3.76 and 4.70 respectively.

\subsection{Measured ancillary parameters\label{sec:anci}}

In this section we list important ancillary parameters, all of which are listed in Table~\ref{tab:anci}. Most of them are determined in the same way as in \cite{Wang2019}, so we only briefly reiterate the procedure here for completeness.

We measure the SFRs in two ways. In both cases we use the same IMF as that used in our SB99 fit. $\sfrir$ is calculated by using the WISE IR data at 12 and 22 $\mu$m \citep{2010AJ....140.1868W} to estimate the rest-frame 24 $\mu$m luminosity, and then using the relation given in \cite{Kennicutt2012}. $\sfrha$ is calculated from extinction-corrected $\ha$ and $\hb$ fluxes, and then using the relation given in \cite{Calzetti2011}.

\begin{figure*}
\gridline{
\fig{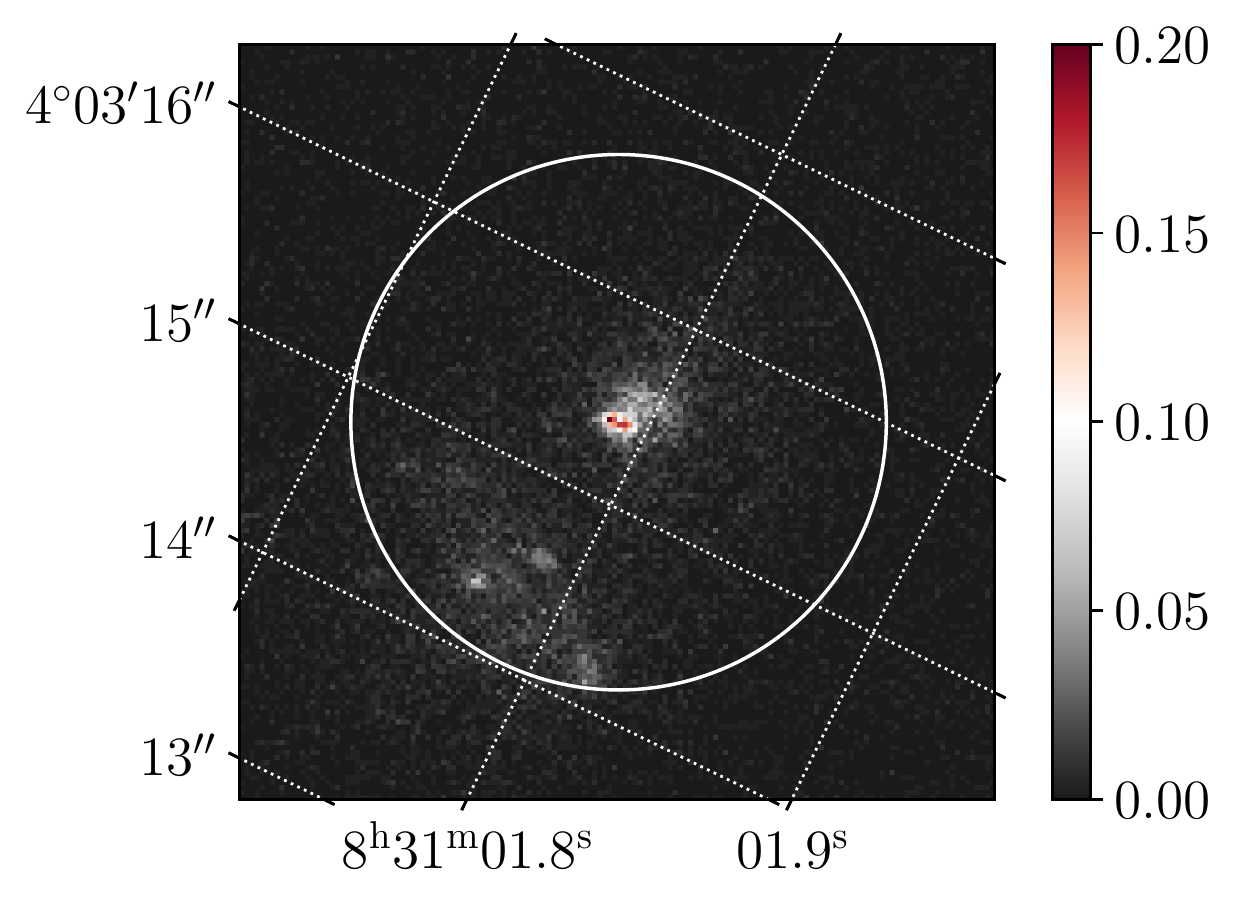}{0.33\textwidth}{(a) J0831(S)}
\fig{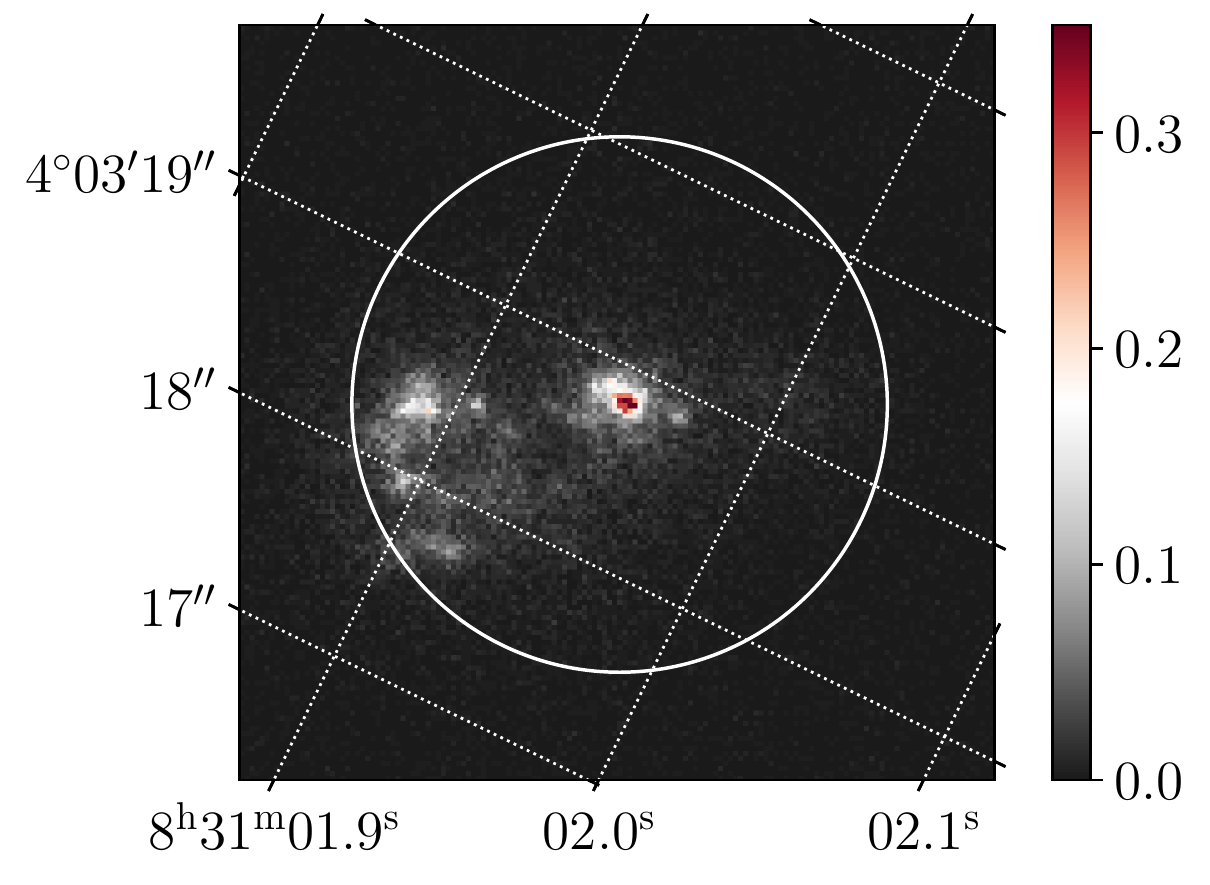}{0.33\textwidth}{(b) J0831(N)}
\fig{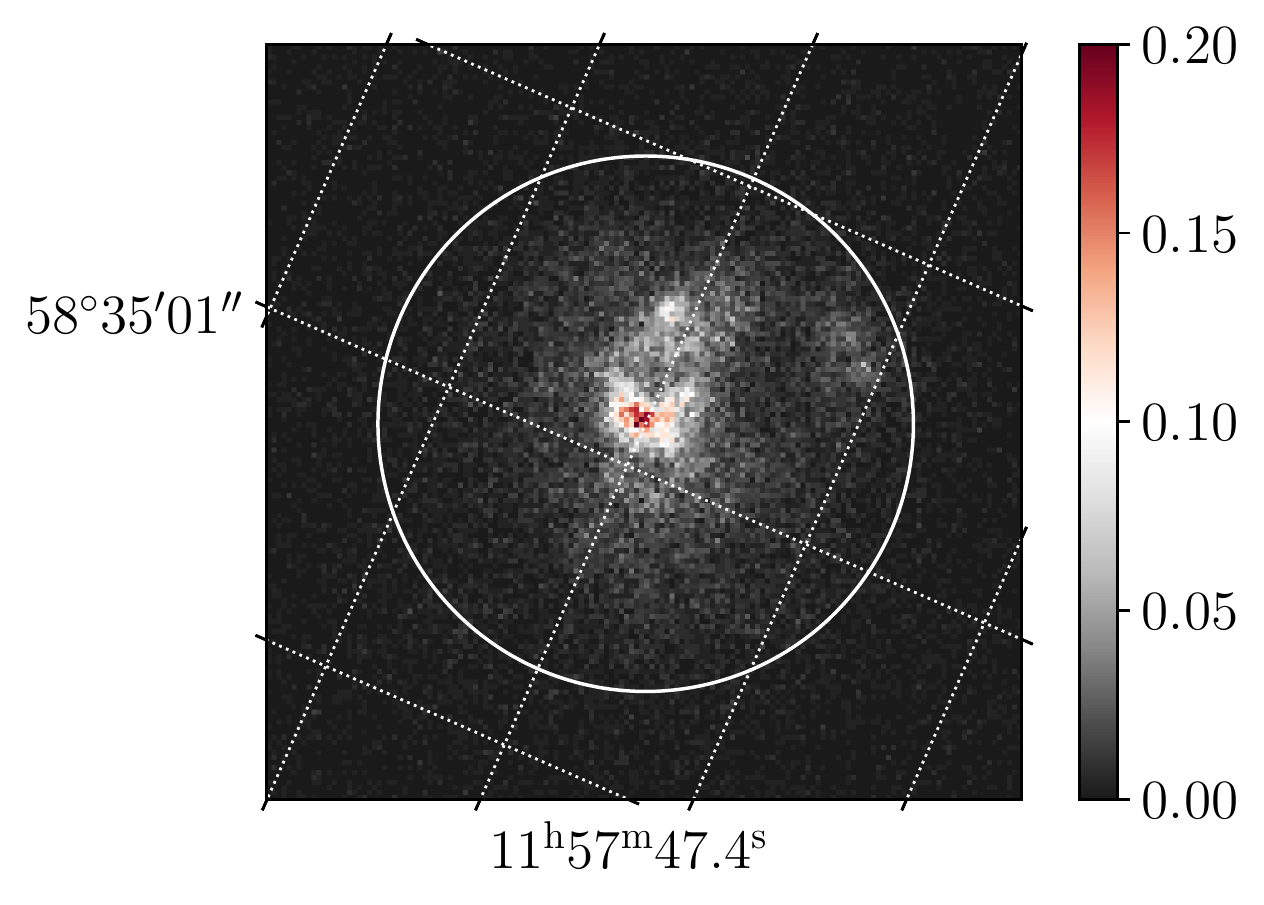}{0.33\textwidth}{(c) J1157}
}
\gridline{
\fig{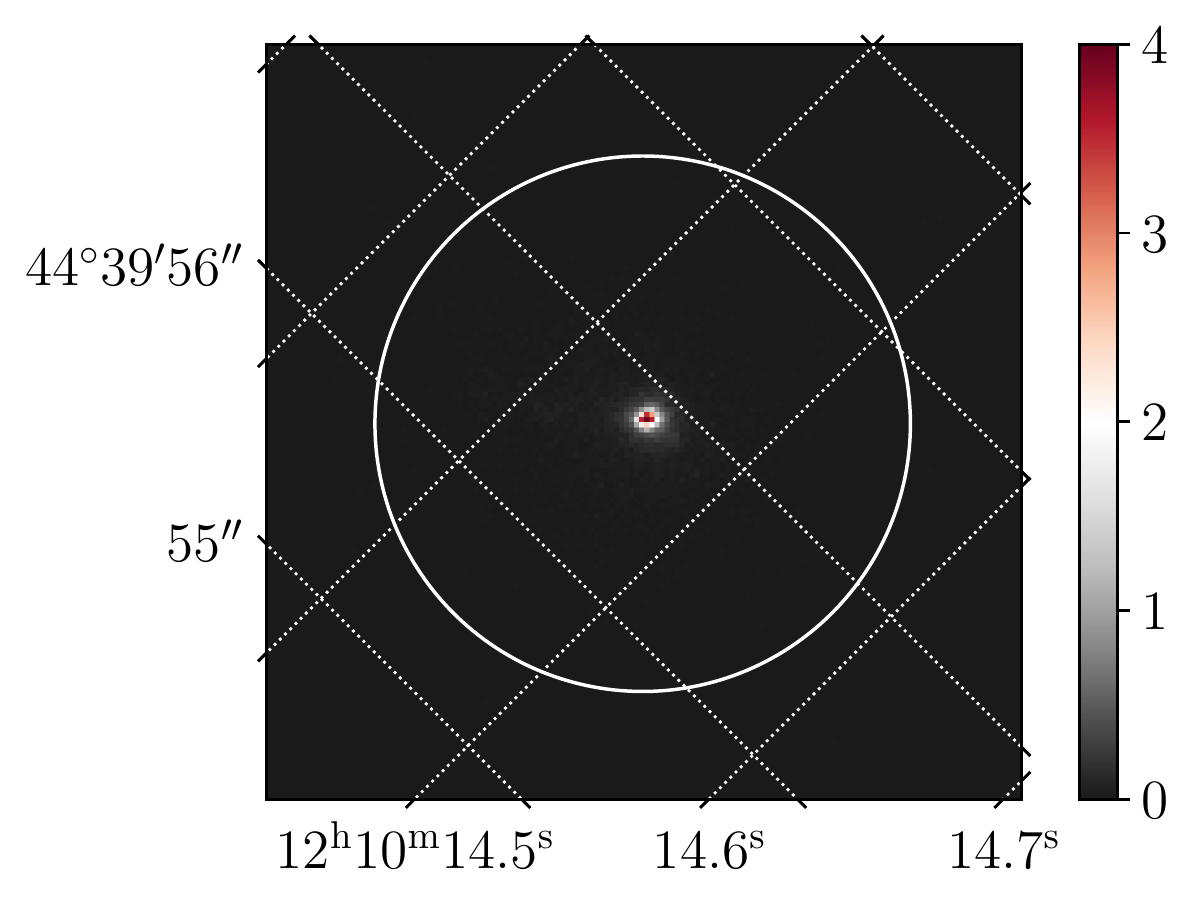}{0.33\textwidth}{(d) J1210}
\fig{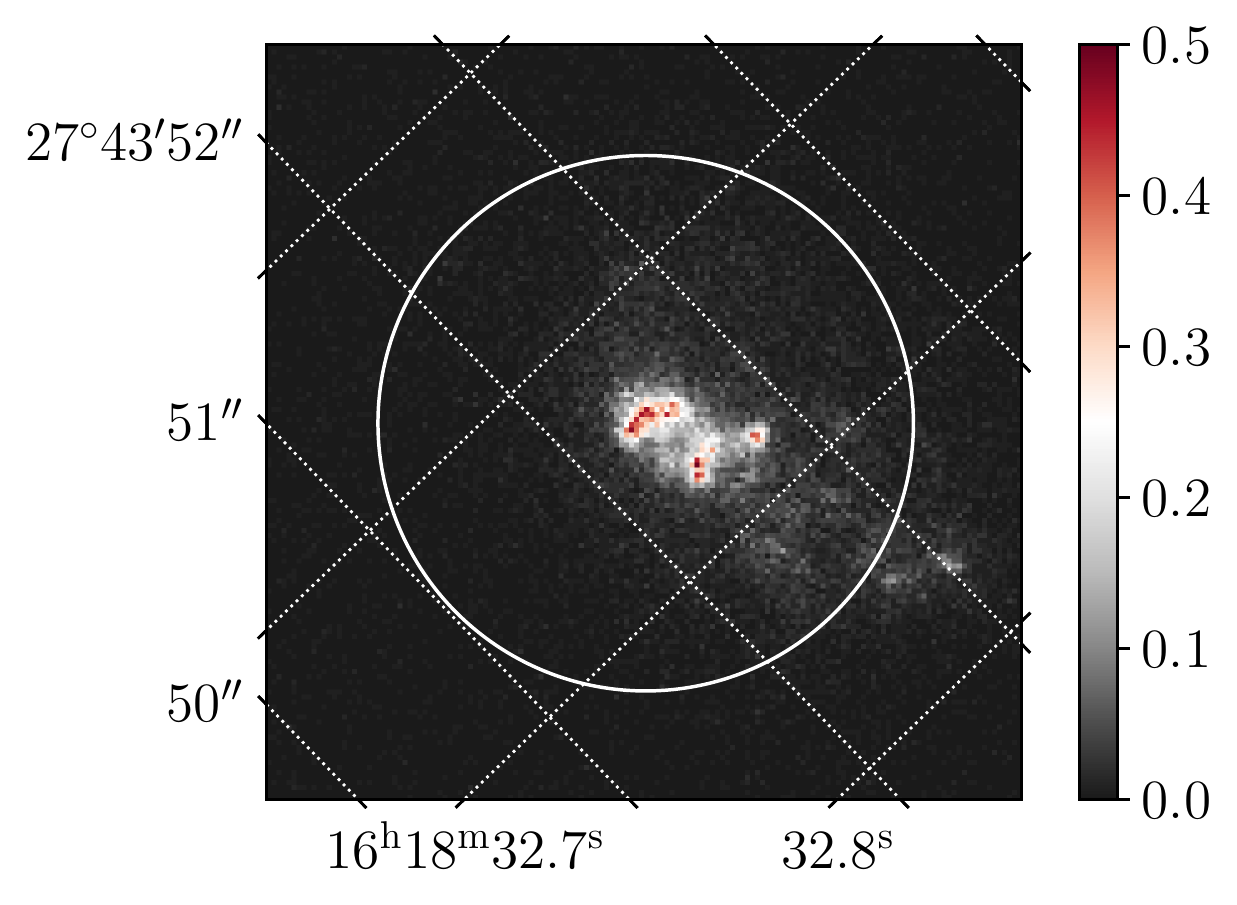}{0.33\textwidth}{(e) J1618}
}
\caption{COS NUV ACQ images. The white circles indicate the COS aperture of radius 1.25\arcsec.\label{fig:nuv}}
\end{figure*}

We also determine the half-light radii, $r_{50}$, in two sets of images. First, the COS NUV ACQ images are used, as shown in Figure~\ref{fig:nuv}. Second, the WFC3 images taken with the F343N filter are used, which maps the UV stellar continuum.
We note that the $r_{50}$'s obtained in the UV continuum images are larger for J1210 and J1618. In the former case, diffuse light outside the field of view of the COS ACQ image is seen in the WFC3 image, while in the latter, extended faint light is detected by WFC3 due to its deeper imaging.

We use the FWHM of the Balmer emission lines in the SDSS to characterize the kinematics of the ionized inter-stellar medium in each galaxy. These widths have been corrected for the SDSS spectral resolution as part of the SDSS pipeline. Lastly, the stellar masses are taken from the median of the corresponding probability density function in the MPA-JHU catalog.

\begin{deluxetable*}{cccccccccc}
\tabletypesize{\footnotesize}
\tablecaption{Measured ancillary parameters for the five galaxies of this paper and LBAs\label{tab:anci}}
\tablecolumns{10}
\tablewidth{0pt}
\tablehead{
\colhead{} &
\colhead{FWHM(Balmer)} &
\colhead{$A_{\ha}$} &
\colhead{$\sfrha$} &
\colhead{SFR(IR,UV)} &
\colhead{$M_\star$} &
\colhead{$12+{\rm log}({\rm O/H})$} &
\colhead{$r_{50}$ (NUV)} &
\colhead{$r_{50}$ (UV cont.)} &
\colhead{$r_{\rm COS}$}\\
\colhead{} &
\colhead{(km/s)} &
\colhead{} &
\colhead{$(M_\odot {\rm {yr}^{-1})}$} &
\colhead{$(M_\odot {\rm {yr}^{-1})}$} &
\colhead{(${\rm log_{10}} M_\odot$)} &
\colhead{} &
\colhead{(kpc)} &
\colhead{(kpc)} &
\colhead{(kpc)}
}
\startdata
J0831(S) & 224 $\pm$ 1 & 1.21 $\pm$ 0.04 & 6.2 & 17.9 $\pm$ 1.1 & 9.88 & 8.43 & 1.04 & 0.98 & 1.61 \\
J0831(N) & 205 $\pm$ 1 & 0.76 $\pm$ 0.03 & 7.0 & 12.0 $\pm$ 1.1 & 9.76 & 8.00 & 0.99 & 1.17 & 1.61 \\
J1157 & 135 $\pm$ 1 & 0.26 $\pm$ 0.03 & 1.7 & 4.6 $\pm$ 0.3 & 9.42 & 8.25 & 0.70 & 0.89 & 1.60 \\
J1210 & 235 $\pm$ 1 & 0.71 $\pm$ 0.03 & 4.6 & 10.4 $\pm$ 0.4 & 9.88 & 8.41 & 0.17 & 1.26 & 1.64 \\
J1618 & 192 $\pm$ 1 & 0.35 $\pm$ 0.03 & 6.8 & 17.4 $\pm$ 0.4 & 9.28 & 8.15 & 0.63 & 1.45 & 1.60 \\
\hline
J0055 & 339 $\pm$ 1 & 0.60 $\pm$ 0.02 & 28.8 & 23.6 & 9.70 & 8.28 & 0.32 & - & 3.70 \\
J0150 & 240 $\pm$ 1 & 0.74 $\pm$ 0.03 & 20.0 & 37.4 & 10.30 & 8.40 & 1.37 & - & 3.31 \\
J0213 & 218 $\pm$ 5 & 0.80 $\pm$ 0.11 & 5.7 & 19.0 & 10.50 & 8.76 & 0.39 & - & 4.57 \\
J0808 & 362 $\pm$ 3 & 0.54 $\pm$ 0.04 & 4.3 & 8.5 & 9.80 & 8.77 & 0.08 & - & 2.20 \\
J0921 & 395 $\pm$ 4 & 0.86 $\pm$ 0.06 & 23.4 & 29.4 & 10.80 & 8.69 & 0.78 & - & 4.82 \\
J0926 & 241 $\pm$ 1 & 0.21 $\pm$ 0.03 & 17.6 & 10.4 & 9.10 & 8.05 & 0.69 & - & 3.93 \\
J0938 & 213 $\pm$ 1 & 0.38 $\pm$ 0.02 & 16.1 & 11.2 & 9.40 & 8.19 & 0.67 & - & 2.43 \\
J2103 & 496 $\pm$ 3 & 0.89 $\pm$ 0.03 & 43.1 & 41.4 & 10.90 & 8.70 & 0.46 & - & 3.13 \\
J0021 & 307 $\pm$ 1 & 0.09 $\pm$ 0.01 & 18.6 & 14.9 & 9.30 & 8.19 & 0.53 & - & 2.35 \\
J0823 & 197 $\pm$ 1 & 0.57 $\pm$ 0.02 & 9.4 & 9.6 & 8.60 & 8.23 & 0.34 & - & 1.20 \\
J1025 & 187 $\pm$ 1 & 0.25 $\pm$ 0.03 & 12.2 & 7.6 & 9.20 & 8.11 & 0.61 & - & 2.92 \\
J1112 & 309 $\pm$ 1 & 0.60 $\pm$ 0.03 & 24.8 & 28.7 & 10.20 & 8.52 & 0.33 & - & 3.03 \\
J1113 & 179 $\pm$ 9 & 0.17 $\pm$ 0.16 & 1.2 & 7.1 & 9.60 & 8.35 & 1.09 & - & 3.83 \\
J1144 & 171 $\pm$ 2 & 0.51 $\pm$ 0.04 & 7.1 & 8.9 & 9.90 & 8.40 & 0.76 & - & 2.93 \\
J1414 & - & -& 6.3 & 5.1 & 8.50 & 8.28 & 0.63 & - & 1.99 \\
J1416 & 257 $\pm$ 1 & 0.65 $\pm$ 0.03 & 19.9 & 23.4 & 10.00 & 8.47 & 0.19 & - & 2.86 \\
J1428 & 214 $\pm$ 1 & 0.37 $\pm$ 0.03 & 19.8 & 13.9 & 9.60 & 8.31 & 0.71 & - & 3.95 \\
J1429 & 300 $\pm$ 1 & 0.13 $\pm$ 0.03 & 36.0 & 26.8 & 9.40 & 8.12 & 0.29 & - & 3.81 \\
J1521 & 258 $\pm$ 1 & 0.32 $\pm$ 0.03 & 6.0 & 5.8 & 9.50 & 8.27 & 0.37 & - & 2.26 \\
J1525 & 213 $\pm$ 1 & 0.52 $\pm$ 0.03 & 6.3 & 9.1 & 9.40 & 8.46 & 0.51 & - & 1.86 \\
J1612 & 290 $\pm$ 1 & 0.77 $\pm$ 0.03 & 32.2 & 36.1 & 10.00 & 8.51 & 0.31 & - & 3.36 \\
\hline
Uncertainty & - & - & $\pm$15\% & - & - & $\pm$0.14 dex & \textless 0.1 dex & \textless 0.1 dex & - \\
\enddata
\tablecomments{The typical uncertainties are estimated following \cite{Heckman2015}.
The uncertainties quoted for $\sfrir$ only account for that from the magnitude measurements of {\it{WISE}}.}
\end{deluxetable*}

\section{Results\label{sec:res}}

\subsection{Spectra\label{sec:res_spec}}

\begin{figure*}
\gridline{
	\fig{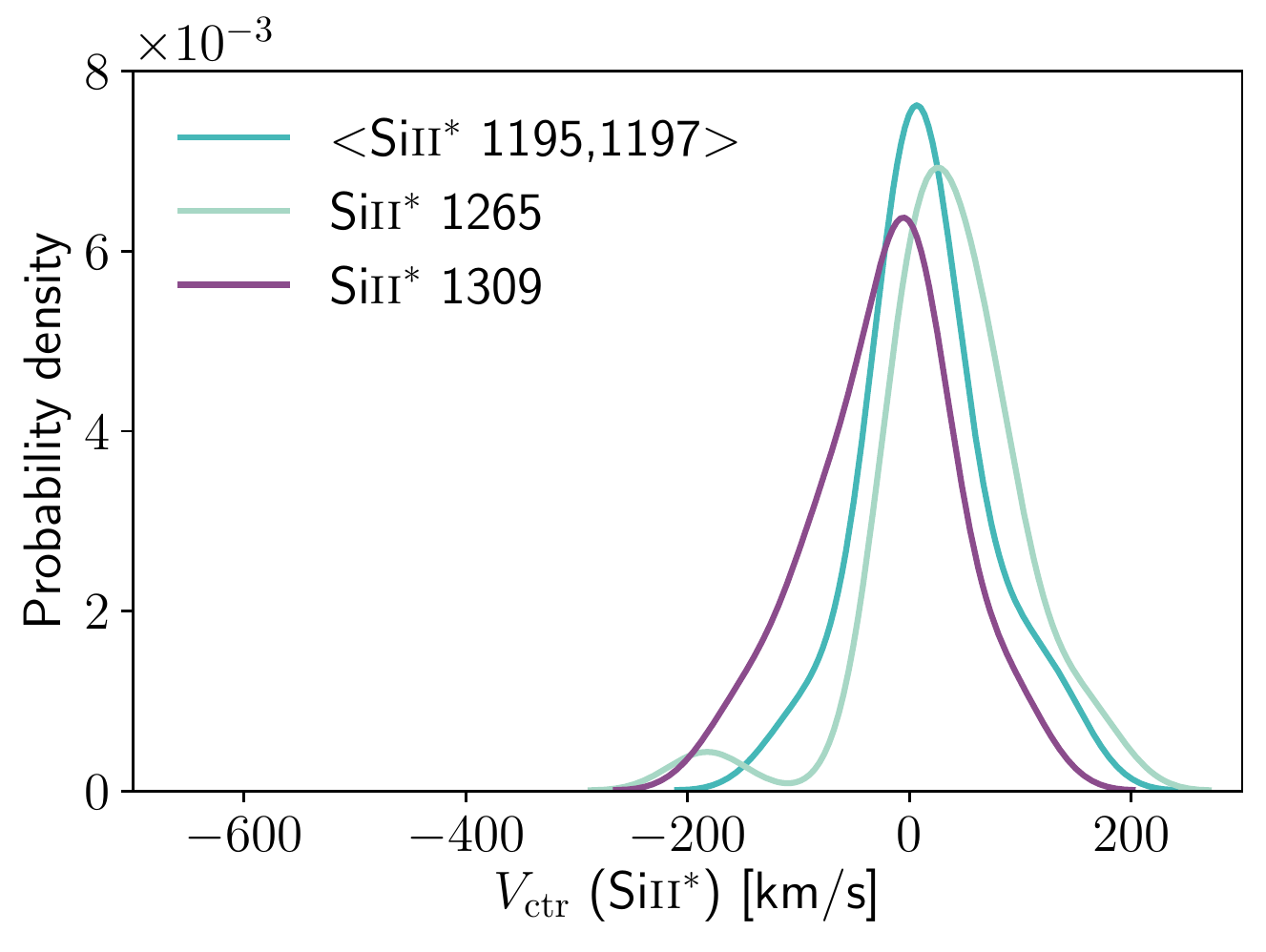}{0.25\textwidth}{}
	\fig{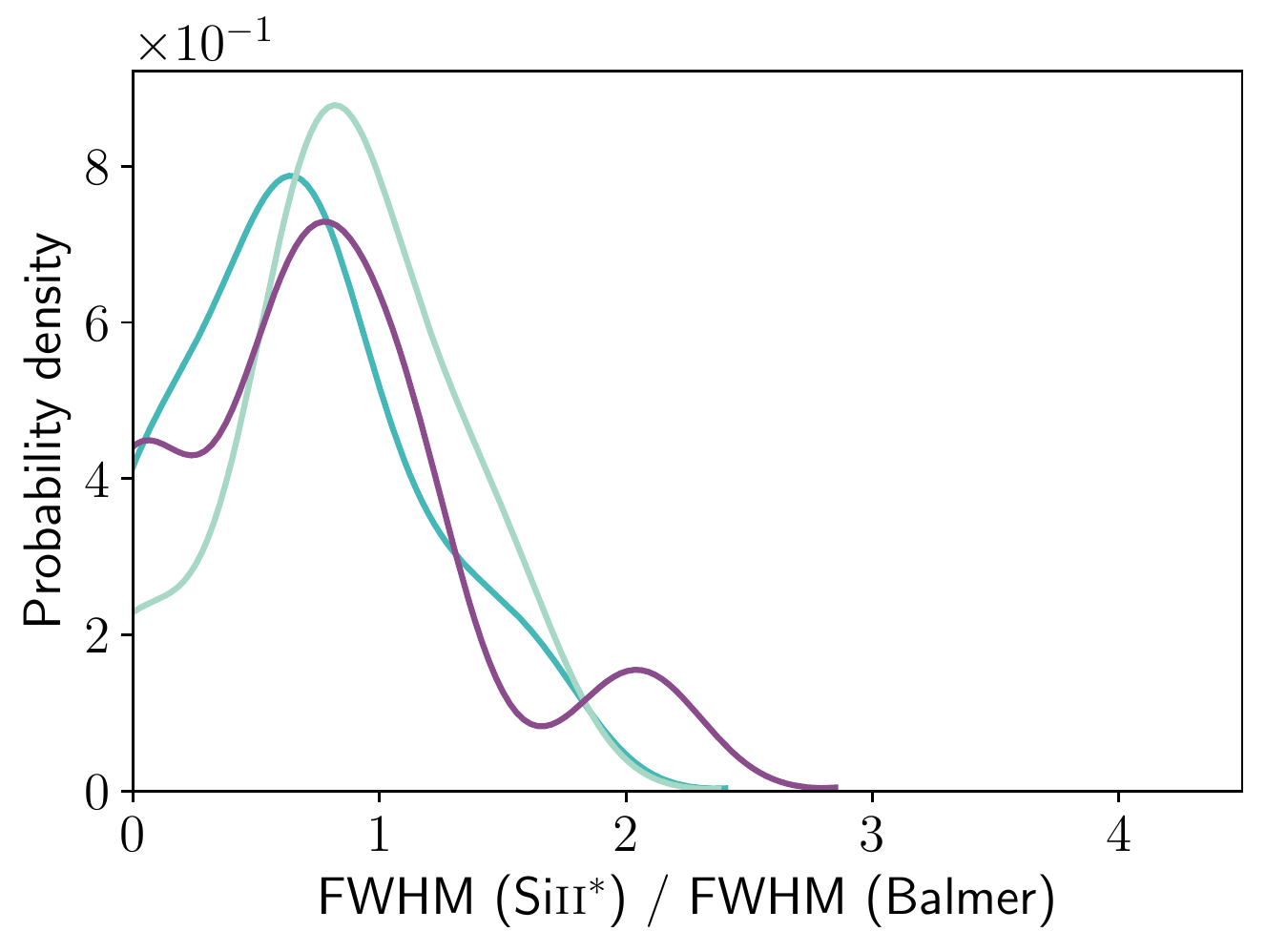}{0.25\textwidth}{}
	\fig{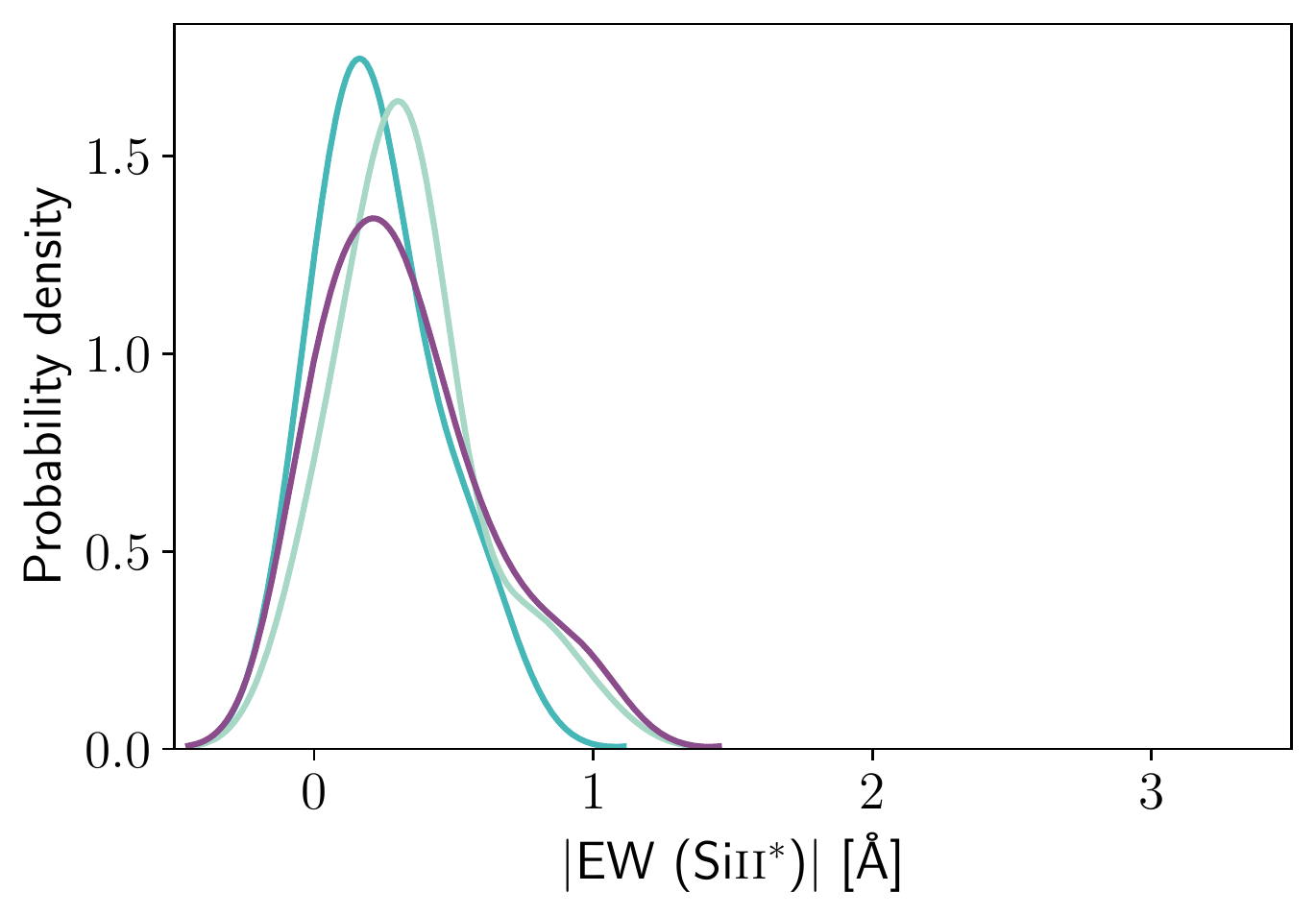}{0.25\textwidth}{}
	\fig{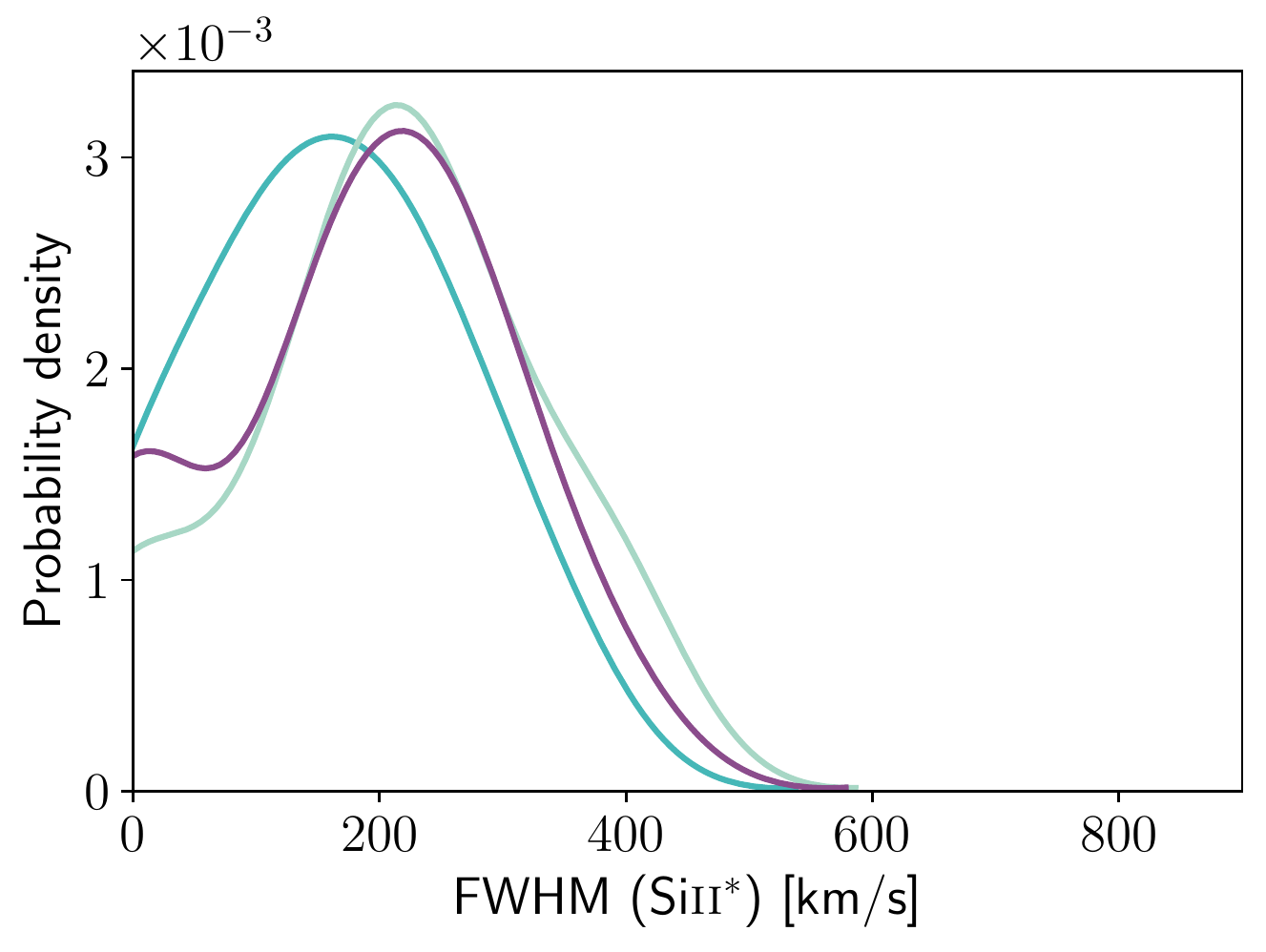}{0.25\textwidth}{}
	}
\gridline{
	\fig{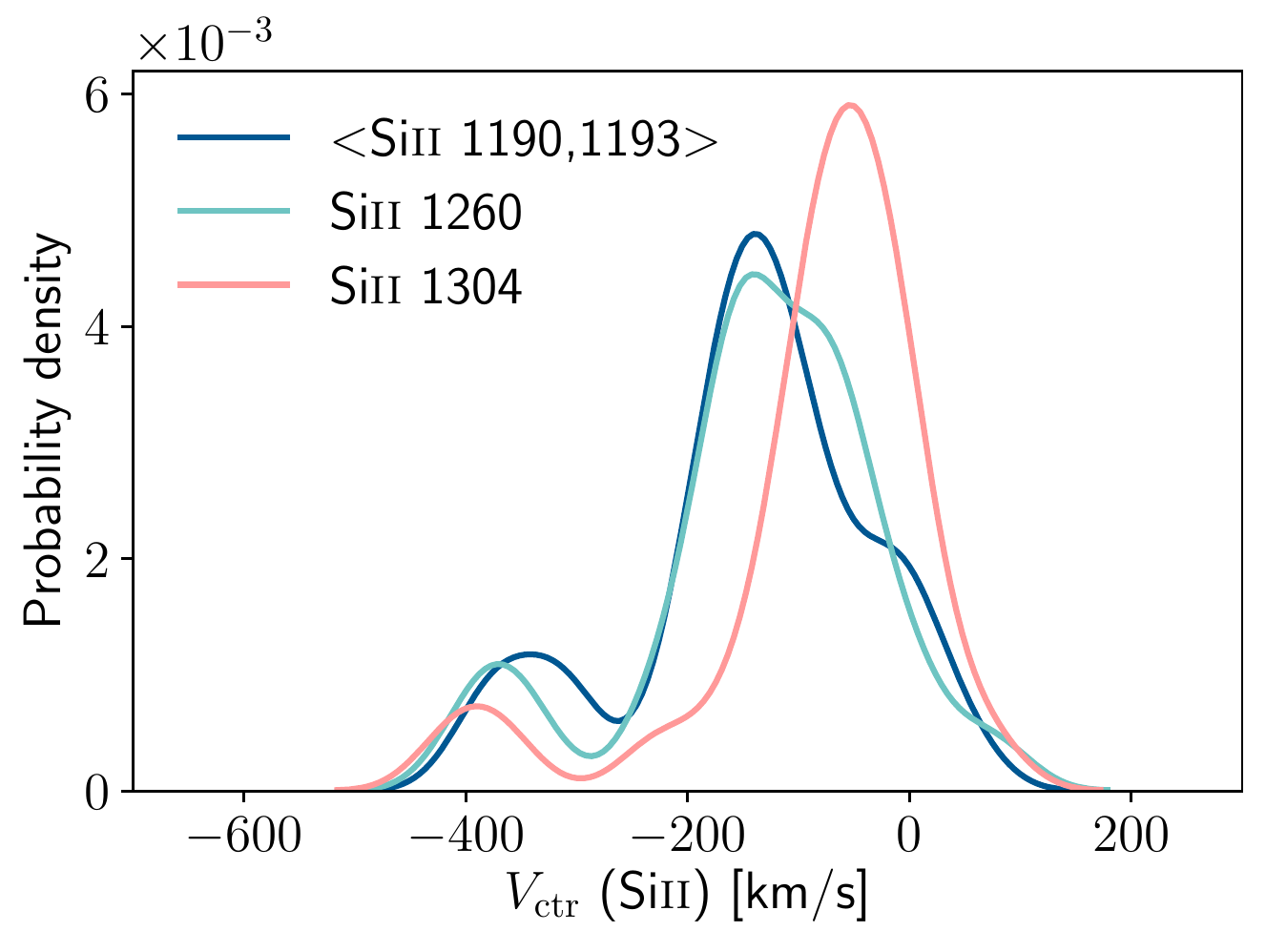}{0.25\textwidth}{}
	\fig{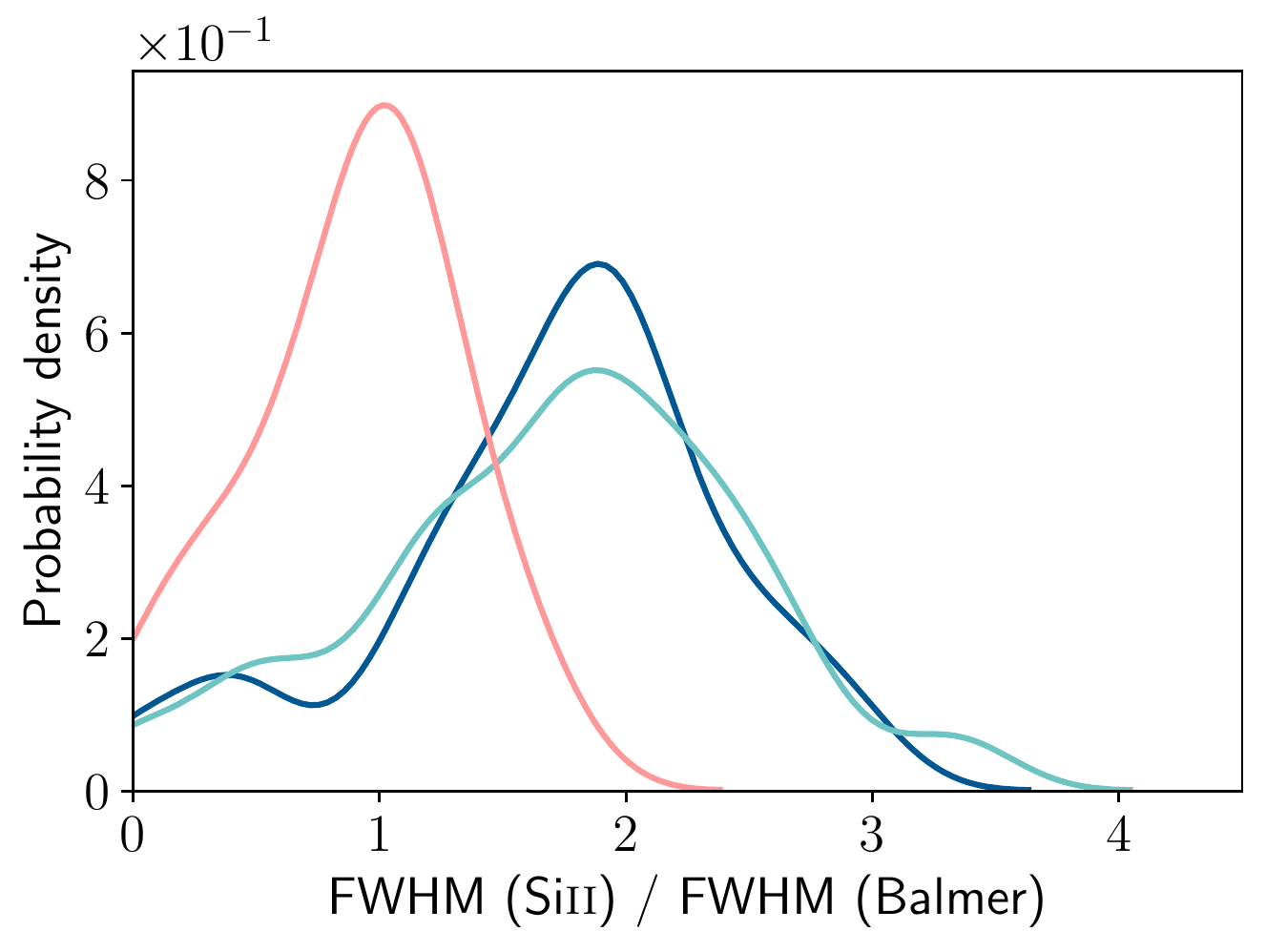}{0.25\textwidth}{}
	\fig{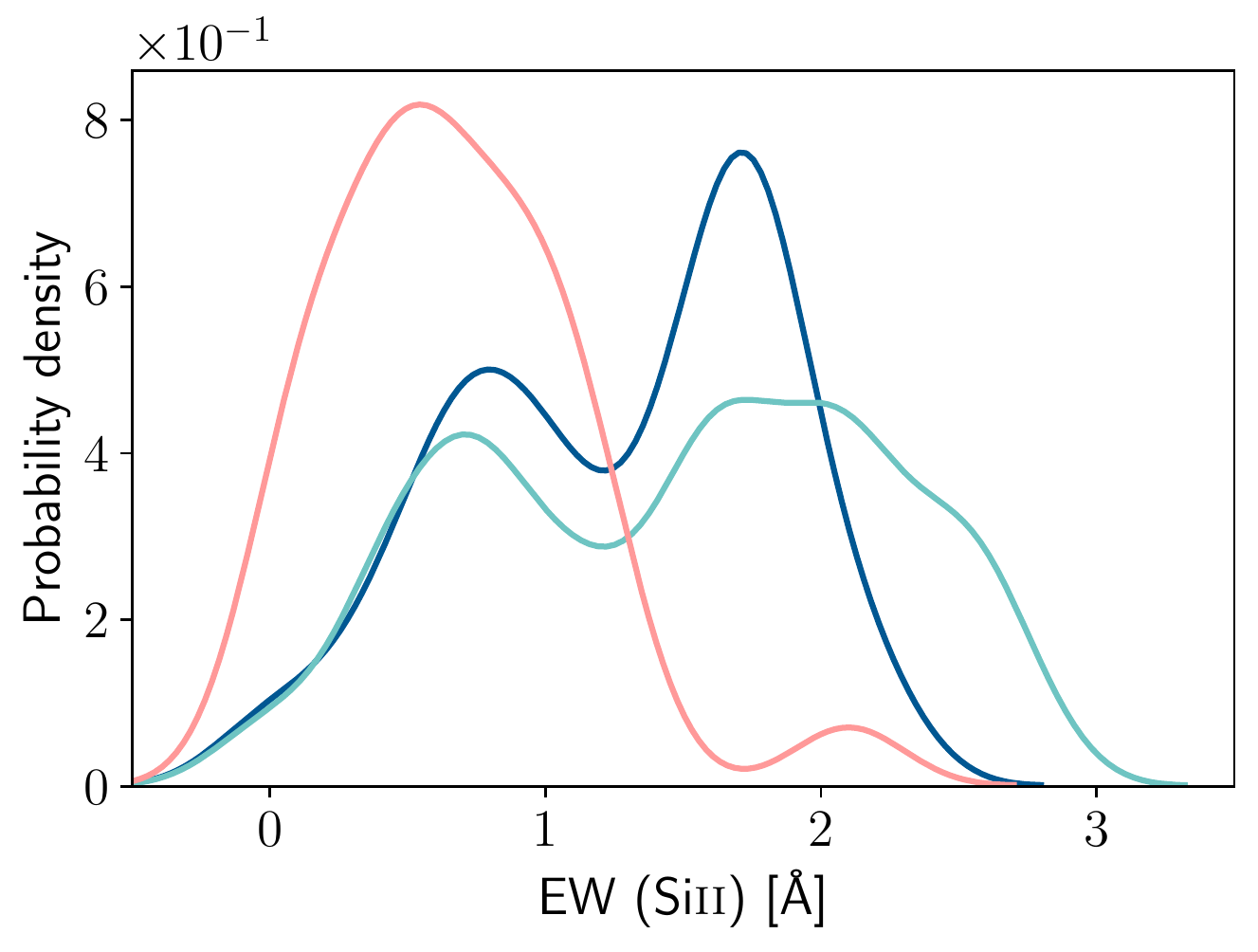}{0.25\textwidth}{}
	\fig{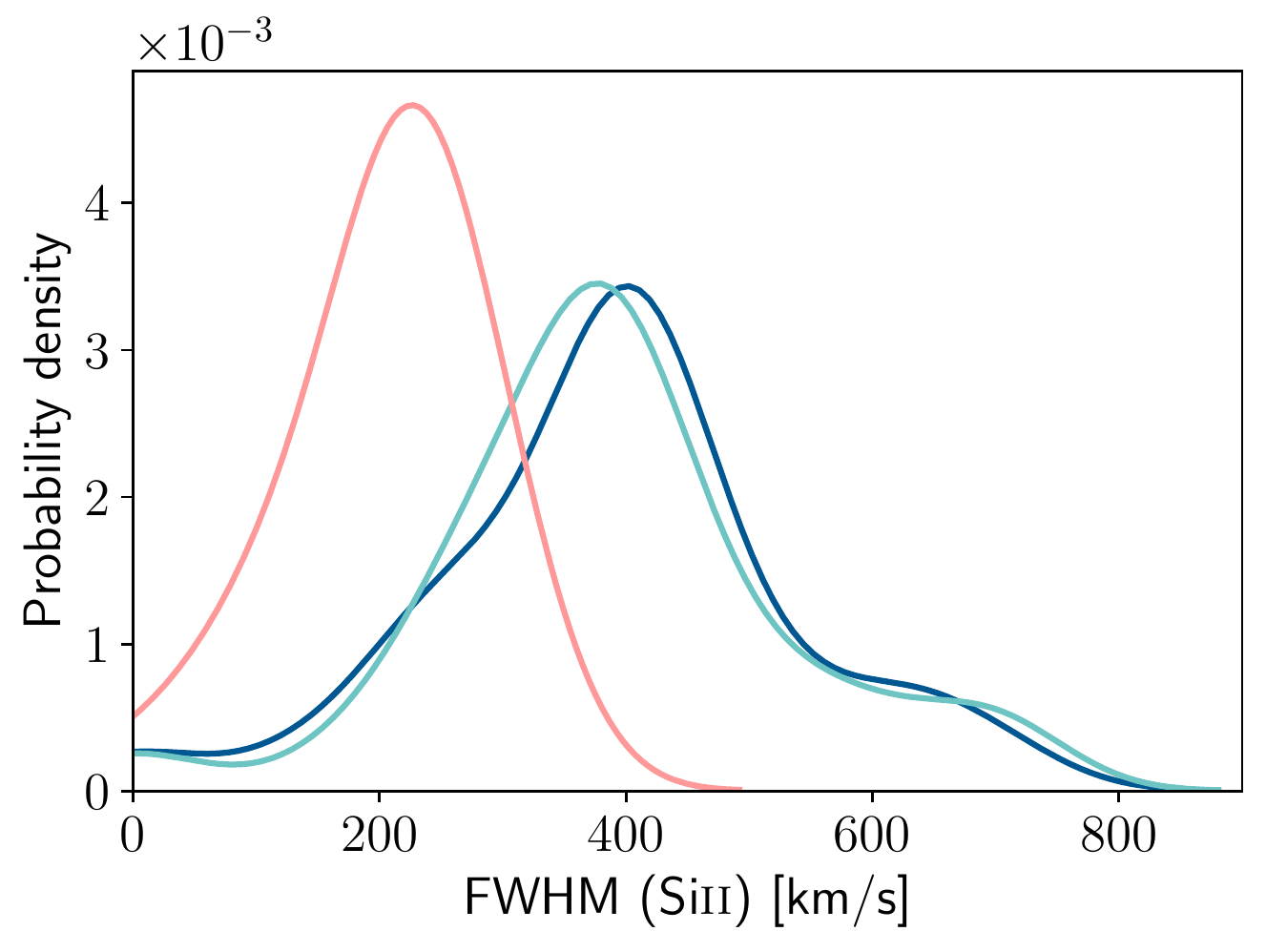}{0.25\textwidth}{}
	}
\gridline{
    \fig{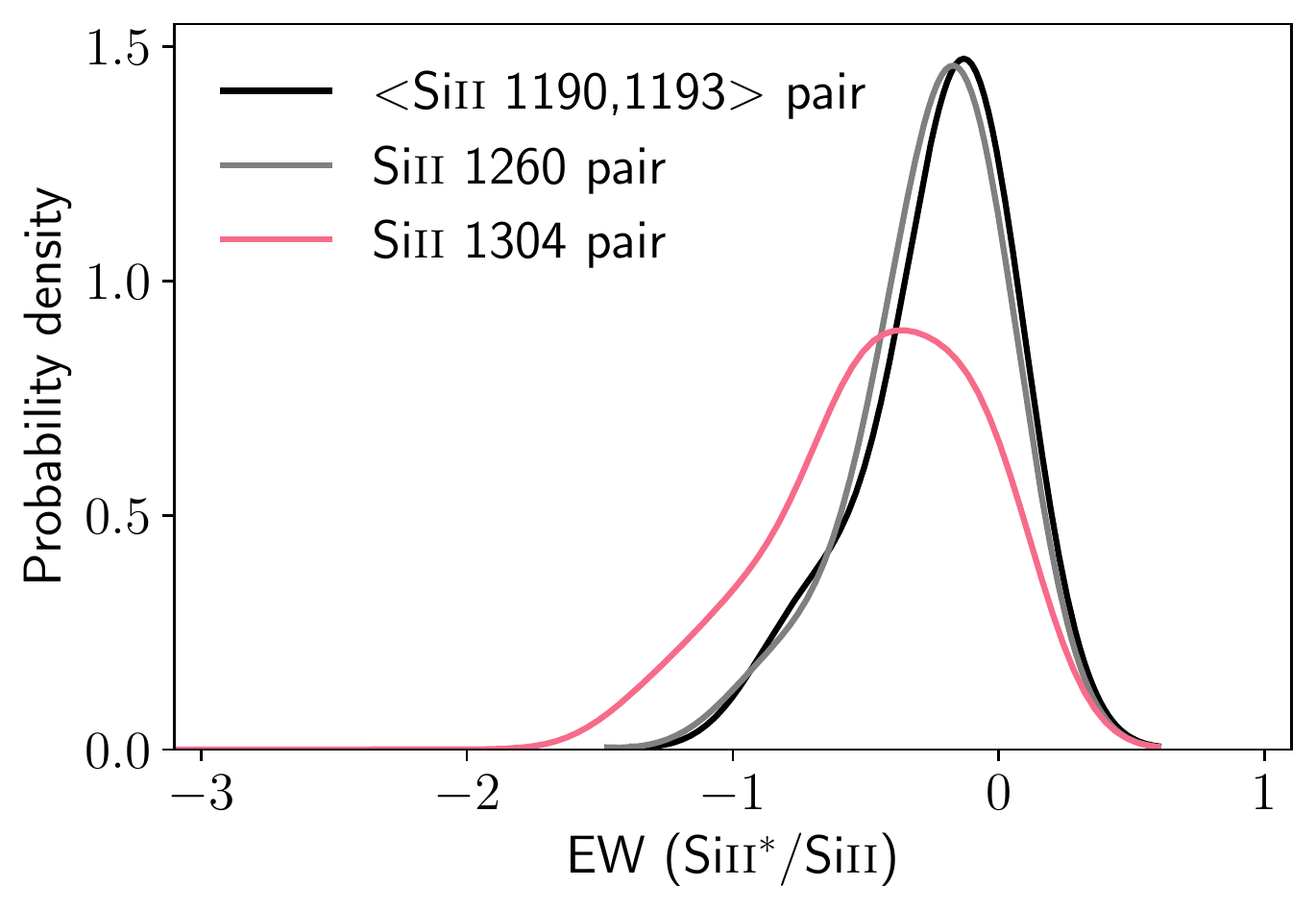}{0.25\textwidth}{}
    \fig{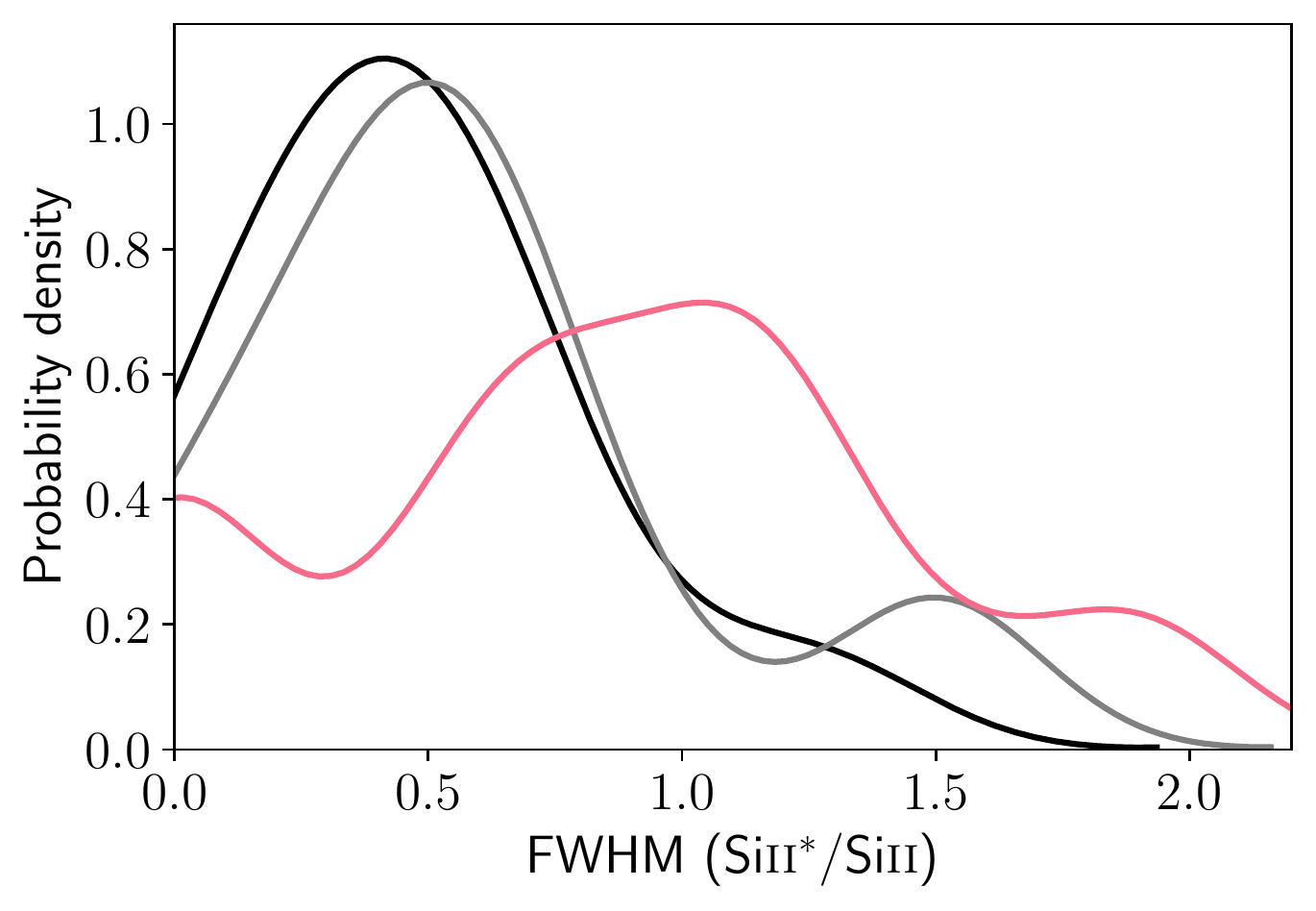}{0.25\textwidth}{}
}
\caption{Kernel density estimates of the distributions of fluorescence lines (upper panel), resonance lines (middle panel), and ratios between the fluorescence and resonance lines (lower panel) for the union of samples considered in this paper. These results imply that a range of conditions is presented, spanning the majority of cases where the fluorescence emission lines are weak and narrow (``ISM-dominated") to a minority of cases where the emission and absorption lines have similar strengths and widths (``wind-dominated").\label{fig:hists}}
\end{figure*}

In Figure~\ref{fig:hists} we show distributions of EW and FWHM of the fluorescence and resonance lines, and their ratios for the union of samples. They are estimated by applying gaussian kernels on the data sets. Four features are evident. First, the emission lines are mostly centered at the systemic velocity of the galaxy, while the absorption lines are blue-shifted. Second, the fluorescence line widths more closely trace the Balmer emission-line widths. Third, the fluorescence emission lines are usually significantly weaker and narrower than the resonance absorption lines. Finally, there is much greater similarity between the widths and strengths of the $\sii$ 1304, $\siistar$ 1309 resonant, fluorescent pair than for the others. This is a significant clue, since the relatively small oscillator strength for the $\sii$ 1304 line means that it will have correspondingly smaller optical depth ($\tau$) at a given velocity than the other resonance lines (e.g., by a factor of $\sim$13 than the $\sii$ 1260 line).

Taken together, these results suggest that in most cases the fluorescence emission lines primarily trace the star-forming ionized ISM and/or the gas with the highest column densities and lowest outflow speeds.

\begin{figure*}
\gridline{
	\fig{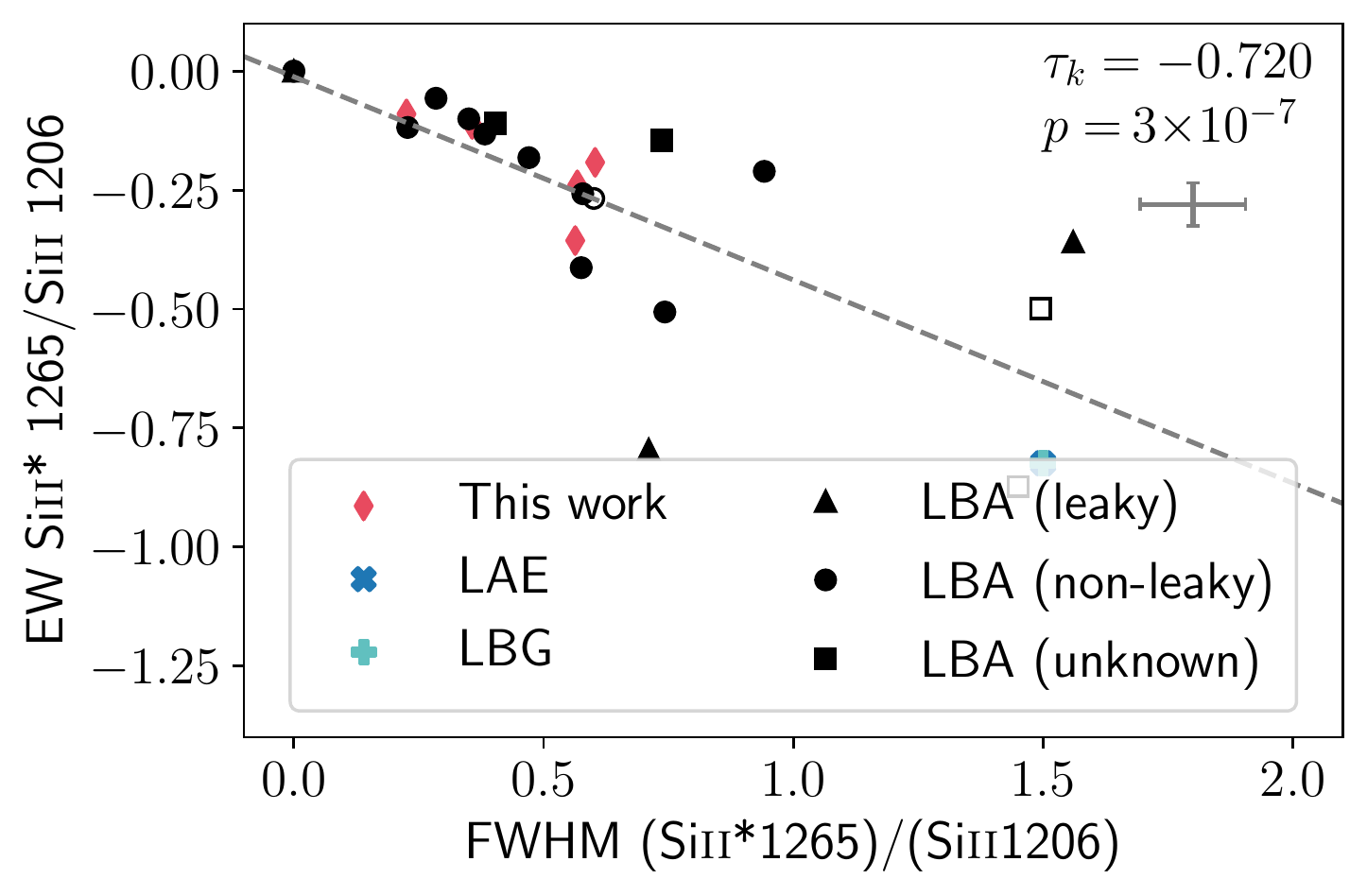}{0.37\textwidth}{}
	\fig{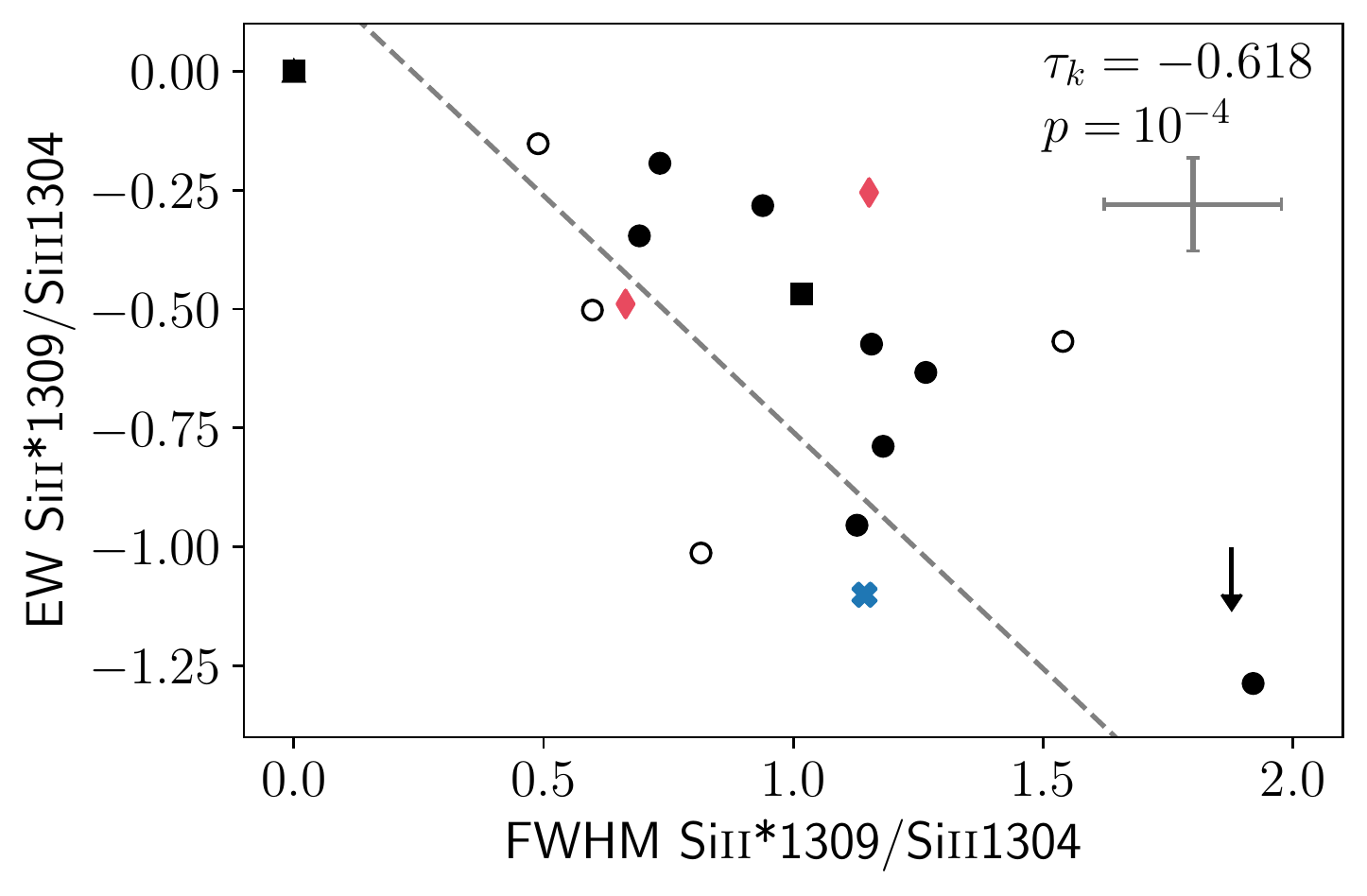}{0.37\textwidth}{}
}
\gridline{
	\fig{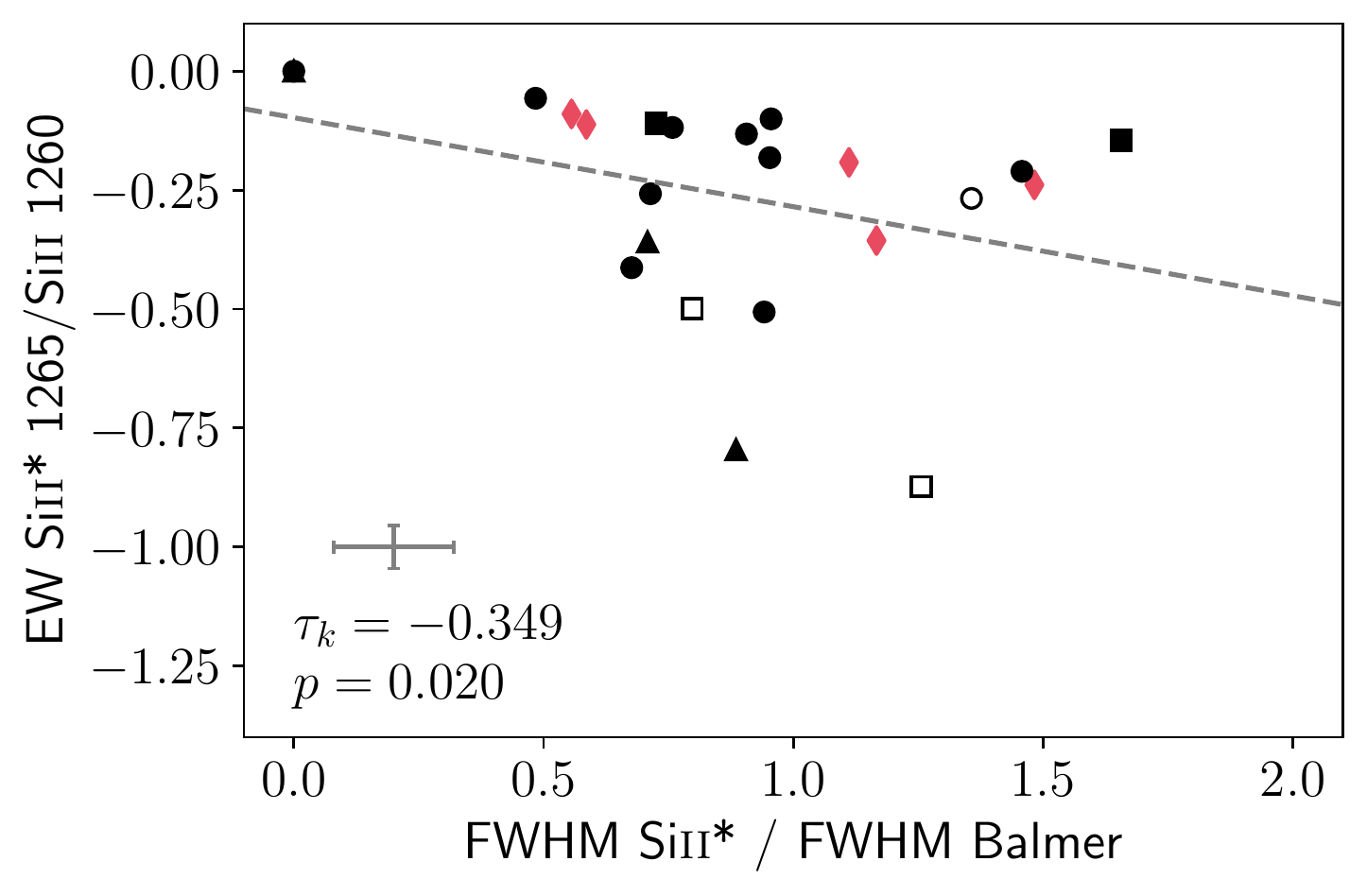}{0.37\textwidth}{}
	\fig{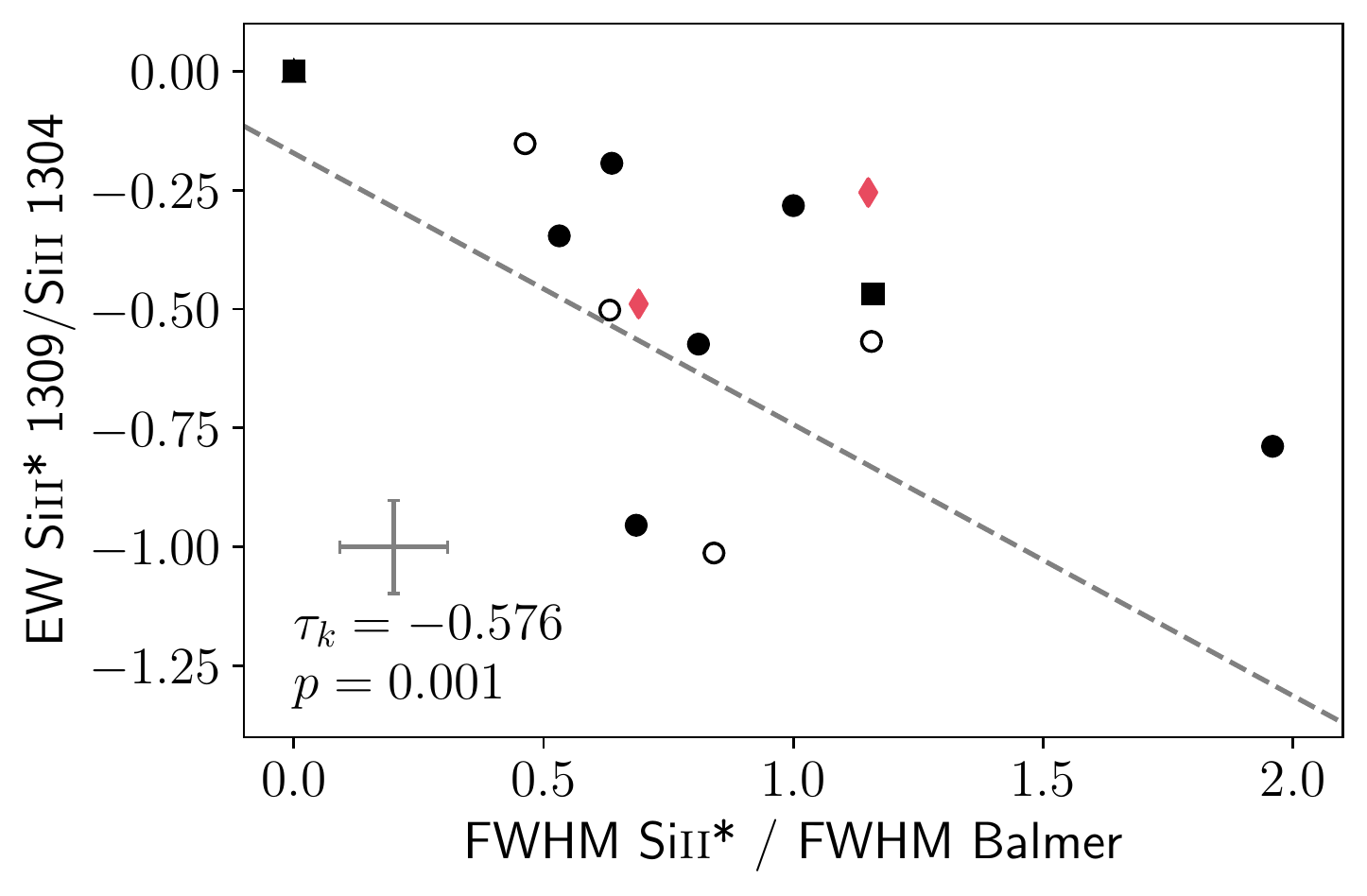}{0.37\textwidth}{}
	}
\gridline{
    \fig{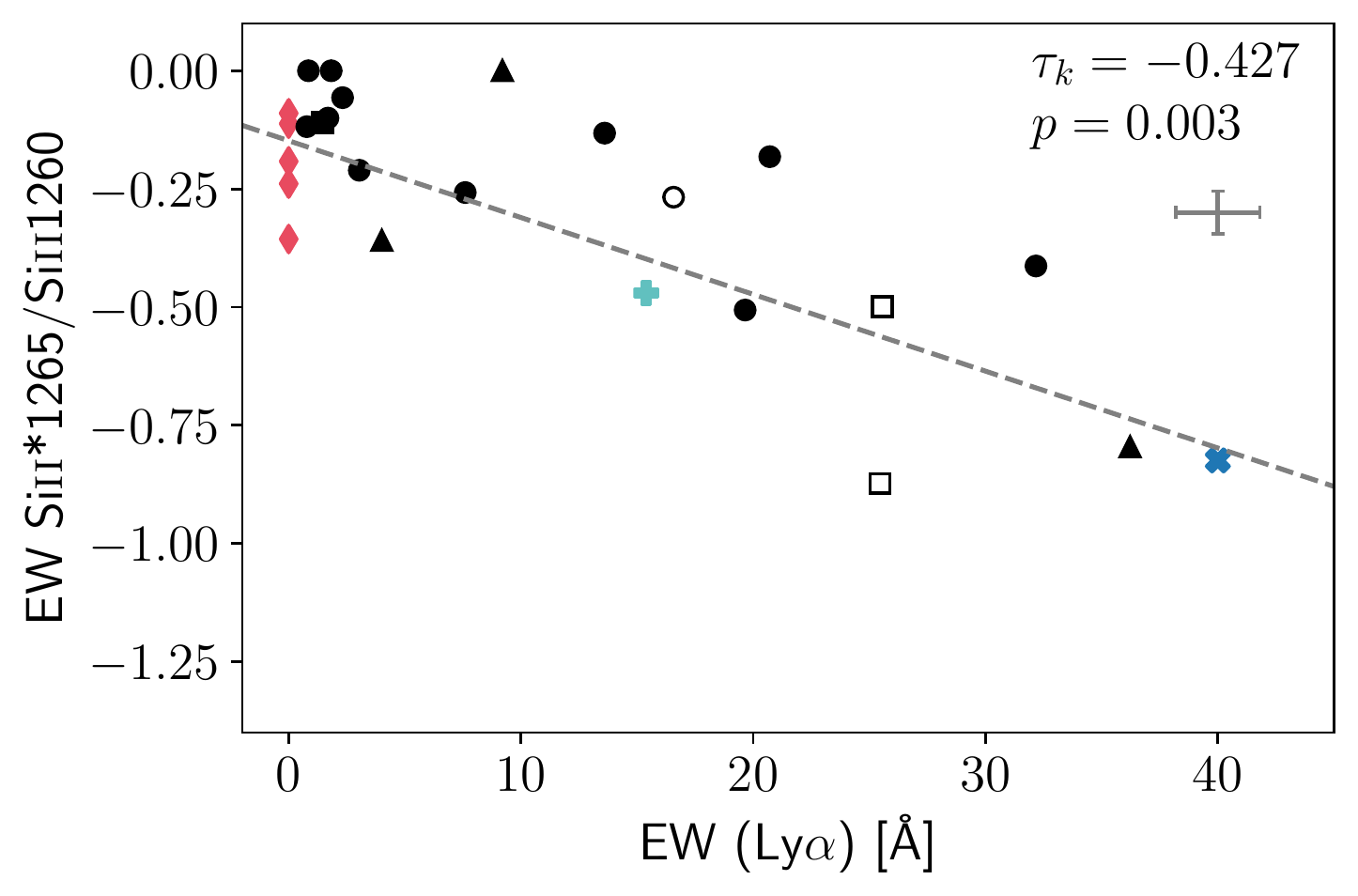}{0.37\textwidth}{}
	\fig{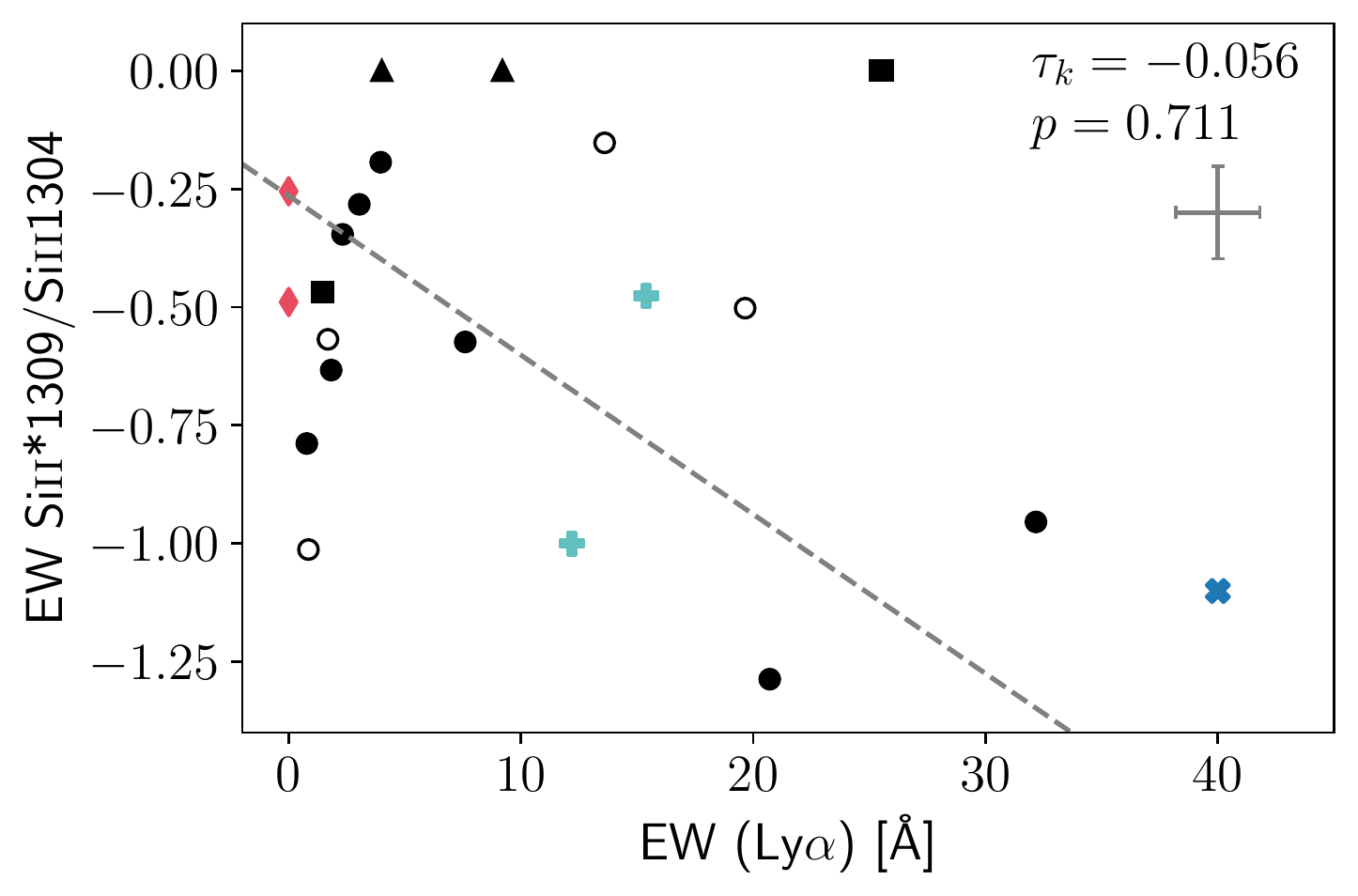}{0.37\textwidth}{}
}
\gridline{
	\fig{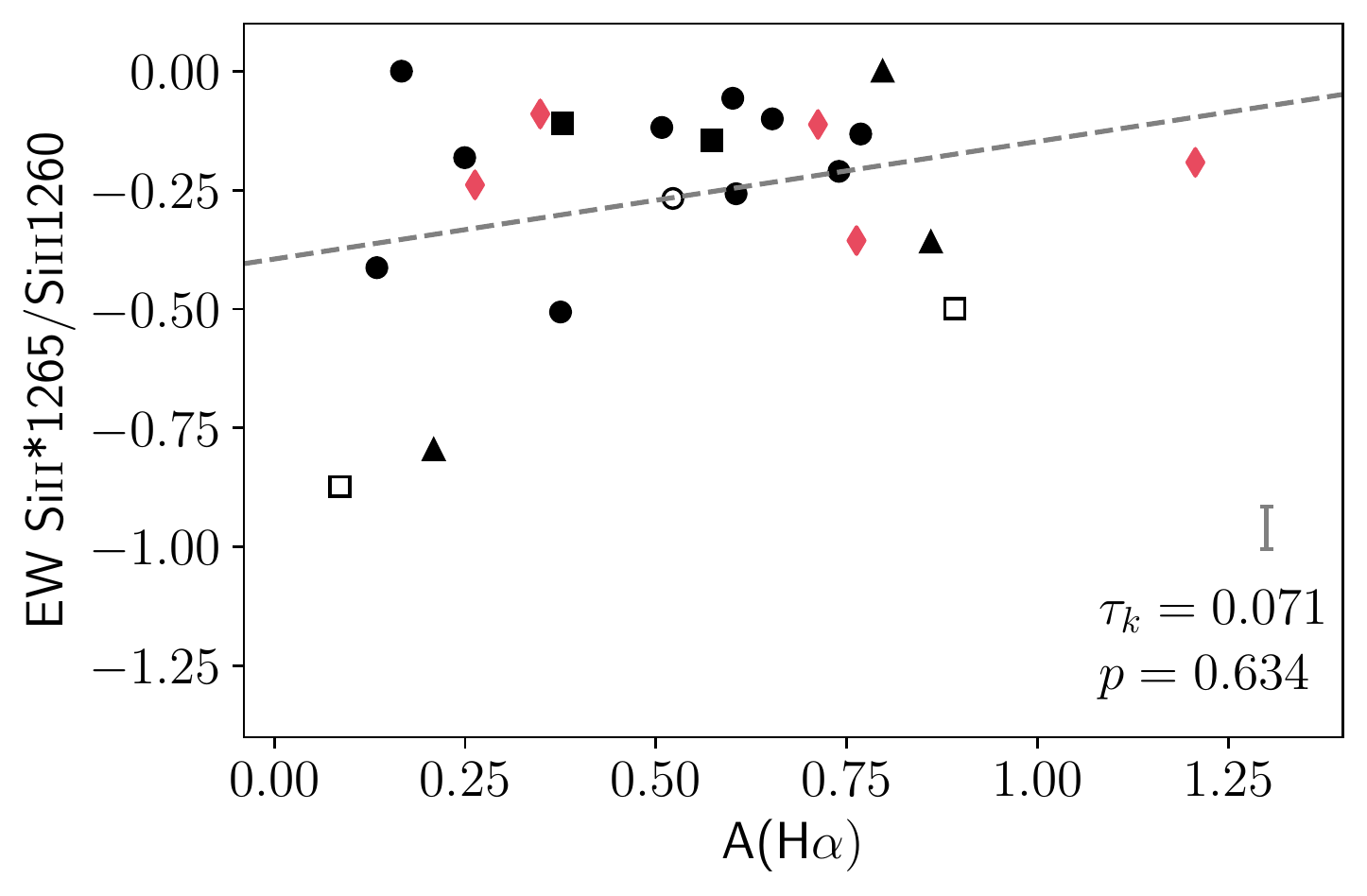}{0.37\textwidth}{}
	\fig{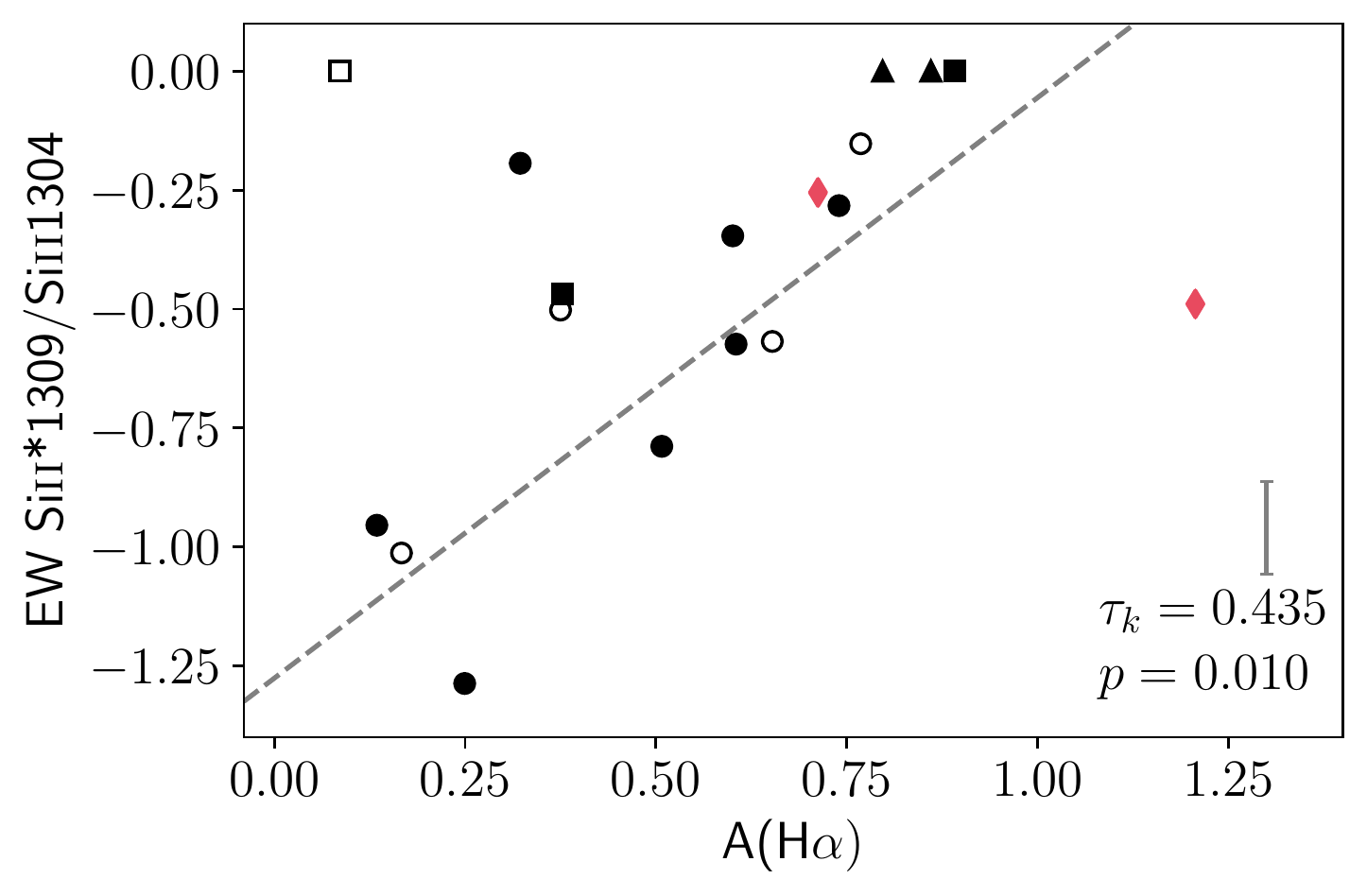}{0.37\textwidth}{}
}
\caption{Scatter plots showing various correlations. Each statistical significance is indicated by Kendall's $\tau$ coefficient and $p$-value. An unfilled marker indicates that the measurement is likely to be affected by additional systematic errors due to a low S/N spectrum or blended lines.  The $\sii$ 1190 pair is quantitatively the same as the $\sii$ 1260 pair, and is hence omitted.\label{fig:corr}}
\end{figure*}

However, we also note that there is a range in the relative strengths of the resonance absorption and fluorescence emission lines, even for the $\sii$ 1260, $\siistar$ 1265 pair (tracing gas with the greatest optical depths). To study any trends that may be manifested in our data, we use the Kendall $\tau$ test in assessing the statistical significance between the ratios of emission/absorption line strength and other parameters. Those correlations, and their corresponding correlation coefficients $\tau_{\rm k}$ and $p$-values, are shown on each of the scatter plot in Figure~\ref{fig:corr}.

Most notable is the correlation between the ratio of fluorescence line EW to resonance line EW with the ratio of the FWHM of the two lines. A weaker correlation (but still significant) is present between the fluorescence to resonance line EW ratio and the fluorescence to Balmer line widths ratio. Both these correlations imply that we observe a range of conditions, spanning the majority of cases where the fluorescence emission lines are weak and narrow to a minority of cases in which the emission and absorption lines have similar strengths and widths.

\subsection{Images\label{sec:res_img}}

The five galaxies for which we have obtained images of the $\feiistar$ fluorescence line emission are all members the majority population of weak and narrow emission lines. In Figure~\ref{fig:contour}---we see that in each case the $\feiistar$ emission is concentrated within the star-bursting regions, which are characterized by high surface-brightness UV continuum and [O{\textsc{ii}}] 3727 line emission.

To further quantify the consistency between the amount of $\siistar$ emission present in the spectra and the amount of $\feiistar$ emission in the images, we translate the photometric data of $\feiistar$ to spectral fluxes via the following relation:
\begin{equation}
	F({\feiistar}) = F_{\lambda}({\feiistar}) \times \Delta W({\rm{F280N}}),
	\label{eq:flux_img}
\end{equation}
where $F_{\lambda}$ is the sum of the flux density inside the aperture of the same size as the COS, and $\Delta W$ is the width of the filter. We take $\Delta W({\rm{F280N}}) = 42.52 \mAA$ from the WFC3 instrument handbook. A similar calculation is done to obtain the [$\oii$] emission-line fluxes.

In Table~\ref{tab:flux} we summarize all the spectral and imaging measurements. Note that the uncertainties quoted for the $\feiistar$ flux include only those from the background fluctuations. It is worth pointing out that ${\feiistar}$ EW$/\lambda$ is subjected to an additional systematic uncertainty induced during the process of estimating its continuum level by an interpolation between the SDSS and COS spectra.

Given that the associated resonance absorption lines will be optically thick, the repeated scattering re-absorption of resonance photons will eventually convert all the absorbed photons into fluorescence emission lines. In this case:

\begin{subequations}
\label{eq:fe_tot}
\begin{eqnarray}
	F({\feiistar, tot}) &=& {\rm{EW_{2587}}}F_{\lambda, {\rm cont}}(2587) + \nonumber \\
	&& {\rm{EW_{2600}}}F_{\lambda, {\rm cont}}(2600),
\end{eqnarray}

To estimate the EWs for these {Fe{\sc{ii}}} absorption lines we assume that they have the same value of EW/$\lambda$ as the {Si{\sc{ii}}} absorption lines measured with COS. More specifically, we use combinations of {Si{\sc{ii}}} line strengths that lead to the same mean oscillator strength $f$ as the corresponding {Fe{\sc{ii}}} lines:

\begin{eqnarray}
	\frac{\rm EW_{2587}}{\lambda_{2587}} &=& \frac{1}{2} \Big( \frac{\rm EW_{1304}}{\lambda_{1304}} + \frac{\rm EW_{1190}}{\lambda_{1190}} \Big)\\
	\frac{\rm EW_{2600}}{\lambda_{2600}} &=& \frac{1}{2} \Big( \frac{\rm EW_{1193}}{\lambda_{1193}} + \frac{\rm EW_{1190}}{\lambda_{1190}} \Big).
\end{eqnarray}
\end{subequations}

Strictly speaking, Equation \ref{eq:fe_tot} applies to photon rates rather than fluxes, but we ignore this, since the fluorescence and resonance photons have nearly the same energy.

While the uncertainties in the measured/estimated fluxes in Table~\ref{tab:flux} are substantial, we conclude that the amount of $\feiistar$ emission is consistent with COS $\siistar$ emission. This implies that a significant fraction of the weak and narrow fluorescence emission seen in the spectra can be ascribed to the emission associated with regions of intense star-formation seen in the images (consistent with an ISM-dominated origin).

\begin{deluxetable*}{cccccc}
	\tablecaption{Line properties inferred from WFC3 imaging\label{tab:flux}}
	\tablecolumns{6}
	\tablewidth{0pt}
	\tablehead{
	\colhead{} &
	\colhead{$\feiistar$} &
	\colhead{$\feiistar$} &
	\colhead{[$\oii$]} &
	\colhead{$\feiistar_{total}$ \tablenotemark{a}} &
	\colhead{}\\
	\colhead{} &
	\colhead{($10^{-16} \, \rm erg \, cm^{-2} \, s^{-1} \, \mAA^{-1}$)} &
	\colhead{($10^{-4} \frac{\rm{EW}}{\lambda 2620}$)} &
	\colhead{($10^{-16} \, \rm{erg \, cm^{-2} \, s^{-1} \, \mAA^{-1}}$)} &
	\colhead{($10^{-16} \, \rm erg \, cm^{-2} \, s^{-1} \, \mAA^{-1}$)} &
	\colhead{}
	}
	\startdata
J0831(S) & 5.1 $\pm$ 7.5 & 6.1 $\pm$ 8.9 & 118 $\pm$ 9 & $20 \pm 13$ \\
J0831(N) & 19.0 $\pm$ 8.3 & 10.1 $\pm$ 4.4 & 224 $\pm$ 16 & $50 \pm 28$ \\
J1157 & 23.0 $\pm$ 7.0 & 12.6 $\pm$ 3.9 & 168 $\pm$ 5 & $30 \pm 16$ \\
J1210 & 31.4 $\pm$ 7.6 & 12.9 $\pm$ 3.1 & 157 $\pm$ 5 & $65 \pm 36$ \\
J1618 & 31.1 $\pm$ 10.5 & 10.8 $\pm$ 3.7 & 203 $\pm$ 3 & $71 \pm 40$ \\
\enddata
\tablenotetext{a}{$\feiistar_{total}$ is the expected total flux of $\feiistar$ emission estimated from $\siistar$ emission, as given in Equation~\ref{eq:fe_tot}.}
\tablecomments{The flux densities of $\feiistar$ are measured in apertures the same size as the COS ($r=1.25$\arcsec), and the uncertainties are the 1$\sigma$ variances from many measurements on the blank sky using the same apertures.}
\end{deluxetable*}

\section{Discussion\label{sec:discuss}}

So far we have presented a continuum of relative emission-line strengths and widths. In the majority of cases the emission lines are much weaker and narrower than the absorption lines. Direct imaging shows that the fluorescence emission in such cases is produced in regions with high surface-brightness in the UV continuum and [$\oii$] 3737 emission lines. These results indicate that in these cases the emission is ``ISM-dominated" (the outflow makes little contribution). In other cases, the presence of stronger and broader emission lines suggest that the outflow is making a significant contribution. In this section, we discuss implications for wind structure based on both these empirical properties and a simple outflow model.

\subsection{A Brief Primer\label{sec:primer}}

The use of the fluorescence emission lines as probes of the structure of galactic outflows has been discussed at length in a number of papers \citep{Prochaska2011,Scarlata2015,Zhu2015}. Here, we want to simply summarize a few salient points that will be used below to interpret our data.

Concerning the resonance absorption lines, the profiles (line depth at a given velocity) are specified by the product of the column density of the relevant ion along the line-of-sight times the oscillator strength of a given transition times the covering factor (the fraction of the background continuum source covered by the foreground gas at a given velocity). It is important to note that the observed $\sii$ resonance transition span a range of 13 in oscillator strength and hence optical depth (see Table ~\ref{tab:atom}).

Once a resonance photon has been absorbed, the excited ion will decay radiatively into either the ground state (emission due to resonance scattering) or into the excited fine-structure level (emission due to fluorescence). The ratio of the number of photons initially produced in these two ways is just given by the ratio of the respective Einstein A's (see Table~\ref{tab:atom}). The most and least effective reprocessing of absorbed resonance photons into fluorescence emission occurs for the $\sii$ 1190, 1193 pair (84\% efficiency) and the $\sii$ 1260, 1265 pair (16\% efficiency). This will be true in the limit where the resonance lines are optically thin (a single absorption occurs). If the resonance lines are optically think, then eventually (via multiple absorption events) all the absorbed photons are converted into fluorescence emission lines.

\begin{figure}
\gridline{
	\fig{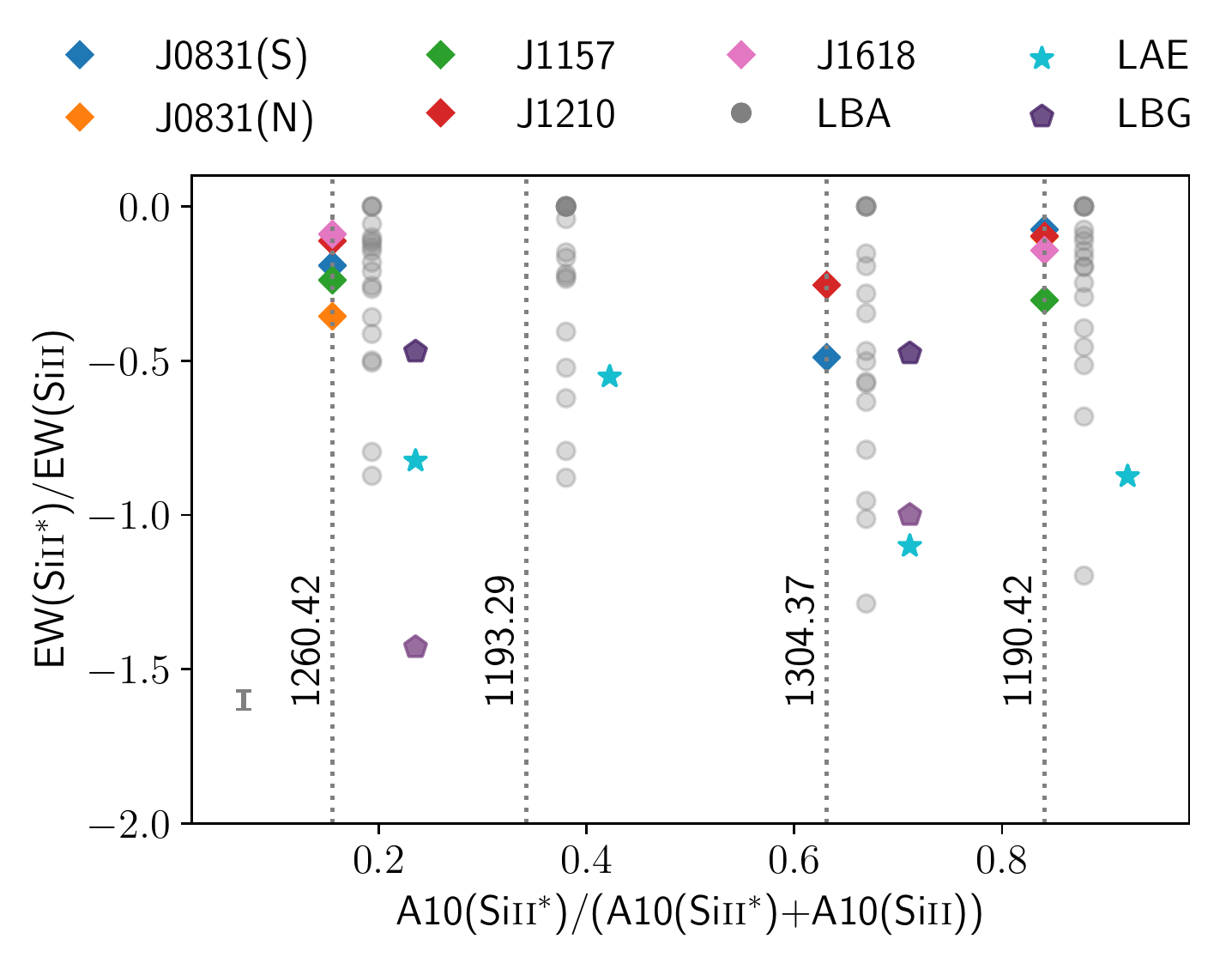}{0.42\textwidth}{(a)}
}
\gridline{
	\fig{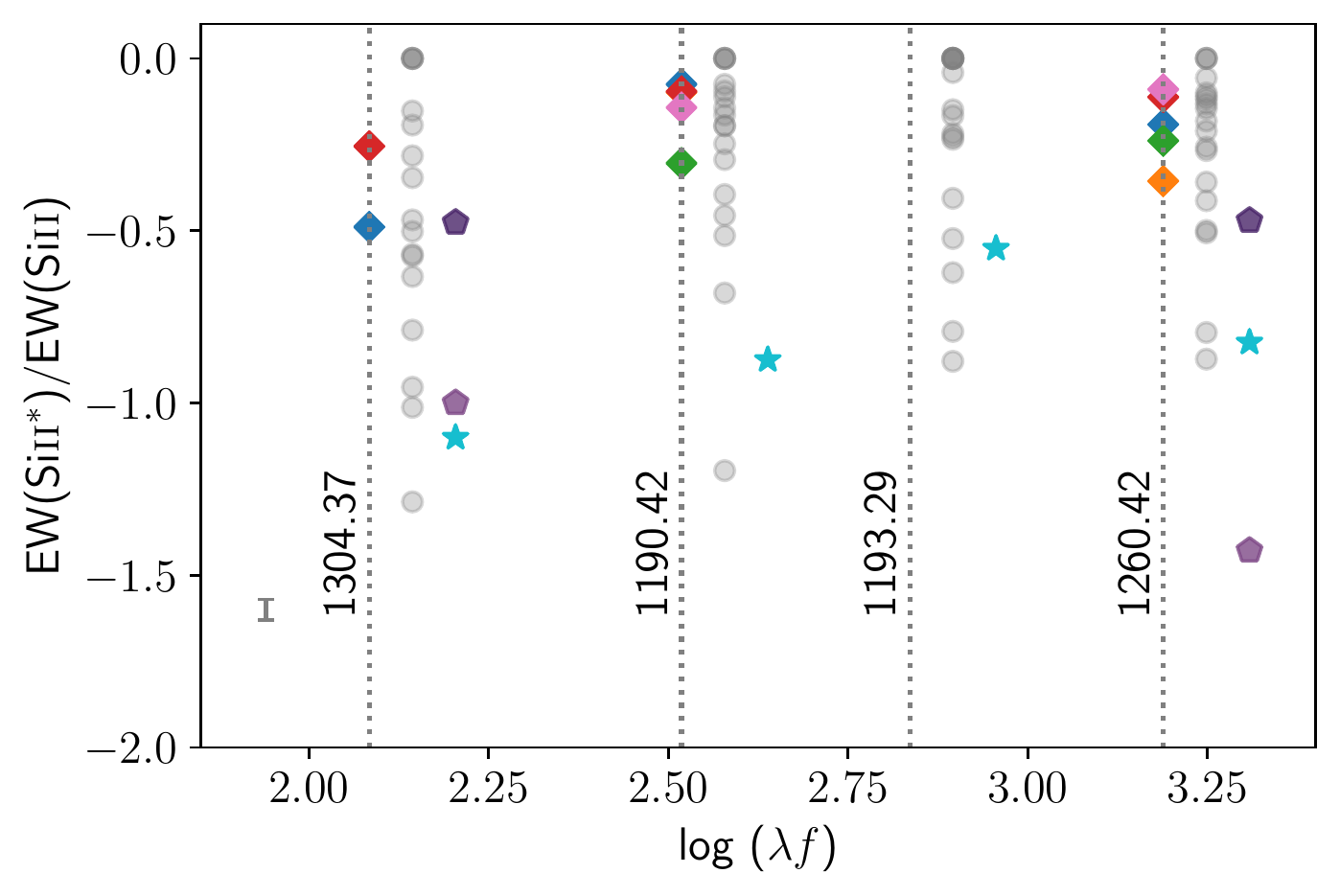}{0.42\textwidth}{(b)}
}
\gridline{
	\fig{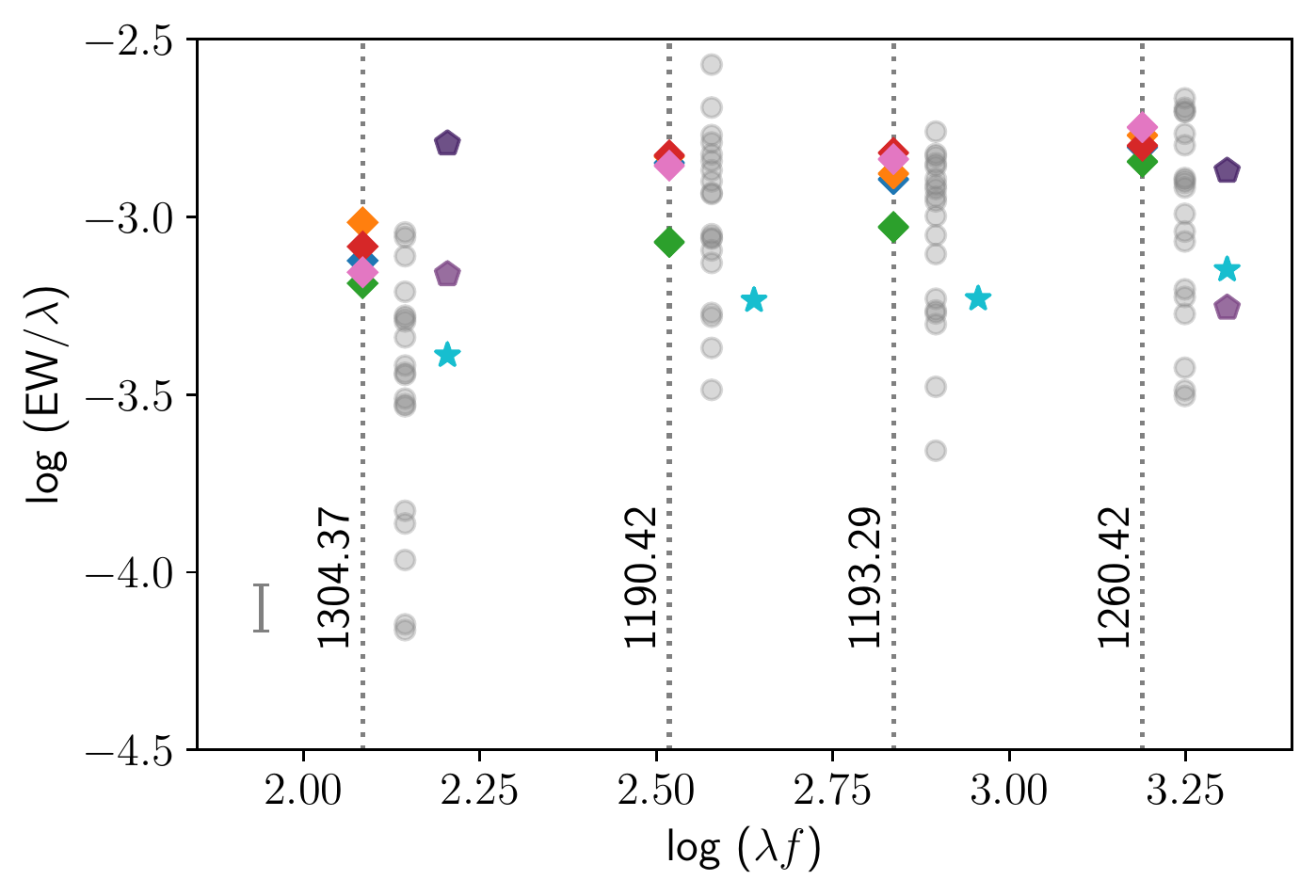}{0.42\textwidth}{(c)}
}
\caption{(a) EW ratios of fluorescence/resonance lines plotted against ratios of Einstein A's. No correlation is shown. (b) EW ratios plotted as a function of oscillator strengths $f$. (c) Evolution of rest-frame EWs of $\sii$ transitions plotted as a function of $f$. The characteristic transition from optically thin absorption (EW/$\lambda \propto f \lambda$) to optically thick absorption (flat) is evident except for J1157 (in green). Data from different samples are offset for clarity. Typical error-bars are shown in gray.\label{fig:flam}}
\end{figure}

With that in mind, we show in Figure \ref{fig:flam} the correlation between the oscillator strength and EW for four of the observed $\sii$ transitions. From this we see that the absorption lines are mainly optically-thick for the $\sii$ 1260, 1193, and 1190 transitions (they show the same EWs), but are starting to become optically thin for the $\sii$ 1304 line.

\begin{figure}
  \centering
    \includegraphics[width=0.47\textwidth]{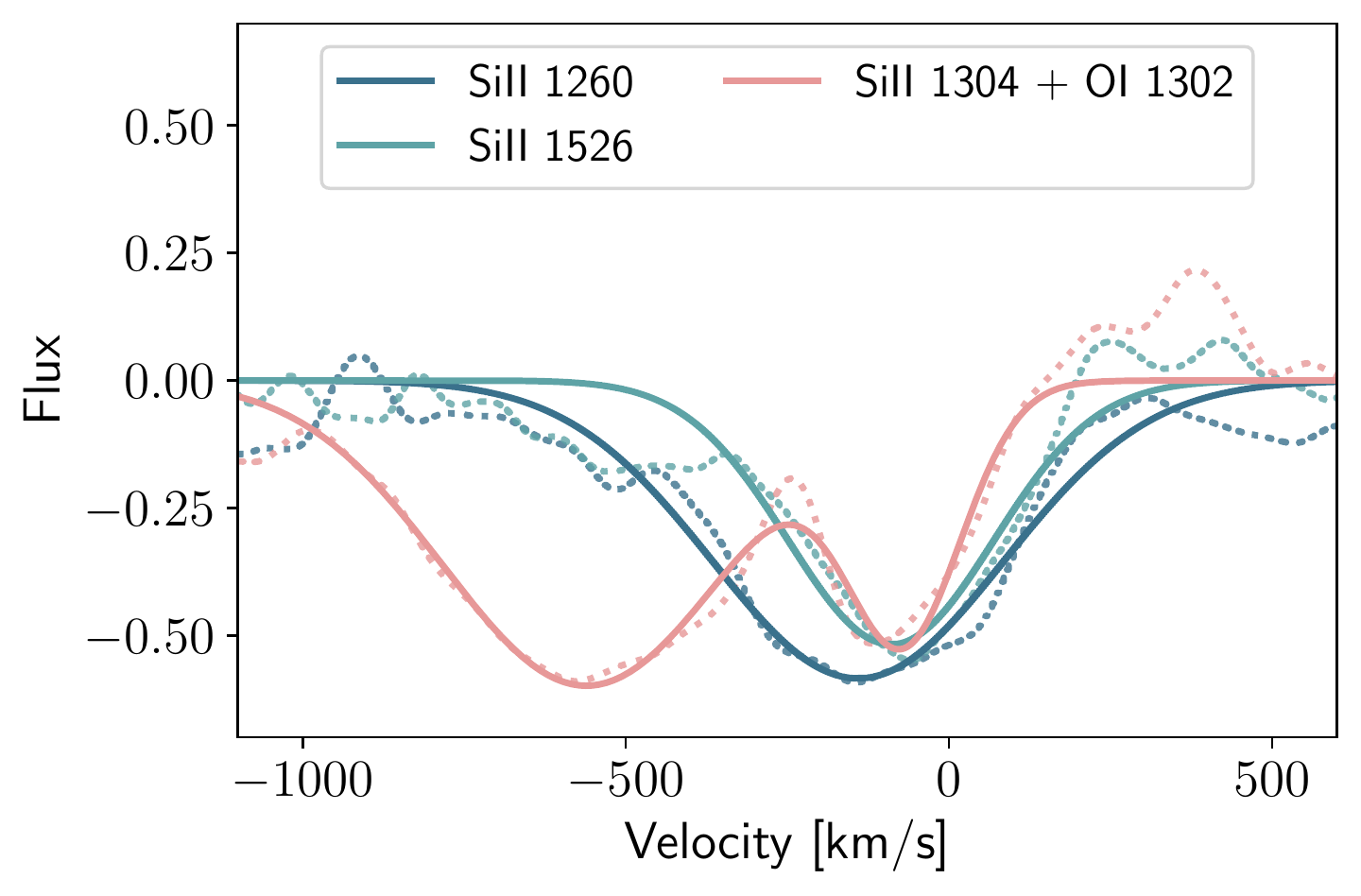}
\caption{Comparison of the absorption-line profiles of {Si\sc{ii}} 1260, 1526, and 1304 from a stacked spectrum including all the LBAs. The dashed lines are the data, while the solid lines are the gaussian fits. Note that the {O\sc{i}} 1302 line is responsible for the strong and broad absorption feature seen centered around -600 km/s blue-ward of {Si\sc{ii}} 1304. The velocities on the x-axis are measured relative to the systemic velocity of the galaxy. The $\sii$ 1260 (1304) line is the most (least) optically thick, showing that the highest velocity outflowing gas has the lowest column density.\label{fig:stack}}
\end{figure}

It is even more instructive to compare the absorption-line profiles for different $\sii$ transitions. To do so, we stack all the COS spectra of the LBAs and fit the resulting profiles gaussians. The results are shown in Figure \ref{fig:stack} for the $\sii$ 1260, 1527, and 1304 lines (in order of decreasing $\tau$), with the lines becoming progressively narrower and less blue-shifted. This shows that the highest velocity material has the lowest column density. Combining this with our previous result showing that the best agreement between the strengths and widths of the resonance and fluorescence lines was for the $\sii$ 1304, 1309 pair (Figure \ref{fig:hists}) implies that the observed fluorescence emission lines are primarily tracing the highest column-density gas (e.g., either a static ISM or the slower parts of the outflow).

Figure \ref{fig:flam} also shows that the ratio of the strengths of the fluorescence emission lines and the resonance absorption lines does not depend on the reprocessing efficiency. This implies that the {\it gas responsible for the observed fluorescence emission} is optically thick for the associated resonance lines (meaning that all the absorbed photons there are being reprocessed into fluorescence emission lines). The fact these emission lines are usually much weaker (and narrower) than the absorption lines returns us to the question raised in the introduction to the paper---why are these emission lines usually so weak and narrow in the majority of the starbursts?

There are several processes/circumstances that could explain this: 1) outflows are confined to small solid angles ($\ll 4 \pi$ sr); 2) the material producing the absorption is located at radii beyond those probed by the spectra; 3) the emission-line photons from the outflow are absorbed by dust in the outflow rather than escaping. The fact that outflows are seen in 100\% of the galaxies we have observed is inconsistent with the first explanation, and so we do not consider it further. To test the other possibilities, and gain more insight into the nature of the emission lines, we examine correlations between their properties and other potentially relevant parameters.

\subsection{Correlations\label{sec:corr}}

Here we examine correlations of properties of fluorescence and resonance absorption lines with properties of the galaxy and its starburst.

To begin, we can further test our inference of a continuum of properties, ranging from spectra in which the fluorescence lines are ISM-dominated (the majority) to more wind-dominated (the minority). We have argued that the key diagnostics are the ratios of both the EWs and line-widths of the fluorescence emission lines and the resonance absorption lines. Indeed, Figure~\ref{fig:corr} shows that these two ratios are correlated for both the most and least optically thick transitions ($\sii$ 1260 and 1304). We also see that as the EW ratio increases, the ratio of the widths of the fluorescence emission lines and the Balmer emission lines (tracing the static ionized ISM) increases.

\begin{figure*}
\gridline{
    \fig{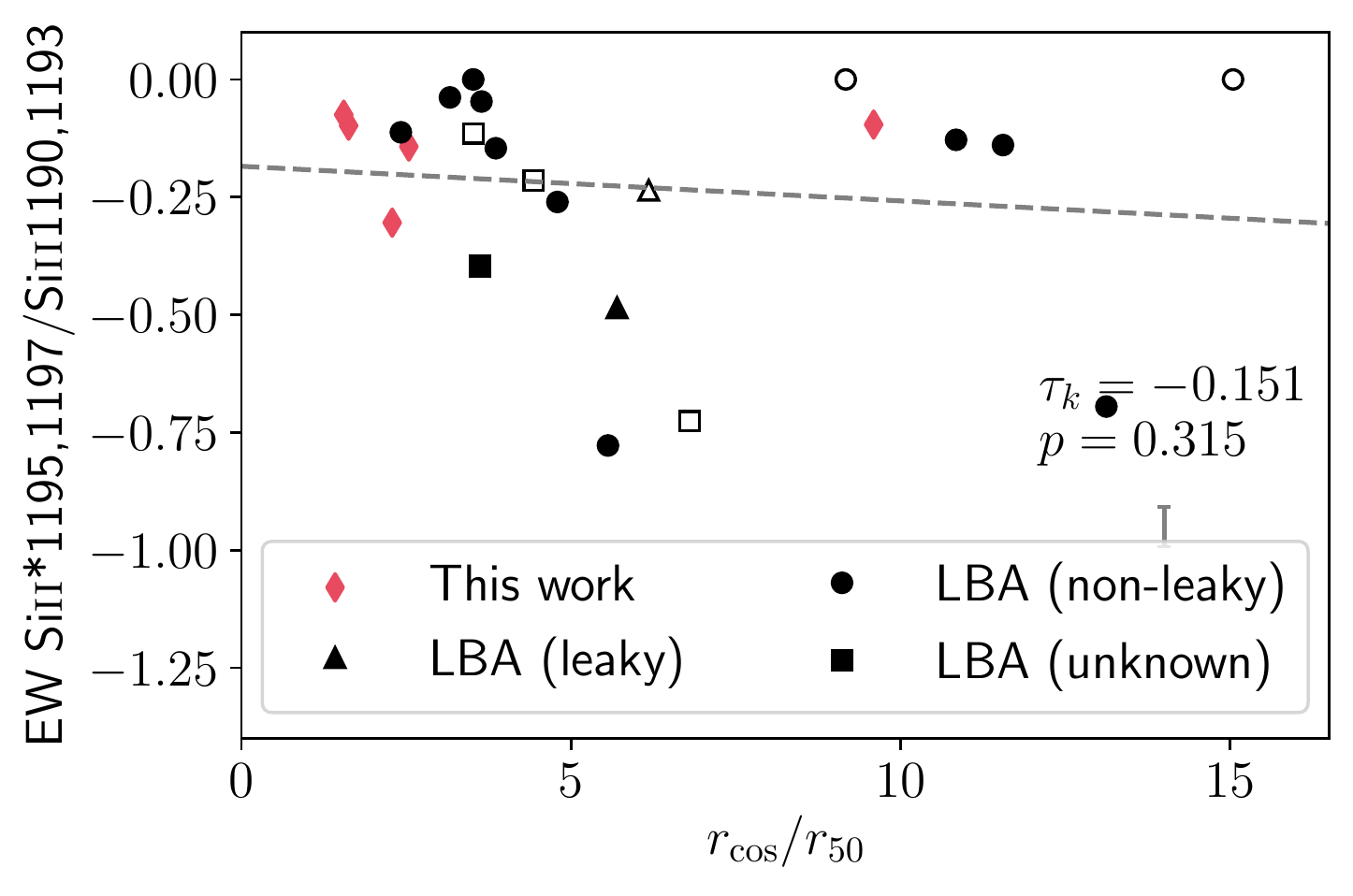}{0.32\textwidth}{}
	\fig{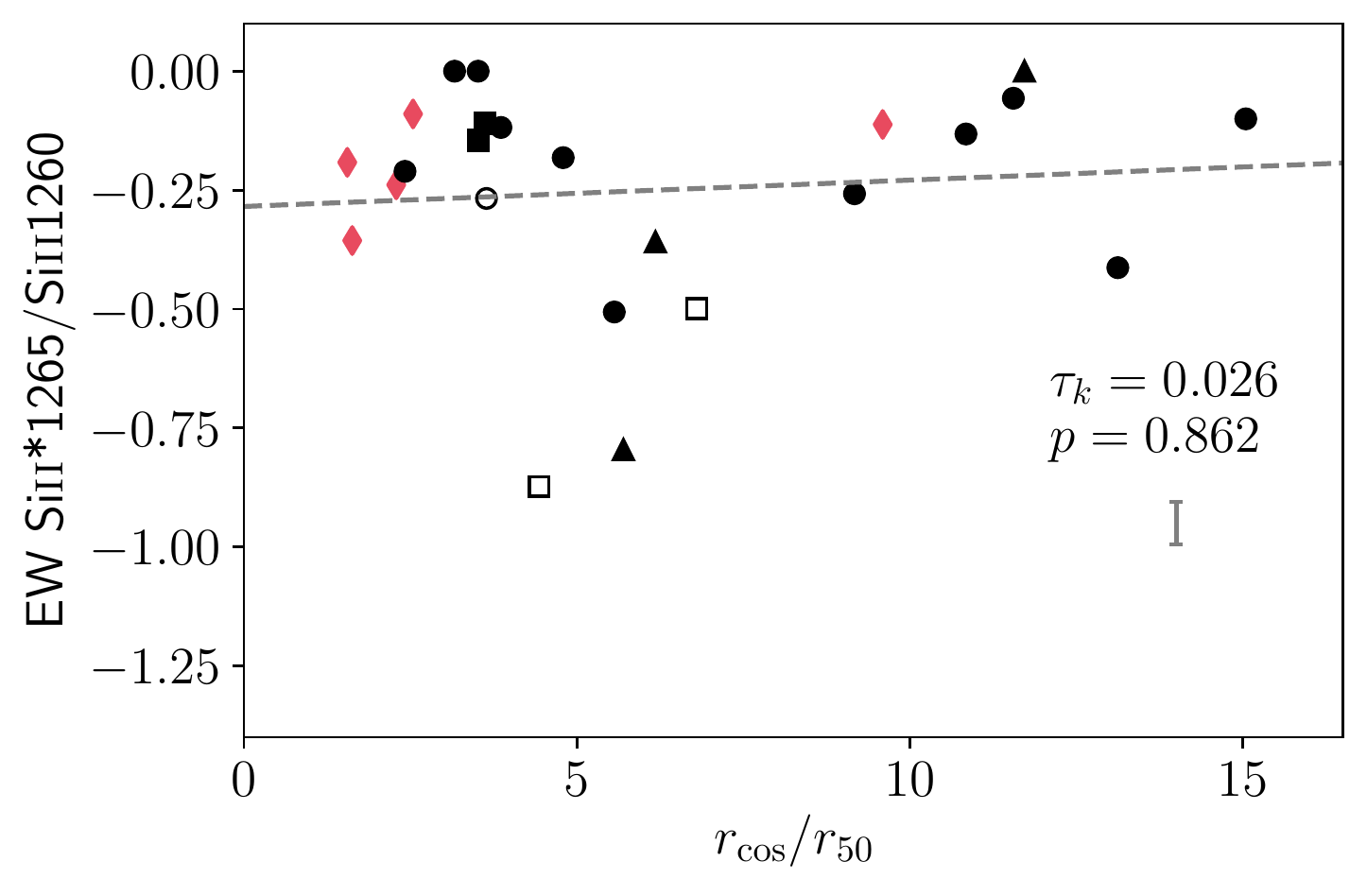}{0.32\textwidth}{}
	\fig{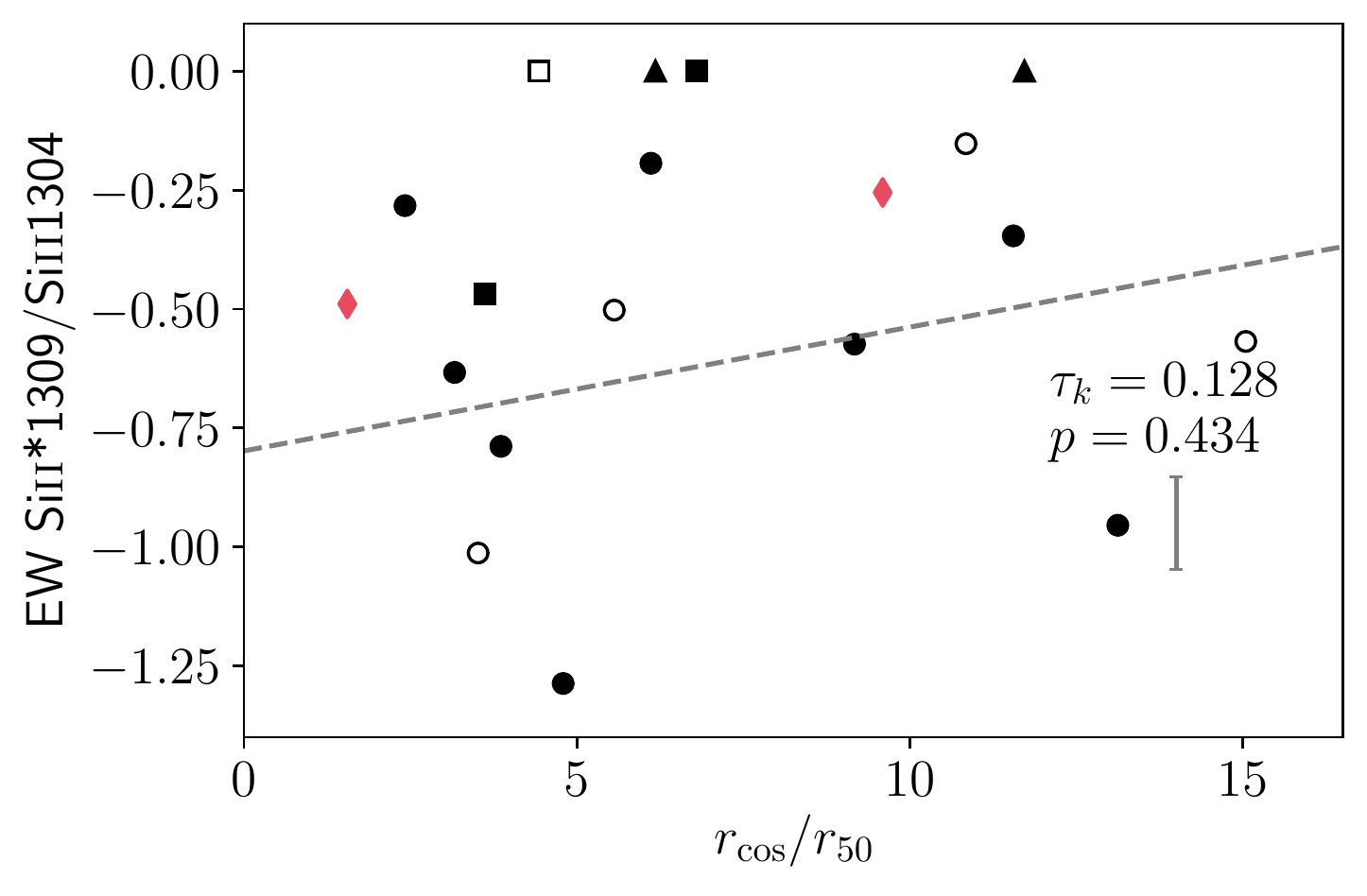}{0.32\textwidth}{}
}
\caption{The relative strengths of the fluorescence emission lines and the resonance absorption lines plotted as functions of the projected size of the COS aperture. Typical error-bars are shown on the right in gray. An unfilled marker indicates that the measurement is likely to be affected by additional systematic errors due to a low S/N spectrum or blended lines. Little correlation is found. \label{fig:r50}}
\end{figure*}

Perhaps most importantly, we find little correlation between relative strengths of the fluorescence emission lines and the resonance absorption lines with the projected size of the COS aperture (e.g., the ratio of radius of the COS aperture and the radius of the starburst measured in the NUV). In fact, Figure~\ref{fig:r50} reveals that in most cases, the fluorescence lines remain weak even when the COS aperture is more than an order-of-magnitude larger than the starburst. We will explore the implications of this quantitatively in the next section.

As noted above, the weakness of the emission lines could be due in part to their absorption by dust that is located in the outflow. To test this, we show the correlations between the relative emission-line strengths and the amount of extinction (from the Balmer decrement in SDSS data). We see no correlation in the case of the $\sii$ 1260, 1265 pair, but a significant correlation in the case of the $\sii$ 1304, 1309 pair. Since the $\sii$ 1260 transition is much more optically thick than $\sii$ 1304 (requiring many more resonance scattering event to escape), and has such a low reprocessing efficiency (low likelihood of producing fluorescence emission), the lack of a correlation between the relative strengths of $\siistar$ 1265 and $\sii$ 1260 features with the Balmer decrement argues against dust absorption in the outflow as the cause of the weak emission. Regarding the $\sii$ 1304, 1309 pair, we lean toward an explanation alluded to earlier: the $\sii$ 1304 absorption line only traces the high column density but low velocity gas, much of which is likely to be the static ISM. Therefore this particular correlation is likely between the strength of absorption from this material and the amount of dust along the line-of-sight (as already seen in \cite{Heckman1998}).

\begin{figure*}
  \centering
    \includegraphics[width=1\textwidth]{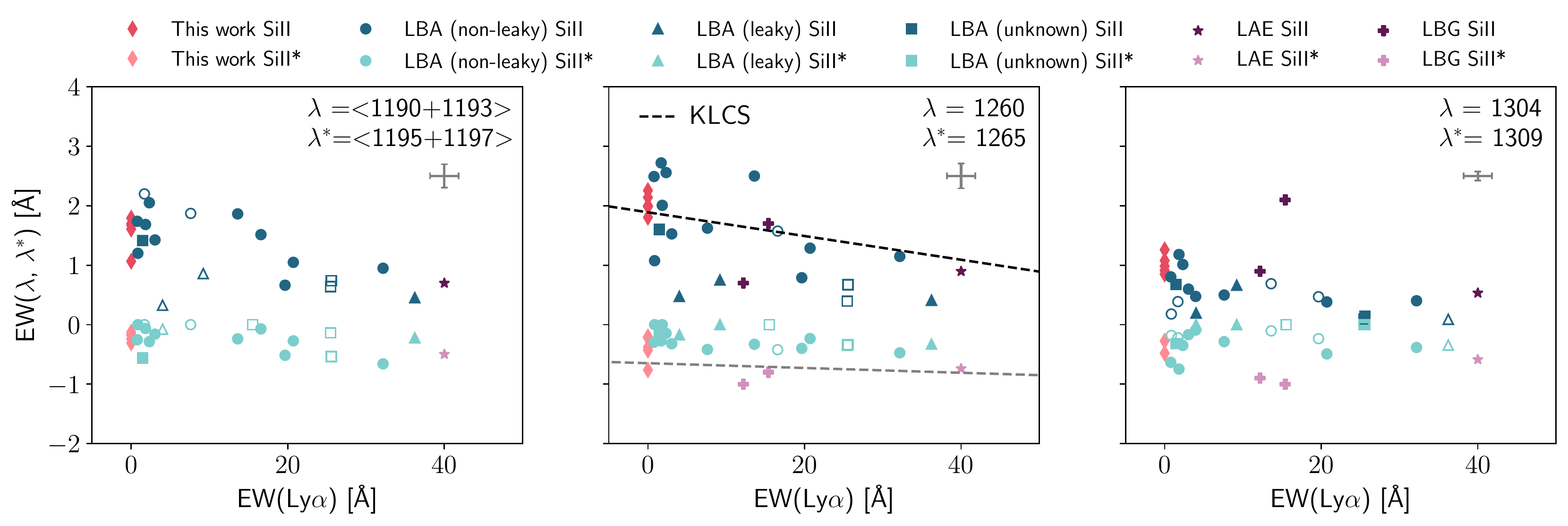}
\caption{EWs of absorption and emission lines plotted as functions of EW($\lya$). The dashed lines in the middle panel are the best-fit relations from KLCS \citep{Steidel2018}. The LAE data points are obtained from a stacked spectrum, of which EW($\lya$) is not measured, and so are placed along the x-axis for illustration purposes only. These plots suggest that the strongest correlation found, which is between EW($\siistar$)/EW($\sii$) and EW($\lya$), is driven mainly by the inverse correlation between the $\lya$ and resonant EWs.\label{fig:lya}}
\end{figure*}

The strongest correlation is found between the fluorescence to resonance line EW ratio and the $\lya$ emission-line EW. This is driven mainly by the inverse correlation between the $\lya$ and resonant EWs (see Figure~\ref{fig:lya}). This finding was also reported at high $z$ in KLCS \citep{Steidel2018}. This correlation is intriguing because both large values of $\lya$ EW and weak low-ionization absorption lines have been empirically linked to the galaxies with significant amounts of escaping Lyman-continuum (LyC) emission in $z \sim 3$ stacked samples (e.g., \citealt{Marchi2018,Steidel2018}) and among individual low-$z$ LyC emitters (e.g., \citealt{Verhamme2017,Chisholm2018,Wang2019}). Taken together, the cases with the relatively strong fluorescence emission perhaps have clearer channels through which ionizing photons can escape the hosting galaxy \citep{Heckman2015}. However, a note of caution is due here since we see no correlations between the relative strengths and widths of the  emission lines and other proposed LyC leakage-diagnostics such as a significantly blue-shifted component in $\lya$ emission or [O{\sc iii}]/[O{\sc ii}] as discussed in \cite{Wang2019}.

\subsection{Implications for the structure of outflows\label{sec:pl}}

From the above discussion, we establish the cause for the weakness of the observed fluorescence emission being that most of the actual emission lying beyond the radii probed by the COS data (``slit-loss"). We therefore conclude by considering the implications of this result for the structure of starburst-driven outflows. We begin with the subsample of our data which has the most constraining power: the group of galaxies observed with the largest apertures as compared to their observed sizes ($r_{\rm COS}/r_{50} \gtrsim 9$; see Figure~\ref{fig:r50}).

To start, it is useful to remind ourselves that most of the fluorescence emission is expected to be generated within the ``photosphere" of the outflow: the region where $\tau \gtrsim 1$ for the associated resonance line. The fact that the observed fluorescence emission lines are much weaker and narrower than the resonance absorption lines then implies that this photosphere is located at a greater distance from the starburst than what is captured in the COS aperture. This is quite plausible as the projected radii of the COS apertures are typically two to four kpc (as shown in Table \ref{tab:anci}). In what follows, we use this idea to constrain the properties of the radial profile of the outflowing material responsible for the observed absorption lines.

We note that any fluorescence emission from the portion of the outflow captured by the COS aperture must be associated with material that is optically thin in the associated resonance line (otherwise the emission-line strength should be similar to that of the absorption-line). Taking then the case where $\tau \ll 1$ inside the region probed by COS, we can approximate the ratio between the observed flux of the fluorescence emission, $F_{\rm fs}$, and that of the resonance absorption, $F_{\rm res}$, as

\begin{equation}
	\frac{F_{\rm fs}}{F_{\rm res}} \simeq f_{\rm reproc} \tau,
	\label{eq:ftau}
\end{equation}
where $f_{\rm reproc}$ is the fraction of the total absorbed photons that have been reprocessed into emission-line photons, and can be simply determined by the corresponding Einstein A's:
\begin{equation}
	f_{\rm reproc} = \frac{A_{\rm fs}}{A_{\rm fs} + A_{\rm res}}.
\end{equation}
We focus on the $\sii$ 1190/$\siistar$ 1195 pair, where its numerical value is about 0.84. This means that by measuring $F_{\rm fs}/F_{\rm res}$, we effectively probe $\tau$ in the resonance line integrated from the starburst to the projected radius of the COS aperture.

For simplicity, we assume that the radial density, $n(r)$, follows a power law for the ion of interest:
\begin{equation}
	n(r) \propto \Big(\frac{r}{r_0}\Big)^{-\alpha},
\label{eq:nr}
\end{equation}
where $\alpha > 0$ and $r$ is the distance to the starburst. We assume this power law extends inward only as far as the starburst radius $r_0$ (i.e., the launch point for the outflow, which we effectively take to be $r_{50}$). Then the column density as a function of $r$, $N(r)$, is simply given as
\begin{equation}
    N(r) = \int_{r_0}^{r} n(r) \, {\rm d}r, \, r > r_0.
\label{eq:Nr}
\end{equation}

Since $\tau(r) \propto N(r)$, we can then relate $N(r)$ to our directly observable $F_{\rm fs}/F_{\rm res}$ so that we have
\begin{equation}
    \frac{N(r_{\rm COS})}{N (\infty)} = \frac{\tau(r_{\rm COS})}{\tau (\infty)} \simeq \frac{F_{\rm fs}}{F_{\rm res}}.
\label{eq:ntau}
\end{equation}

As seen in Figure~\ref{fig:r50}, we find $F_{\rm fs}/F_{\rm res} \lesssim 0.25$ even when $r_{\rm COS}/r_0 \gtrsim 9$. Equations \ref{eq:nr} to \ref{eq:ntau} then imply that $\alpha \lesssim 1.13$.

\begin{figure}
  \centering
    \includegraphics[width=0.47\textwidth]{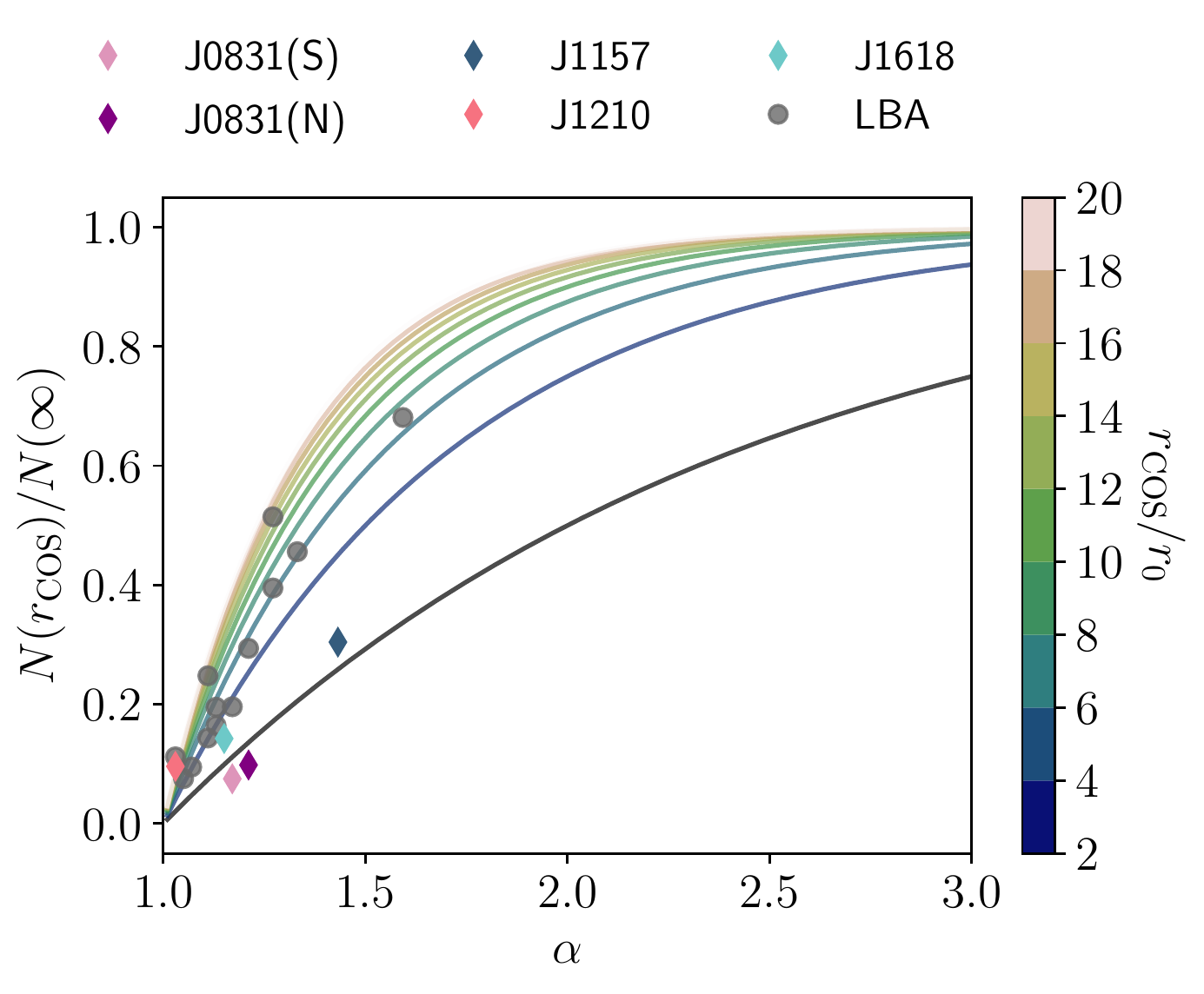}
\caption{The ratio between the column density of absorbing material inside the radius probed by COS and the total column density (as inferred from the ratio of the fluxes in the fluorescence emission and that in resonance absorption) plotted as a function of the power-law index $\alpha$ of the radial density profile for the absorbing gas. The color shades denote the ratio of the radius of the COS aperture ($r_{\rm COS}$) relative to the starburst radius $r_0$. We show the locations of the individual galaxies in our sample, and find that in nearly all cases, a shallow density profile is required ($\alpha < 1.5$).\label{fig:alpha}}
\end{figure}

More generally, we show the $\alpha$-dependence of Equation~\ref{eq:ntau} in Figure~\ref{fig:alpha} for a range of $r_{\rm COS}/r_0$. Taking Equation~\ref{eq:ntau}, we can plot each galaxy in our sample in this figure. From this, we see those galaxies observed with smaller effective aperture sizes (even most of the ``wind-dominated" cases), require $\alpha < 1.5$ (the only exception is the LBA J1428).

These results have some important implications. Let us adopt an outflow velocity that scales with radius as a power law:

\begin{equation}
	v(r) \propto r^{-\beta},
\end{equation}
and adopt a mass-outflow rate that varies with radius as

\begin{equation}
    \dot{M} \propto r^\gamma.
\end{equation}
Then from
\begin{equation}
    \dot{M} \propto n(r) \, v(r) \, r^2,
\end{equation}
we arrive at the result that
\begin{equation}
	\gamma = 2 - \alpha - \beta.
\end{equation}

For the outflow to be mass-conserving ($\gamma = 0$), the requirement of $0 < \alpha < 1.5$ means the outflow must be strongly decelerating as it travels out ($2 > \beta > 0.5$). This would require that the fastest moving material is closest to the starburst. This is inconsistent with the both the narrowness of the fluorescence emission lines observed within the COS aperture, and with direct observations of the nebular emission-line gas in outflows \citep{Heckman1990,Shopbell1998}.

As an example, for the simple case of a constant wind velocity ($\beta = 0$), $\alpha = 1.5$ yields $\dot{M} \propto r^{1/2}$. Thus, the wind would not be mass-conserving, and most of the outflowing material would be added at radii much larger than the starburst. Such a ``mass-growing" scenario is consistent with several recent studies which have found that the gas producing the absorption lines may form {\it in situ} via radiative cooling instabilities in a much hotter wind fluid \citep{Thompson2016,Schneider2018}. The latter would be too hot to produce the observed absorption lines. In particular, \cite{Gronke2018,Gronke2019} have shown that absorbing clouds that exceed a critical size can not only survive destruction by the wind fluid, they enable new gas to condense out of the wind. This could increase the mass-flux in absorbing gas with increasing radius through the depletion of the hot gas (e.g., the outflow as-a-whole could still be mass-conserving). Alternatively, models of wind-blown bubbles driven by starbursts (e.g., \citealt{Lochhaas2018}) imply an increasing mass in swept-up material with increasing radius (are not mass-conserving).

Those new models are certainly encouraging, but more studies are needed to bridge the theoretical and observational fronts---e.g., simulations of mass growth in a full galactic context, as well as more sensitive imaging observations of emission from large-scale galactic outflows.

\subsection{Implications for deriving outflow rates\label{sec:rate}}

As emphasized in the introduction, one of the main motivations for better understanding the structure of the outflow of the absorbing material is to use this to improve estimates of the outflow rates as derived from the analysis of the absorption-line data. A simple argument showing the importance of the size/structure of the outflow is as follows:

The mass-outflow rate is given by the outflowing mass divided by the outflow time. The latter is just the outflow size divided by the outflow velocity. What can be estimated from the absorption-line data is actually the total column density ($N$), which is essentially (half) the total mass divided by the cross-sectional area of the outflow. For a spherically-symmetric outflow this implies that

\begin{equation}
\dot{M} \sim 4 \pi N \mu \, v \, r,
\end{equation}
where $\mu$ is the mean mass per particle and $r$ is some kind of column-density weighted mean outflow radius. This is usually taken to be only a few times larger than the starburst radius $r_0$ (e.g., \citealt{Heckman2000,Heckman2015}). Taken at face value, our results imply that the effective value for $r$ is much larger than the starburst radius, and therefore the mass outflow rates will be correspondingly larger. More fundamentally, the outflow is unlikely to be mass-conserving, so there is no single well-defined value for $\dot{M}$.

Finally, we wish to emphasize a further complication. Galactic outflows are known to be multiphase, with hot, warm, and cold gas spanning about five orders-of-magnitude in temperature \citep{Heckman2017}. Whatever information about mass-outflow rates are derived from the resonance absorption-line data will only pertain to the warm phase ($T \sim 10^4$ to $10^5$ K).

\section{Conclusions\label{sec:conclusion}}

We have reported on the analysis of HST imaging and spectroscopic data for a sample of low-$z$ starburst galaxies, focusing on exploiting the diagnostic power of the UV emission lines that are created via fluorescent reprocessing of absorbed resonance-line photons in galactic outflows. Our principal conclusions are as follows:

\begin{enumerate}
\item We find that in the majority of cases the $\siistar$ emission lines in the COS spectra have significantly smaller EWs than the associated resonance absorption lines, and are usually much narrower. The resonance absorption lines are strongly blue-shifted, tracing an outflow, while the emission lines are usually centered near the systemic velocity of the galaxy.

\item Direct imaging of the $\feiistar$ emission in five of these galaxies shows that the emission arises in or near regions of intense star-formation (regions with high surface-brightness in the UV continuum and the [$\oii$] 3727 emission line).

\item By comparing the properties of the five different $\sii$ resonance lines, which span a range of 13 in oscillator strength (optical depth), we find that the highest column density gas has the lowest outflow velocity. The weakest absorption line ($\sii$ 1304) also most strongly resembles its associated fluorescence line ($\siistar$ 1309) in terms of EW and FWHM.

\item These results all imply that in the majority of cases, the observed fluorescence emission is associated with the static ISM and/or the slow-moving, central-most region in the outflow.

\item We do however, see a range in these properties, and find a good correlation between the ratio of the EWs of the fluorescence emission and resonance absorption lines and the ratio of their line widths. We interpret this as sequence from a majority of cases in which the fluorescence lines are ``ISM-dominated" to a minority of cases in which the outflow contributes significantly.

\item The strongest correlation we see is between the ratio of the fluorescence and resonance EWs and the EW of the $\lya$ emission line. This is driven largely by an inverse correlation between the $\lya$ and $\sii$ EWs. This is very similar to what is observed in high-$z$ star-forming galaxies. Intriguingly, both strong $\lya$ emission and weak absorption from low-ions like $\sii$ have been empirically related to the escape of LyC radiation. This suggests a link between the outflow structure and LyC leakage.

\item We conclude that the relative weakness of fluorescence from the outflow (as traced by the absorption lines) means that the bulk of the wind emission arises on larger scales than what is probed by the COS aperture. We use a simple model to show that the radial fall-off in the outflow density cannot be significantly steeper than $r^{-\alpha}$, where $\alpha \sim$ 1 to 1.5.

\item Unless the outflow is rapidly decelerating with radius (which is inconsistent with other observations of outflows), the shallow radial density profile implies that the mass-flux in the outflowing absorbing material increases with radius. This is consistent with some recent models in which either relatively cool absorbing gas condenses out of a hotter wind fluid at large radii, or in which the wind fluid sweeps up more and more ambient cool gas as it travels outward.

\item These results imply that existing estimates of the outflow rates in starbursts are unlikely to be accurate, may systematically underestimate the true values, and do not capture any radial dependence in the rates.
\end{enumerate}

\acknowledgments
{We thank Max Gronke and Claudia Scarlata for valuable input to the paper. We also thank Claudia Scarlata for sharing the stacked spectrum of LAEs.
B.W. thanks Caroline Huang, David Thilker, and Wenlong Yuan for helpful discussions on data analysis.
This work is supported by {\it{HST}}-GO-15340, provided by NASA through a grant from the Space Telescope Science Institute, which is operated by the Association of Universities for Research in Astronomy, Inc., under NASA contract NAS5-26555.

This publication made use of data products from the Wide-field Infrared Survey Explorer, which is a joint project of the University of California, Los Angeles, and the Jet Propulsion Laboratory/California Institute of Technology, funded by NASA; the NASA/IPAC Extragalactic Database, which is operated by the Jet Propulsion Laboratory, California Institute of Technology, under contract with NASA; and the NASA Astrophysical Data System for bibliographic information.
This project also made use of SDSS data. Funding for the Sloan Digital Sky Survey IV has been provided by the Alfred P. Sloan Foundation, the U.S. Department of Energy Office of Science, and the Participating Institutions. SDSS-IV acknowledges support and resources from the Center for High-Performance Computing at the University of Utah. The SDSS web site is www.sdss.org.
SDSS-IV is managed by the Astrophysical Research Consortium for the
Participating Institutions of the SDSS Collaboration including the
Brazilian Participation Group, the Carnegie Institution for Science,
Carnegie Mellon University, the Chilean Participation Group, the French Participation Group, Harvard-Smithsonian Center for Astrophysics,
Instituto de Astrof\'isica de Canarias, The Johns Hopkins University, Kavli Institute for the Physics and Mathematics of the Universe (IPMU) /
University of Tokyo, the Korean Participation Group, Lawrence Berkeley National Laboratory,
Leibniz Institut f\"ur Astrophysik Potsdam (AIP),
Max-Planck-Institut f\"ur Astronomie (MPIA Heidelberg),
Max-Planck-Institut f\"ur Astrophysik (MPA Garching),
Max-Planck-Institut f\"ur Extraterrestrische Physik (MPE),
National Astronomical Observatories of China, New Mexico State University,
New York University, University of Notre Dame,
Observat\'ario Nacional / MCTI, The Ohio State University,
Pennsylvania State University, Shanghai Astronomical Observatory,
United Kingdom Participation Group,
Universidad Nacional Aut\'onoma de M\'exico, University of Arizona,
University of Colorado Boulder, University of Oxford, University of Portsmouth,
University of Utah, University of Virginia, University of Washington, University of Wisconsin,
Vanderbilt University, and Yale University.
}
 
\facilities{{\it{HST}} (COS, WFC3), {\it{GALEX}}, Sloan, {\it{WISE}}}
\software{Astropy v. 3.3.2 \citep{astropy2013,astropy2018}, DrizzlePac v. 2.2.6, Matplotlib v. 3.1.1 \citep{Matplotlib}, NumPy v. 1.17.2 \citep{numpy}, SciPy v. 1.3.1 \citep{scipy}}

\bibliography{fs.bib,refs.bib}

\begin{thebibliography}{}
\expandafter\ifx\csname natexlab\endcsname\relax\def\natexlab#1{#1}\fi
\providecommand{\url}[1]{\href{#1}{#1}}
\providecommand{\dodoi}[1]{doi:~\href{http://doi.org/#1}{\nolinkurl{#1}}}
\providecommand{\doeprint}[1]{\href{http://ascl.net/#1}{\nolinkurl{http://ascl.net/#1}}}
\providecommand{\doarXiv}[1]{\href{https://arxiv.org/abs/#1}{\nolinkurl{https://arxiv.org/abs/#1}}}

\bibitem[{Alexandroff {et~al.}(2015)Alexandroff, Heckman, Borthakur, Overzier,
  \& Leitherer}]{Alexandroff2015}
Alexandroff, R.~M., Heckman, T.~M., Borthakur, S., Overzier, R., \& Leitherer,
  C. 2015, \apj, 810, 104, \dodoi{10.1088/0004-637X/810/2/104}

\bibitem[{{Armus} {et~al.}(1990){Armus}, {Heckman}, \& {Miley}}]{Armus1990}
{Armus}, L., {Heckman}, T.~M., \& {Miley}, G.~K. 1990, \apj, 364, 471,
  \dodoi{10.1086/169431}

\bibitem[{{Astropy Collaboration} {et~al.}(2013){Astropy Collaboration},
  {Robitaille}, {Tollerud}, {Greenfield}, {Droettboom}, {Bray}, {Aldcroft},
  {Davis}, {Ginsburg}, {Price-Whelan}, {Kerzendorf}, {Conley}, {Crighton},
  {Barbary}, {Muna}, {Ferguson}, {Grollier}, {Parikh}, {Nair}, {Unther},
  {Deil}, {Woillez}, {Conseil}, {Kramer}, {Turner}, {Singer}, {Fox}, {Weaver},
  {Zabalza}, {Edwards}, {Azalee Bostroem}, {Burke}, {Casey}, {Crawford},
  {Dencheva}, {Ely}, {Jenness}, {Labrie}, {Lim}, {Pierfederici}, {Pontzen},
  {Ptak}, {Refsdal}, {Servillat}, \& {Streicher}}]{astropy2013}
{Astropy Collaboration}, {Robitaille}, T.~P., {Tollerud}, E.~J., {et~al.} 2013,
  \aap, 558, A33, \dodoi{10.1051/0004-6361/201322068}

\bibitem[{{Astropy Collaboration} {et~al.}(2018){Astropy Collaboration},
  {Price-Whelan}, {Sip{\H o}cz}, {G{\"u}nther}, {Lim}, {Crawford}, {Conseil},
  {Shupe}, {Craig}, {Dencheva}, {Ginsburg}, {VanderPlas}, {Bradley},
  {P{\'e}rez-Su{\'a}rez}, {de Val-Borro}, {Aldcroft}, {Cruz}, {Robitaille},
  {Tollerud}, {Ardelean}, {Babej}, {Bach}, {Bachetti}, {Bakanov}, {Bamford},
  {Barentsen}, {Barmby}, {Baumbach}, {Berry}, {Biscani}, {Boquien}, {Bostroem},
  {Bouma}, {Brammer}, {Bray}, {Breytenbach}, {Buddelmeijer}, {Burke},
  {Calderone}, {Cano Rodr{\'{\i}}guez}, {Cara}, {Cardoso}, {Cheedella},
  {Copin}, {Corrales}, {Crichton}, {D'Avella}, {Deil}, {Depagne}, {Dietrich},
  {Donath}, {Droettboom}, {Earl}, {Erben}, {Fabbro}, {Ferreira}, {Finethy},
  {Fox}, {Garrison}, {Gibbons}, {Goldstein}, {Gommers}, {Greco}, {Greenfield},
  {Groener}, {Grollier}, {Hagen}, {Hirst}, {Homeier}, {Horton}, {Hosseinzadeh},
  {Hu}, {Hunkeler}, {Ivezi{\'c}}, {Jain}, {Jenness}, {Kanarek}, {Kendrew},
  {Kern}, {Kerzendorf}, {Khvalko}, {King}, {Kirkby}, {Kulkarni}, {Kumar},
  {Lee}, {Lenz}, {Littlefair}, {Ma}, {Macleod}, {Mastropietro}, {McCully},
  {Montagnac}, {Morris}, {Mueller}, {Mumford}, {Muna}, {Murphy}, {Nelson},
  {Nguyen}, {Ninan}, {N{\"o}the}, {Ogaz}, {Oh}, {Parejko}, {Parley}, {Pascual},
  {Patil}, {Patil}, {Plunkett}, {Prochaska}, {Rastogi}, {Reddy Janga},
  {Sabater}, {Sakurikar}, {Seifert}, {Sherbert}, {Sherwood-Taylor}, {Shih},
  {Sick}, {Silbiger}, {Singanamalla}, {Singer}, {Sladen}, {Sooley},
  {Sornarajah}, {Streicher}, {Teuben}, {Thomas}, {Tremblay}, {Turner},
  {Terr{\'o}n}, {van Kerkwijk}, {de la Vega}, {Watkins}, {Weaver}, {Whitmore},
  {Woillez}, {Zabalza}, \& {Astropy Contributors}}]{astropy2018}
{Astropy Collaboration}, {Price-Whelan}, A.~M., {Sip{\H o}cz}, B.~M., {et~al.}
  2018, \aj, 156, 123, \dodoi{10.3847/1538-3881/aabc4f}

\bibitem[{Calzetti(2011)}]{Calzetti2011}
Calzetti, D. 2011, in Secul. Evol. Galaxies, ed. J.~Falcon-Barroso \& J.~H.
  Knapen (Cambridge: Cambridge University Press), 419--458,
  \dodoi{10.1017/CBO9781139547420.008}

\bibitem[{{Chisholm} {et~al.}(2017){Chisholm}, {Tremonti}, {Leitherer}, \&
  {Chen}}]{Chisholm2017}
{Chisholm}, J., {Tremonti}, C.~A., {Leitherer}, C., \& {Chen}, Y. 2017, \mnras,
  469, 4831, \dodoi{10.1093/mnras/stx1164}

\bibitem[{Chisholm {et~al.}(2018)Chisholm, Gazagnes, Schaerer, Verhamme, Rigby,
  Bayliss, Sharon, Gladders, \& Dahle}]{Chisholm2018}
Chisholm, J., Gazagnes, S., Schaerer, D., {et~al.} 2018, \aap, 616, A30,
  \dodoi{10.1051/0004-6361/201832758}

\bibitem[{Erb {et~al.}(2012)Erb, Quider, Henry, \& Martin}]{Erb2012}
Erb, D.~K., Quider, A.~M., Henry, A.~L., \& Martin, C.~L. 2012, \apj, 759, 26,
  \dodoi{10.1088/0004-637X/759/1/26}

\bibitem[{Finley {et~al.}(2017)Finley, Bouch{\'{e}}, Contini, Epinat, Bacon,
  Brinchmann, Cantalupo, Erroz-Ferrer, Marino, Maseda, Richard, Schroetter,
  Verhamme, Weilbacher, Wendt, \& Wisotzki}]{Finley2017a}
Finley, H., Bouch{\'{e}}, N., Contini, T., {et~al.} 2017, \aap, 605, A118,
  \dodoi{10.1051/0004-6361/201730428}

\bibitem[{{Finley} {et~al.}(2017){Finley}, {Bouch{\'e}}, {Contini}, {Paalvast},
  {Boogaard}, {Maseda}, {Bacon}, {Blaizot}, {Brinchmann}, {Epinat}, {Feltre},
  {Marino}, {Muzahid}, {Richard}, {Schaye}, {Verhamme}, {Weilbacher}, \&
  {Wisotzki}}]{Finley2017b}
{Finley}, H., {Bouch{\'e}}, N., {Contini}, T., {et~al.} 2017, \aap, 608, A7,
  \dodoi{10.1051/0004-6361/201731499}

\bibitem[{{Gronke} \& {Oh}(2018)}]{Gronke2018}
{Gronke}, M., \& {Oh}, S.~P. 2018, \mnras, 480, L111,
  \dodoi{10.1093/mnrasl/sly131}

\bibitem[{{Gronke} \& {Oh}(2019)}]{Gronke2019}
---. 2019, \mnras, 2995, \dodoi{10.1093/mnras/stz3332}

\bibitem[{Heckman {et~al.}(2015)Heckman, Alexandroff, Borthakur, Overzier, \&
  Leitherer}]{Heckman2015}
Heckman, T.~M., Alexandroff, R.~M., Borthakur, S., Overzier, R., \& Leitherer,
  C. 2015, \apj, 809, 147, \dodoi{10.1088/0004-637X/809/2/147}

\bibitem[{{Heckman} {et~al.}(1990){Heckman}, {Armus}, \& {Miley}}]{Heckman1990}
{Heckman}, T.~M., {Armus}, L., \& {Miley}, G.~K. 1990, \apjs, 74, 833,
  \dodoi{10.1086/191522}

\bibitem[{Heckman {et~al.}(2000)Heckman, Lehnert, Strickland, \&
  Armus}]{Heckman2000}
Heckman, T.~M., Lehnert, M.~D., Strickland, D.~K., \& Armus, L. 2000, \apjs,
  129, 493, \dodoi{10.1086/313421}

\bibitem[{{Heckman} {et~al.}(1998){Heckman}, {Robert}, {Leitherer}, {Garnett},
  \& {van der Rydt}}]{Heckman1998}
{Heckman}, T.~M., {Robert}, C., {Leitherer}, C., {Garnett}, D.~R., \& {van der
  Rydt}, F. 1998, \apj, 503, 646, \dodoi{10.1086/306035}

\bibitem[{{Heckman} \& {Thompson}(2017)}]{Heckman2017}
{Heckman}, T.~M., \& {Thompson}, T.~A. 2017, arXiv e-prints, arXiv:1701.09062.
\newblock \doarXiv{1701.09062}

\bibitem[{Heckman {et~al.}(2011)Heckman, Borthakur, Overzier, Kauffmann,
  Basu-Zych, Leitherer, Sembach, Martin, Rich, Schiminovich, \&
  Seibert}]{Heckman2011a}
Heckman, T.~M., Borthakur, S., Overzier, R., {et~al.} 2011, \apj, 730, 5,
  \dodoi{10.1088/0004-637X/730/1/5}

\bibitem[{{Hunter}(2007)}]{Matplotlib}
{Hunter}, J.~D. 2007, Computing in Science Engineering, 9, 90,
  \dodoi{10.1109/MCSE.2007.55}

\bibitem[{Jones {et~al.}(2001)Jones, Oliphant, Peterson, {et~al.}}]{scipy}
Jones, E., Oliphant, T., Peterson, P., {et~al.} 2001, {SciPy}: Open source
  scientific tools for {Python}.
\newblock \url{http://www.scipy.org/}

\bibitem[{Jones {et~al.}(2012)Jones, Stark, \& Ellis}]{Jones2012}
Jones, T., Stark, D.~P., \& Ellis, R.~S. 2012, \apj, 751, 51,
  \dodoi{10.1088/0004-637X/751/1/51}

\bibitem[{Kennicutt \& Evans(2012)}]{Kennicutt2012}
Kennicutt, R.~C., \& Evans, N.~J. 2012, \araa, 50, 531,
  \dodoi{10.1146/annurev-astro-081811-125610}

\bibitem[{{Kornei} {et~al.}(2013){Kornei}, {Shapley}, {Martin}, {Coil}, {Lotz},
  \& {Weiner}}]{Kornei2013}
{Kornei}, K.~A., {Shapley}, A.~E., {Martin}, C.~L., {et~al.} 2013, \apj, 774,
  50, \dodoi{10.1088/0004-637X/774/1/50}

\bibitem[{Kroupa(2001)}]{Kroupa2001}
Kroupa, P. 2001, \mnras, 322, 231, \dodoi{10.1046/j.1365-8711.2001.04022.x}

\bibitem[{Leitherer {et~al.}(1999)Leitherer, Schaerer, Goldader, Delgado,
  Robert, Kune, de~Mello, Devost, \& Heckman}]{Leitherer1999}
Leitherer, C., Schaerer, D., Goldader, J.~D., {et~al.} 1999, \apjs, 123, 3,
  \dodoi{10.1086/313233}

\bibitem[{{Lochhaas} {et~al.}(2018){Lochhaas}, {Thompson}, {Quataert}, \&
  {Weinberg}}]{Lochhaas2018}
{Lochhaas}, C., {Thompson}, T.~A., {Quataert}, E., \& {Weinberg}, D.~H. 2018,
  \mnras, 481, 1873, \dodoi{10.1093/mnras/sty2421}

\bibitem[{{Marchi} {et~al.}(2018){Marchi}, {Pentericci}, {Guaita}, {Schaerer},
  {Verhamme}, {Castellano}, {Ribeiro}, {Garilli}, {Le F{\`e}vre}, {Amorin},
  {Bardelli}, {Cassata}, {Durkalec}, {Grazian}, {Hathi}, {Lemaux}, {Maccagni},
  {Vanzella}, \& {Zucca}}]{Marchi2018}
{Marchi}, F., {Pentericci}, L., {Guaita}, L., {et~al.} 2018, \aap, 614, A11,
  \dodoi{10.1051/0004-6361/201732133}

\bibitem[{McCully {et~al.}(2018)McCully, Crawford, Kovacs, Tollerud, Betts,
  Bradley, Craig, Turner, Streicher, Sipocz, Robitaille, \&
  Deil}]{curtis_mccully_2018_1482019}
McCully, C., Crawford, S., Kovacs, G., {et~al.} 2018, astropy/astroscrappy:
  v1.0.5 Zenodo Release, v1.0.5,  Zenodo, \dodoi{10.5281/zenodo.1482019}

\bibitem[{{Planck Collaboration} {et~al.}(2018){Planck Collaboration},
  {Aghanim}, {Akrami}, {Ashdown}, {Aumont}, {Baccigalupi}, {Ballardini},
  {Banday}, {Barreiro}, {Bartolo}, {Basak}, {Battye}, {Benabed}, {Bernard},
  {Bersanelli}, {Bielewicz}, {Bock}, {Bond}, {Borrill}, {Bouchet}, {Boulanger},
  {Bucher}, {Burigana}, {Butler}, {Calabrese}, {Cardoso}, {Carron},
  {Challinor}, {Chiang}, {Chluba}, {Colombo}, {Combet}, {Contreras}, {Crill},
  {Cuttaia}, {de Bernardis}, {de Zotti}, {Delabrouille}, {Delouis}, {Di
  Valentino}, {Diego}, {Dor{\'e}}, {Douspis}, {Ducout}, {Dupac}, {Dusini},
  {Efstathiou}, {Elsner}, {En{\ss}lin}, {Eriksen}, {Fantaye}, {Farhang},
  {Fergusson}, {Fernandez-Cobos}, {Finelli}, {Forastieri}, {Frailis},
  {Fraisse}, {Franceschi}, {Frolov}, {Galeotta}, {Galli}, {Ganga},
  {G{\'e}nova-Santos}, {Gerbino}, {Ghosh}, {Gonz{\'a}lez-Nuevo}, {G{\'o}rski},
  {Gratton}, {Gruppuso}, {Gudmundsson}, {Hamann}, {Handley}, {Hansen},
  {Herranz}, {Hildebrandt}, {Hivon}, {Huang}, {Jaffe}, {Jones}, {Karakci},
  {Keih{\"a}nen}, {Keskitalo}, {Kiiveri}, {Kim}, {Kisner}, {Knox},
  {Krachmalnicoff}, {Kunz}, {Kurki-Suonio}, {Lagache}, {Lamarre}, {Lasenby},
  {Lattanzi}, {Lawrence}, {Le Jeune}, {Lemos}, {Lesgourgues}, {Levrier},
  {Lewis}, {Liguori}, {Lilje}, {Lilley}, {Lindholm}, {L{\'o}pez-Caniego},
  {Lubin}, {Ma}, {Mac{\'\i}as-P{\'e}rez}, {Maggio}, {Maino}, {Mandolesi},
  {Mangilli}, {Marcos-Caballero}, {Maris}, {Martin}, {Martinelli},
  {Mart{\'\i}nez-Gonz{\'a}lez}, {Matarrese}, {Mauri}, {McEwen}, {Meinhold},
  {Melchiorri}, {Mennella}, {Migliaccio}, {Millea}, {Mitra},
  {Miville-Desch{\^e}nes}, {Molinari}, {Montier}, {Morgante}, {Moss}, {Natoli},
  {N{\o}rgaard-Nielsen}, {Pagano}, {Paoletti}, {Partridge}, {Patanchon},
  {Peiris}, {Perrotta}, {Pettorino}, {Piacentini}, {Polastri}, {Polenta},
  {Puget}, {Rachen}, {Reinecke}, {Remazeilles}, {Renzi}, {Rocha}, {Rosset},
  {Roudier}, {Rubi{\~n}o-Mart{\'\i}n}, {Ruiz-Granados}, {Salvati}, {Sandri},
  {Savelainen}, {Scott}, {Shellard}, {Sirignano}, {Sirri}, {Spencer},
  {Sunyaev}, {Suur-Uski}, {Tauber}, {Tavagnacco}, {Tenti}, {Toffolatti},
  {Tomasi}, {Trombetti}, {Valenziano}, {Valiviita}, {Van Tent}, {Vibert},
  {Vielva}, {Villa}, {Vittorio}, {Wand elt}, {Wehus}, {White}, {White},
  {Zacchei}, \& {Zonca}}]{Planck2018}
{Planck Collaboration}, {Aghanim}, N., {Akrami}, Y., {et~al.} 2018, arXiv
  e-prints, arXiv:1807.06209.
\newblock \doarXiv{1807.06209}

\bibitem[{Prochaska {et~al.}(2011)Prochaska, Kasen, \& Rubin}]{Prochaska2011}
Prochaska, J.~X., Kasen, D., \& Rubin, K. 2011, \apj, 734, 24,
  \dodoi{10.1088/0004-637X/734/1/24}

\bibitem[{{Rubin} {et~al.}(2011){Rubin}, {Prochaska}, {M{\'e}nard}, {Murray},
  {Kasen}, {Koo}, \& {Phillips}}]{Rubin2011}
{Rubin}, K. H.~R., {Prochaska}, J.~X., {M{\'e}nard}, B., {et~al.} 2011, \apj,
  728, 55, \dodoi{10.1088/0004-637X/728/1/55}

\bibitem[{Scarlata \& Panagia(2015)}]{Scarlata2015}
Scarlata, C., \& Panagia, N. 2015, \apj, 801, 43,
  \dodoi{10.1088/0004-637X/801/1/43}

\bibitem[{{Schneider} {et~al.}(2018){Schneider}, {Robertson}, \&
  {Thompson}}]{Schneider2018}
{Schneider}, E.~E., {Robertson}, B.~E., \& {Thompson}, T.~A. 2018, \apj, 862,
  56, \dodoi{10.3847/1538-4357/aacce1}

\bibitem[{{Shopbell} \& {Bland-Hawthorn}(1998)}]{Shopbell1998}
{Shopbell}, P.~L., \& {Bland-Hawthorn}, J. 1998, \apj, 493, 129,
  \dodoi{10.1086/305108}

\bibitem[{Somerville \& Dav{\'{e}}(2015)}]{Somerville2015}
Somerville, R.~S., \& Dav{\'{e}}, R. 2015, \araa, 53, 51,
  \dodoi{10.1146/annurev-astro-082812-140951}

\bibitem[{{Steidel} {et~al.}(2011){Steidel}, {Bogosavljevi{\'c}}, {Shapley},
  {Kollmeier}, {Reddy}, {Erb}, \& {Pettini}}]{Steidel2011}
{Steidel}, C.~C., {Bogosavljevi{\'c}}, M., {Shapley}, A.~E., {et~al.} 2011,
  \apj, 736, 160, \dodoi{10.1088/0004-637X/736/2/160}

\bibitem[{Steidel {et~al.}(2018)Steidel, Bogosavljevi{\'{c}}, Shapley, Reddy,
  Rudie, Pettini, Trainor, \& Strom}]{Steidel2018}
Steidel, C.~C., Bogosavljevi{\'{c}}, M., Shapley, A.~E., {et~al.} 2018, \apj,
  869, 123, \dodoi{10.3847/1538-4357/aaed28}

\bibitem[{Tang {et~al.}(2014)Tang, Giavalisco, Guo, \& Kurk}]{Tang2014}
Tang, Y., Giavalisco, M., Guo, Y., \& Kurk, J. 2014, \apj, 793, 92,
  \dodoi{10.1088/0004-637X/793/2/92}

\bibitem[{{Thompson} {et~al.}(2016){Thompson}, {Quataert}, {Zhang}, \&
  {Weinberg}}]{Thompson2016}
{Thompson}, T.~A., {Quataert}, E., {Zhang}, D., \& {Weinberg}, D.~H. 2016,
  \mnras, 455, 1830, \dodoi{10.1093/mnras/stv2428}

\bibitem[{{van der Walt} {et~al.}(2011){van der Walt}, {Colbert}, \&
  {Varoquaux}}]{numpy}
{van der Walt}, S., {Colbert}, S.~C., \& {Varoquaux}, G. 2011, Computing in
  Science Engineering, 13, 22, \dodoi{10.1109/MCSE.2011.37}

\bibitem[{{van Dokkum}(2001)}]{2001PASP..113.1420V}
{van Dokkum}, P.~G. 2001, \pasp, 113, 1420, \dodoi{10.1086/323894}

\bibitem[{Veilleux {et~al.}(2005)Veilleux, Cecil, \&
  Bland-Hawthorn}]{Veilleux2005}
Veilleux, S., Cecil, G., \& Bland-Hawthorn, J. 2005, \araa, 43, 769,
  \dodoi{10.1146/annurev.astro.43.072103.150610}

\bibitem[{{Verhamme} {et~al.}(2017){Verhamme}, {Orlitov{\'a}}, {Schaerer},
  {Izotov}, {Worseck}, {Thuan}, \& {Guseva}}]{Verhamme2017}
{Verhamme}, A., {Orlitov{\'a}}, I., {Schaerer}, D., {et~al.} 2017, \aap, 597,
  A13, \dodoi{10.1051/0004-6361/201629264}

\bibitem[{Wang {et~al.}(2019)Wang, Heckman, Leitherer, Alexandroff, Borthakur,
  \& Overzier}]{Wang2019}
Wang, B., Heckman, T.~M., Leitherer, C., {et~al.} 2019, \apj, 885, 57,
  \dodoi{10.3847/1538-4357/ab418f}

\bibitem[{{Wright} {et~al.}(2010){Wright}, {Eisenhardt}, {Mainzer}, {Ressler},
  {Cutri}, {Jarrett}, {Kirkpatrick}, {Padgett}, {McMillan}, {Skrutskie},
  {Stanford}, {Cohen}, {Walker}, {Mather}, {Leisawitz}, {Gautier}, {McLean},
  {Benford}, {Lonsdale}, {Blain}, {Mendez}, {Irace}, {Duval}, {Liu}, {Royer},
  {Heinrichsen}, {Howard}, {Shannon}, {Kendall}, {Walsh}, {Larsen}, {Cardon},
  {Schick}, {Schwalm}, {Abid}, {Fabinsky}, {Naes}, \&
  {Tsai}}]{2010AJ....140.1868W}
{Wright}, E.~L., {Eisenhardt}, P. R.~M., {Mainzer}, A.~K., {et~al.} 2010, \aj,
  140, 1868, \dodoi{10.1088/0004-6256/140/6/1868}

\bibitem[{Zhu {et~al.}(2015)Zhu, Comparat, Kneib, Delubac, Raichoor, Dawson,
  Newman, Y{\`{e}}che, Zhou, \& Schneider}]{Zhu2015}
Zhu, G.~B., Comparat, J., Kneib, J.-P., {et~al.} 2015, \apj, 815, 48,
  \dodoi{10.1088/0004-637X/815/1/48}

\end{thebibliography}

\appendix
\label{sec:app}
Measured $\sii$ and $\siistar$ line profiles of the five galaxies of this paper and LBAs are listed in Table~\ref{tab:spec_lines} and \ref{tab:spec_lines_lba} respectively.

\startlongtable
\begin{deluxetable}{clcccc}
\tablecaption{Measured spectral line properties of the five galaxies of this paper\label{tab:spec_lines}}
\tablecolumns{6}
\tablewidth{0pt}
\tablehead{
\colhead{Galaxy} &
\colhead{Line} &
\colhead{$V_{\rm ctr}$} &
\colhead{EW} &
\colhead{FWHM} &
\colhead{FW90\tablenotemark{a}}\\
\colhead{} &
\colhead{} &
\colhead{(km/s)} &
\colhead{($\mAA$)} &
\colhead{(km/s)} &
\colhead{(km/s)}
}
\startdata
J0831(S) \\
 & $\sii$ 1190 & -106 & 1.7 & 403 & 737 \\
 & $\sii$ 1193 & -66 & 1.5 & 343 & 625 \\
 & $\siistar$ 1195 & 98 & -0.1 & 89 & 162 \\
 & $\siistar$ 1197\tablenotemark{b} & - & - & - & - \\
 & $\sii$ 1260 & -94 & 2.0 & 412 & 748 \\
 & $\siistar$ 1265 & 80 & -0.4 & 248 & 450 \\
 & $\sii$ 1304 & -41 & 1.0 & 232 & 422 \\
 & $\siistar$ 1309 & 82 & -0.5 & 154 & 283 \\
\hline
J0831(N) \\
 & $\sii$ 1190 & -82 & 1.8 & 403 & 733 \\
 & $\sii$ 1193 & -72 & 1.6 & 383 & 703 \\
 & $\siistar$ 1195 & -9 & -0.2 & 129 & 237 \\
 & $\siistar$ 1197\tablenotemark{b} & - & - & - & - \\
 & $\sii$ 1260 & -76 & 2.1 & 425 & 777 \\
 & $\siistar$ 1265 & -12 & -0.8 & 240 & 439 \\
 & $\sii$ 1304 & -20 & 1.3 & 312 & 568 \\
 & $\siistar$ 1309\tablenotemark{c} & - & - & - & - \\
\hline
J1157 \\
 & $\sii$ 1190 & -75 & 1.0 & 250 & 457 \\
 & $\sii$ 1193 & -69 & 1.1 & 278 & 508 \\
 & $\siistar$ 1195 & -23 & -0.3 & 183 & 329 \\
 & $\siistar$ 1197\tablenotemark{b} & - & - & - & - \\
 & $\sii$ 1260 & -105 & 1.8 & 352 & 642 \\
 & $\siistar$ 1265 & 57 & -0.4 & 200 & 360 \\
 & $\sii$ 1304 & -55 & 0.8 & 185 & 338 \\
 & $\siistar$ 1309\tablenotemark{c} & - & - & - & - \\
\hline
J1210 \\
 & $\sii$ 1190 & -117 & 1.8 & 374 & 680 \\
 & $\sii$ 1193 & -60 & 1.8 & 439 & 796 \\
 & $\siistar$ 1195 & 4 & -0.2 & 115 & 204 \\
 & $\siistar$ 1197\tablenotemark{b} & - & - & - & - \\
 & $\sii$ 1260 & -80 & 2.0 & 385 & 701 \\
 & $\siistar$ 1265 & 43 & -0.2 & 137 & 248 \\
 & $\sii$ 1304 & -35 & 1.1 & 234 & 423 \\
 & $\siistar$ 1309 & -22 & -0.3 & 270 & 490 \\
\hline
J1618 \\
 & $\sii$ 1190 & -133 & 1.7 & 387 & 703 \\
 & $\sii$ 1193 & -138 & 1.7 & 402 & 736 \\
 & $\siistar$ 1195 & 21 & -0.2 & 176 & 322 \\
 & $\siistar$ 1197\tablenotemark{b} & - & - & - & - \\
 & $\sii$ 1260 & -168 & 2.3 & 472 & 862 \\
 & $\siistar$ 1265 & 66 & -0.2 & 107 & 193 \\
 & $\sii$ 1304 & -87 & 0.9 & 209 & 379 \\
 & $\siistar$ 1309\tablenotemark{c} & - & - & - & - \\
\enddata
\tablenotetext{a}{FW90: FW at 90\% continuum.}
\tablenotetext{b}{Line falls on the detector gap.}
\tablenotetext{c}{Line is likely contaminated by the MW absorption $\rm{Si\textsc{iv}}$ 1402.8.}
\tablecomments{All measurements can be assumed to have errors on the order of 10-15\% dominated by systematics in the polynomial fit to the continuum and subtraction of the SB99 models.
}
\end{deluxetable}

$\quad$

\startlongtable
\begin{deluxetable}{clcccc}
\tablecaption{Measured spectral line properties of LBAs\label{tab:spec_lines_lba}}
\tablecolumns{6}
\tablewidth{0pt}
\tablehead{
\colhead{Galaxy} &
\colhead{Line} &
\colhead{$V_{\rm ctr}$} &
\colhead{EW} &
\colhead{FWHM} &
\colhead{FW90}\\
\colhead{} &
\colhead{} &
\colhead{(km/s)} &
\colhead{($\mAA$)} &
\colhead{(km/s)} &
\colhead{(km/s)}
}
\startdata
J0055 & & & & &  \\
 & $\sii$ 1190 & -145 & 2.0 & 477 & 868 \\
 & $\sii$ 1193 & -110 & 2.1 & 457 & 834 \\
 & $\siistar$ 1195 & 59 & -0.2 & 266 & 484 \\
 & $\siistar$ 1197 & -32 & -0.3 & 271 & 496 \\
 & $\sii$ 1260 & -169 & 2.6 & 576 & 1047 \\
 & $\siistar$ 1265 & 6 & -0.1 & 164 & 298 \\
 & $\sii$ 1304 & -59 & 1.0 & 260 & 472 \\
 & $\siistar$ 1309 & 41 & -0.4 & 180 & 331 \\
\hline
J0150  & & & & &  \\
 & $\sii$ 1190 & -131 & 1.4 & 368 & 670 \\
 & $\sii$ 1193 & -98 & 1.5 & 433 & 789 \\
 & $\siistar$ 1195 & 0 & 0.0 & 0 & 0 \\
 & $\siistar$ 1197 & 55 & -0.3 & 233 & 422 \\
 & $\sii$ 1260 & -139 & 1.5 & 371 & 670 \\
 & $\siistar$ 1265 & 151 & -0.3 & 349 & 633 \\
 & $\sii$ 1304 & -85 & 0.6 & 255 & 466 \\
 & $\siistar$ 1309 & 96 & -0.2 & 240 & 435 \\
\hline
J0921 & & & & & \\
 & $\sii$ 1190 & 42 & 0.4 & 152 & 277 \\
 & $\sii$ 1193 & 48 & 0.3 & 142 & 261 \\
 & $\siistar$ 1195 & 142 & -0.2 & 191 & 351 \\
 & $\siistar$ 1197 & 0 & 0.0 & 0 & 0 \\
 & $\sii$ 1260 & 73 & 0.5 & 179 & 330 \\
 & $\siistar$ 1265 & 84 & -0.2 & 279 & 509 \\
 & $\sii$ 1304 & 67 & 0.2 & 84 & 148 \\
 & $\siistar$ 1309 & 0 & 0.0 & 0 & 0 \\
\hline
J0926 & & & & & \\
 & $\sii$ 1190 & -329 & 0.5 & 402 & 733 \\
 & $\sii$ 1193 & -282 & 0.4 & 286 & 522 \\
 & $\siistar$ 1195 & -20 & -0.2 & 161 & 292 \\
 & $\siistar$ 1197 & -18 & -0.2 & 142 & 258 \\
 & $\sii$ 1260 & -371 & 0.4 & 301 & 547 \\
 & $\siistar$ 1265 & 4 & -0.3 & 214 & 388 \\
 & $\sii$ 1304 & -100 & 0.1 & 124 & 224 \\
 & $\siistar$ 1309 & -11 & -0.3 & 233 & 425 \\
\hline
J0938 & & & & & \\
 & $\sii$ 1190 & -52 & 1.4 & 410 & 745 \\
 & $\sii$ 1193 & 2 & 1.4 & 434 & 788 \\
 & $\siistar$ 1195 & 70 & -0.2 & 150 & 277 \\
 & $\siistar$ 1197 & -49 & -0.9 & 566 & 1028 \\
 & $\sii$ 1260 & -3 & 1.6 & 383 & 697 \\
 & $\siistar$ 1265 & 58 & -0.2 & 154 & 283 \\
 & $\sii$ 1304 & -1 & 0.7 & 242 & 444 \\
 & $\siistar$ 1309 & -11 & -0.3 & 247 & 448 \\
\hline
J2103 & & & & & \\
 & $\sii$ 1190 & -337 & 0.9 & 303 & 552 \\
 & $\sii$ 1193 & -299 & 0.6 & 185 & 337 \\
 & $\siistar$ 1195 & -46 & -0.6 & 189 & 343 \\
 & $\siistar$ 1197 & -119 & -0.5 & 369 & 672 \\
 & $\sii$ 1260 & -168 & 0.7 & 265 & 481 \\
 & $\siistar$ 1265 & 83 & -0.3 & 396 & 720 \\
 & $\sii$ 1304 & -74 & 0.1 & 104 & 191 \\
 & $\siistar$ 1309 & 0 & 0.0 & 0 & 0 \\
\hline
\hline
J0021 & & & & & \\
 & $\sii$ 1190 & -212 & 0.6 & 410 & 746 \\
 & $\sii$ 1193 & -212 & 0.6 & 410 & 746 \\
 & $\siistar$ 1195 & 115 & -0.1 & 162 & 296 \\
 & $\siistar$ 1197 & -49 & -0.2 & 311 & 568 \\
 & $\sii$ 1260 & -227 & 0.4 & 265 & 483 \\
 & $\siistar$ 1265 & 19 & -0.3 & 385 & 701 \\
 & $\sii$ 1304 & -95 & 0.1 & 105 & 187 \\
 & $\siistar$ 1309 & 0 & 0.0 & 0 & 0 \\
\hline
J0823 & & & & & \\
 & $\sii$ 1190 & -38 & 1.6 & 378 & 690 \\
 & $\sii$ 1193 & 8 & 1.8 & 413 & 753 \\
 & $\siistar$ 1195 & 126 & -0.3 & 212 & 389 \\
 & $\siistar$ 1197 & 124 & -0.1 & 90 & 167 \\
 & $\sii$ 1260 & -64 & 2.2 & 443 & 808 \\
 & $\siistar$ 1265 & 165 & -0.3 & 327 & 593 \\
 & $\sii$ 1304 & 16 & 1.1 & 232 & 396 \\
 & $\siistar$ 1309 & - & - & - & - \\
\hline
J1025 & & & & & \\
 & $\sii$ 1190 & -147 & 1.0 & 332 & 608 \\
 & $\sii$ 1193 & -147 & 1.1 & 371 & 674 \\
 & $\siistar$ 1195 & 60 & -0.3 & 184 & 338 \\
 & $\siistar$ 1197 & -35 & -0.2 & 246 & 450 \\
 & $\sii$ 1260 & -139 & 1.3 & 379 & 690 \\
 & $\siistar$ 1265 & 18 & -0.2 & 178 & 323 \\
 & $\sii$ 1304 & -130 & 0.4 & 207 & 377 \\
 & $\siistar$ 1309 & 10 & -0.5 & 398 & 727 \\
\hline
J1112 & & & & & \\
 & $\sii$ 1190 & -94 & 2.4 & 657 & 1194 \\
 & $\sii$ 1193 & -175 & 1.3 & 409 & 748 \\
 & $\siistar$ 1195 & 0 & 0.0 & 0 & 0 \\
 & $\siistar$ 1197 & 0 & 0.0 & 0 & 0 \\
 & $\sii$ 1260 & -151 & 1.6 & 381 & 696 \\
 & $\siistar$ 1265 & 43 & -0.4 & 221 & 407 \\
 & $\sii$ 1304 & -71 & 0.5 & 216 & 393 \\
 & $\siistar$ 1309 & -43 & -0.3 & 250 & 458 \\
\hline
J1113 & & & & & \\
 & $\sii$ 1190 & -236 & 1.0 & 363 & 662 \\
 & $\sii$ 1193 & -118 & 1.4 & 482 & 876 \\
 & $\siistar$ 1195 & 0 & 0.0 & 0 & 0 \\
 & $\siistar$ 1197 & 0 & 0.0 & 0 & 0 \\
 & $\sii$ 1260 & -84 & 1.1 & 354 & 647 \\
 & $\siistar$ 1265 & 0 & 0.0 & 0 & 0 \\
 & $\sii$ 1304 & -39 & 0.2 & 184 & 338 \\
 & $\siistar$ 1309 & -133 & -0.2 & 150 & 274 \\
\hline
J1144 & & & & & \\
 & $\sii$ 1190 & -196 & 1.8 & 439 & 798 \\
 & $\sii$ 1193 & -181 & 1.7 & 427 & 778 \\
 & $\siistar$ 1195 & 4 & -0.3 & 186 & 340 \\
 & $\siistar$ 1197 & -61 & -0.3 & 210 & 385 \\
 & $\sii$ 1260 & -151 & 2.5 & 570 & 1042 \\
 & $\siistar$ 1265 & 41 & -0.3 & 130 & 239 \\
 & $\sii$ 1304 & -158 & 0.8 & 285 & 517 \\
 & $\siistar$ 1309 & -88 & -0.6 & 336 & 610 \\
\hline
J1414 & & & & & \\
 & $\sii$ 1190 & -78 & 1.7 & 389 & 712 \\
 & $\sii$ 1193 & -65 & 1.7 & 433 & 789 \\
 & $\siistar$ 1195 & 19 & -0.1 & 159 & 289 \\
 & $\siistar$ 1197 & 0 & 0.0 & 0 & 0 \\
 & $\sii$ 1260 & -62 & 2.0 & 383 & 699 \\
 & $\siistar$ 1265 & 0 & 0.0 & 0 & 0 \\
 & $\sii$ 1304 & -14 & 1.2 & 260 & 476 \\
 & $\siistar$ 1309 & 12 & -0.7 & 329 & 596 \\
\hline
J1416 & & & & & \\
 & $\sii$ 1190 & -39 & 3.2 & 968 & 1768 \\
 & $\sii$ 1193 & 33 & 1.2 & 426 & 778 \\
 & $\siistar$ 1195 & 0 & 0.0 & 0 & 0 \\
 & $\siistar$ 1197 & 0 & 0.0 & 0 & 0 \\
 & $\sii$ 1260 & -49 & 2.7 & 701 & 1279 \\
 & $\siistar$ 1265 & 61 & -0.3 & 245 & 447 \\
 & $\sii$ 1304 & -36 & 0.4 & 193 & 351 \\
 & $\siistar$ 1309 & -49 & -0.2 & 297 & 542 \\\hline
J1428 & & & & & \\
 & $\sii$ 1190 & -181 & 0.6 & 227 & 416 \\
 & $\sii$ 1193 & -152 & 0.7 & 273 & 496 \\
 & $\siistar$ 1195 & 167 & -0.7 & 485 & 881 \\
 & $\siistar$ 1197 & -41 & -0.3 & 184 & 334 \\
 & $\sii$ 1260 & -152 & 0.8 & 271 & 495 \\
 & $\siistar$ 1265 & 1 & -0.4 & 201 & 366 \\
 & $\sii$ 1304 & -411 & 0.5 & 226 & 412 \\
 & $\siistar$ 1309 & -79 & -0.2 & 135 & 249 \\
\hline
J1429 & & & & & \\
 & $\sii$ 1190 & -179 & 1.0 & 319 & 580 \\
 & $\sii$ 1193 & -147 & 0.9 & 316 & 574 \\
 & $\siistar$ 1195 & 2 & -0.5 & 171 & 313 \\
 & $\siistar$ 1197 & 38 & -0.8 & 251 & 455 \\
 & $\sii$ 1260 & -139 & 1.1 & 352 & 640 \\
 & $\siistar$ 1265 & 30 & -0.5 & 203 & 368 \\
 & $\sii$ 1304 & -100 & 0.4 & 182 & 336 \\
 & $\siistar$ 1309 & 16 & -0.4 & 205 & 373 \\
\hline
J1521 & & & & & \\
 & $\sii$ 1190 & - & - & - & - \\
 & $\sii$ 1193 & - & - & - & - \\
 & $\siistar$ 1195 & - & - & - & - \\
 & $\siistar$ 1197 & 0 & 0.0 & 0 & 0 \\
 & $\sii$ 1260 & - & - & - & - \\
 & $\siistar$ 1265 & - & - & - & - \\
 & $\sii$ 1304 & -69 & 0.5 & 224 & 404 \\
 & $\siistar$ 1309 & -86 & -0.1 & 164 & 302 \\
\hline
J1525 & & & & & \\
 & $\sii$ 1190 & -375 & 1.5 & 471 & 860 \\
 & $\sii$ 1193 & -358 & 1.5 & 498 & 905 \\
 & $\siistar$ 1195 & 1 & -0.1 & 121 & 221 \\
 & $\siistar$ 1197 & 0 & 0.0 & 0 & 0 \\
 & $\sii$ 1260 & -343 & 1.6 & 482 & 879 \\
 & $\siistar$ 1265 & 102 & -0.4 & 290 & 529 \\
 & $\sii$ 1304 & - & - & - & - \\
 & $\siistar$ 1309 & - & - & - & - \\
\hline
J1612 & & & & & \\
 & $\sii$ 1190 & -410 & 1.9 & 636 & 1160 \\
 & $\sii$ 1193 & -354 & 1.8 & 573 & 1044 \\
 & $\siistar$ 1195 & 270 & -0.5 & 459 & 836 \\
 & $\siistar$ 1197 & 0 & 0.0 & 0 & 0 \\
 & $\sii$ 1260 & -395 & 2.5 & 687 & 1249 \\
 & $\siistar$ 1265 & 3 & -0.3 & 263 & 480 \\
 & $\sii$ 1304 & -369 & 0.7 & 274 & 504 \\
 & $\siistar$ 1309 & -70 & -0.1 & 134 & 246 \\
\enddata
\tablecomments{``-" indicates that data is not valid due to MW absorption or detector gap. All measurements can be assumed to have errors on the order of 10-15\% dominated by systematics in the polynomial fit to the continuum and subtraction of the SB99 models.}
\end{deluxetable}

$\quad$

\end{document}